\documentclass[prc,twocolumn,superscriptaddress,showpacs,amssymb,amsmath,amsfonts,aps]{revtex4}
\usepackage{graphicx}
\usepackage{dcolumn}
\usepackage{epsfig}
\usepackage{latexsym}
\usepackage{amsmath} 
\usepackage[dvips]{color}
\begin{document}

\date{\today}
\title{\large New Results from the Studies of the $N(1440)1/2^+$, $N(1520)3/2^-$, and $\Delta(1620)1/2^-$ 
Resonances in Exclusive $ep \to e'p' \pi^+ \pi^-$ Electroproduction with the CLAS Detector  \\ }
\newcommand*{\JLAB}{Thomas Jefferson National Accelerator Facility, Newport News, Virginia 23606}
\newcommand*{\JLABindex}{33}
\affiliation{\JLAB}
\newcommand*{\MSU}{Skobeltsyn Nuclear Physics Institute and Physics Department at Moscow State University, 
119899 Moscow, Russia}
\newcommand*{\MSUindex}{31}
\affiliation{\MSU}
\newcommand*{\SCAROLINA}{University of South Carolina, Columbia, South Carolina 29208}
\newcommand*{\SCAROLINAindex}{32}
\affiliation{\SCAROLINA}
\newcommand*{\OHIOU}{Ohio University, Athens, Ohio  45701}
\newcommand*{\OHIOUindex}{27}
\affiliation{\OHIOU}

\author {V.I.~Mokeev} 
\altaffiliation[Current address: ]{\JLAB}
\affiliation{\JLAB}
\affiliation{\MSU}
\author {V.D.~Burkert} 
\affiliation{\JLAB}
\author {D.S.~Carman} 
\affiliation{\JLAB}
\author {L.~Elouadrhiri} 
\affiliation{\JLAB}
\author {G.V.~Fedotov} 
\affiliation{\SCAROLINA}
\affiliation{\MSU}
\author {E.N.~Golovatch} 
\affiliation{\MSU}
\author {R.W.~Gothe} 
\affiliation{\SCAROLINA}
\author {K.~Hicks} 
\affiliation{\OHIOU}
\author {B.S.~Ishkhanov} 
\affiliation{\MSU}
\author {E.L.~Isupov} 
\affiliation{\MSU}
\author {Iu.~Skorodumina} 
\affiliation{\SCAROLINA}
\affiliation{\MSU}



\begin{abstract}
{ The transition helicity amplitudes from the proton ground state to the  $N(1440)1/2^+$,  $N(1520)3/2^-$, 
and $\Delta(1620)1/2^-$  resonances ($\gamma_vpN^*$ electrocouplings) were determined from the analysis of 
nine independent one-fold differential $\pi^+ \pi^- p$ electroproduction cross sections off a proton 
target, taken with CLAS at photon virtualities 0.5~GeV$^2$ $< Q^2 <$ 1.5~GeV$^2$. The phenomenological 
reaction model employed for separation of the resonant and non-resonant contributions to this exclusive 
channel was further developed. The $N(1440)1/2^+$, $N(1520)3/2^-$, and $\Delta(1620)1/2^-$ electrocouplings 
were obtained from the resonant amplitudes of charged double-pion electroproduction off the proton in the 
aforementioned area of photon virtualities for the first time. Consistent results on $\gamma_vpN^*$ 
electrocouplings available from independent analyses of several $W$-intervals with different non-resonant 
contributions offer clear evidence for the reliable extraction of these fundamental quantities. These studies 
also improved the knowledge on hadronic branching ratios for the $N(1440)1/2^+$, $N(1520)3/2^-$, and 
$\Delta(1620)1/2^-$ decays to the $\pi \Delta$ and $\rho N$ final states. These new results provide a substantial 
impact on the QCD-based approaches that describe the $N^*$ structure and demonstrate the capability to explore 
fundamental ingredients of the non-perturbative strong interaction that are behind the excited nucleon state 
formation.} 
\end{abstract}

\pacs{ 11.55.Fv, 13.40.Gp, 13.60.Le, 14.20.Gk  }

\maketitle

\section{Introduction}
\label{intro}

The studies of transition amplitudes from the ground to excited nucleon states off the proton ($\gamma_vpN^*$ 
electrocouplings) offer insight into the $N^*$ structure and allow the exploration of the non-perturbative strong 
interaction mechanisms that are responsible for the resonance formation as relativistic bound systems of 
quarks and gluons~\cite{Bu12,Az13,Cr14}. The data on $\gamma_vpN^*$ electrocouplings represent the only 
source of information on different manifestations of the non-perturbative strong interaction in the generation 
of excited nucleon states of different quantum numbers. 

The CLAS detector at Jefferson Lab is a unique large-acceptance instrument designed for the comprehensive 
exploration of exclusive meson electroproduction. It offers excellent opportunities for the study of 
electroexcitation of nucleon resonances in detail and with precision. The CLAS detector has provided the 
dominant portion of all data on meson electroproduction in the resonance excitation region. The studies of 
transition helicity amplitudes from the proton ground state to its excited states represent a key aspect of
the $N^*$ program with CLAS~\cite{Mo11,Bu12,Mo14,Bu14}.

Meson-electroproduction data off nucleons in the $N^*$ region obtained with CLAS open up an opportunity to 
determine the $Q^2$-evolution of the $\gamma_vNN^*$ electrocouplings in a combined analysis of various 
meson-electroproduction channels for the first time. A variety of measurements of $\pi^+n$ and $\pi^0p$ 
single-pion electroproduction off the proton, including polarization measurements, have been performed with CLAS 
in the range of $Q^2$ from 0.16 to 6~GeV$^2$ and in the area of invariant masses of the final hadrons 
$W < 2.0$~GeV~\cite{Joo,Joo2,Joo3,Egiyan,Ungaro,Smith,Park,Biselli,Park15}. Exclusive $\eta p$ electroproduction 
off the proton was studied with CLAS for $W < 2.3$~GeV and $Q^2$ from 0.2 to 3.1~GeV$^2$~\cite{Den07}. Furthermore, 
differential cross section and polarization asymmetries in exclusive $KY$ electroproduction channels were 
obtained for $W$ from threshold to 2.6~GeV and for $Q^2 < 5.4$~GeV$^2$~\cite{Car12,Car08,Car07,Gab14,Car09,Car03}. 
These experiments were complemented by the measurements of nine independent $\pi^+\pi^-p$ electroproduction 
cross sections off the proton. The data on charged double-pion electroproduction covered the area of $W < 1.6$~GeV 
at photon virtualities from 0.25 to 0.55~GeV$^2$~\cite{Fe09}. They are also available from earlier measurements 
with CLAS for $W$ from 1.40~GeV to 2.10~GeV and 0.5~GeV$^2$ $< Q^2 <$ 1.5~GeV$^2$~\cite{Ri03}.

The electroexcitation amplitudes for the low-lying resonances $\Delta(1232)3/2^+$, $N(1440)1/2^+$, $N(1520)3/2^-$,
and $N(1535)1/2^-$ were determined over a wide range of $Q^2$ in a comprehensive analysis of JLab-CLAS data 
on differential cross sections, longitudinally polarized beam asymmetries, and longitudinal target and 
beam-target asymmetries~\cite{Az09}. Recently $\gamma_vNN^*$ electrocouplings of several higher-lying nucleon 
resonances: $N(1675)5/2^-$, $N(1680)5/2^+$, and $N(1710)1/2^+$ have become available for the first time for 
1.5~GeV$^2$ $< Q^2 <$ 4.5~GeV$^2$ from analysis of exclusive $\pi^+n$ electroproduction off the proton~\cite{Park15}. 
Electrocouplings for the $N(1440)1/2^+$ and $N(1520)3/2^-$ resonances for $Q^2 < 0.6$~GeV$^2$ have been 
determined from the data on exclusive $\pi^+\pi^-p$ electroproduction off the proton~\cite{Mo12} together with the 
preliminary results on the electrocouplings of several resonances in the mass range from 1.6~GeV to 1.75~GeV 
available for the first time from this exclusive channel at 0.5~GeV$^2$ $< Q^2 <$ 1.5~GeV$^2$~\cite{Az13,Mo14}. 
The CLAS results on the $\gamma_vpN^*$ electrocouplings~\cite{Az09,Bu12,Mo12,Mo14,Park15} have had a stimulating 
impact on the theory of the excited nucleon state structure, in particular, on the QCD-based approaches.

The light cone sum rule (LCSR) approach~\cite{Br09,Br15} for the first time provided access to the quark 
distribution amplitudes (DA) inside the $N(1535)1/2^-$ resonance from analysis of the CLAS results on the
$\gamma_vpN^*$ electrocouplings of this state~\cite{Az09}. Confronting the quark DA's of excited nucleon states 
determined from the experimental results on the $\gamma_vpN^*$ electrocouplings to the LQCD expectations, makes 
it possible to explore the emergence of the resonance structure starting from the QCD Lagrangian. The moments 
of the $N(1535)1/2^-$ quark DA's derived from the CLAS data are consistent with the LQCD expectations~\cite{Br14}. 

The Dyson-Schwinger Equations of QCD (DSEQCD) provide a conceptually different avenue for relating the 
$\gamma_vpN^*$ electrocouplings to the fundamental QCD Lagrangian~\cite{Cr15,Cr15b,Cr15a,Eich12}. The DSEQCD 
approach allows for the evaluation of the contribution of the three bound dressed quarks, the so-called quark 
core, to the structure of excited nucleon states starting from the QCD Lagrangian. A successful description of 
the nucleon elastic form factors and the CLAS results on the $N \to \Delta$, $N \to N(1440)1/2^+$ transition 
electromagnetic form factors~\cite{Az09,Mo12,Bu12,Mo14} at photon virtualities $Q^2 > 2.0$~GeV$^2$ has been 
recently achieved within the DSEQCD framework~\cite{Cr13,Cr15,Cr15a}. However, at smaller photon virtualities 
$Q^2 < 1.0$~GeV$^2$, the DSEQCD approach failed to describe the CLAS results~\cite{Az09,Mo12,Mo14} on the 
$\Delta(1232)3/2^+$ and $N(1440)1/2^+$ $\gamma_vpN^*$ electrocouplings~\cite{Cr15,Cr15a}. 

Furthermore, most quark models~\cite{San15,Met12,Ram14,Ram10} that take into account the contributions from 
quark degrees of freedom only, have substantial shortcomings in describing resonance electrocouplings at 
$Q^2 < 1.0$~GeV$^2$ even if they provide a reasonable description of the experimental results at higher photon 
virtualities. These are the indications for the contributions of degrees of freedom other than dressed quarks 
to the structure of excited nucleon states, contributions that become more relevant at small photon virtualities.

A successful description of the CLAS results on the $N(1440)1/2^+$ $\gamma_vpN^*$ electrocouplings 
\cite{Az09,Mo12,Mo14} has been recently achieved at small photon virtualities up to 0.5~GeV$^2$ within the 
framework of effective field theory employing pions, $\rho$ mesons, the nucleon, and the Roper $N(1440)1/2^+$
resonance as the effective degrees of freedom~\cite{Tia14}. This success emphasizes the importance of 
meson-baryon degrees of freedom for the structure of excited nucleon states at small photon virtualities. 
Furthermore, a general unitarity requirement imposes meson-baryon contributions to both resonance electromagnetic 
excitations and hadronic decay amplitudes. Studies of meson-baryon dressing contributions to the $\gamma_vpN^*$ 
electrocouplings from the global analysis of the $N\pi$ photo-, electro-, and hadroproduction data carried out 
by Argonne-Osaka Collaboration~\cite{Lee10,Lee091,Lee08} within the framework of a coupled channel approach, 
conclusively demonstrated the contributions from both meson-baryon and quark degrees of freedom to the structure 
of nucleon resonances.

Some quark models that have been developed~\cite{Az15,Az12,Ob14,Si14,Si09} take into account the contribution from 
both meson-baryon and quark degrees of freedom to the structure of excited nucleon states. Implementation of 
meson-baryon degrees of freedom allowed for a considerably improved description of the CLAS results on the
$N(1440)1/2^+$ and $N(1520)3/2^-$ $\gamma_vpN^*$ electrocouplings at photon virtualities $Q^2 < 1.0$~GeV$^2$,
while simultaneously retaining a good description of these results for $Q^2 > 2.0$~GeV$^2$.

Physics analyses of the CLAS results~\cite{Az09,Mo12,Mo14} on the $\gamma_vpN^*$ electrocouplings revealed the 
structure of excited nucleon states at photon virtualities $Q^2 < 5.0$~GeV$^2$ as a complex interplay between 
meson-baryon and quark degrees of freedom. The relative contributions from the meson-baryon cloud and the quark 
core are strongly dependent on the quantum numbers of the excited nucleons. Analyses of the $A_{1/2}$ 
electrocouplings of the $N(1520)3/2^-$ resonance~\cite{Lee08,Sa12} demonstrated that this amplitude is dominated 
by quark core contributions in the entire range of $Q^2 < 5.0$~GeV$^2$ measured by CLAS. However, the recent 
analysis~\cite{Az15} of the first CLAS results~\cite{Park15} on the $N(1675)5/2^-$ $\gamma_vpN^*$ electrocouplings 
suggested a dominance of the meson-baryon cloud. The experimental results on the $\gamma_vpN^*$ electrocouplings 
for all prominent resonances obtained in a wide range of photon virtualities are of particular importance in order to 
explore the contributions from different degrees of freedom to the resonance structure. 

Analyses of different exclusive channels are essential for a reliable extraction of the resonance parameters over 
the full spectrum of excited nucleon states. Currently the separation of the resonant and non-resonant parts of the 
electroproduction amplitudes can be done only within phenomenological reaction models. Therefore, the credibility of 
any resonance parameters extracted from the meson electroproduction data fit within the framework of any particular 
reaction model should be further examined. Non-resonant mechanisms in various meson-electroproduction channels are 
completely different, while the $\gamma_vNN^*$ electrocouplings are the same. Consistent results for the 
$\gamma_vpN^*$ electrocouplings of the  $N(1440)1/2^+$ and $N(1520)3/2^-$ resonances that were determined from 
independent analyses of the major meson electroproduction channels, $\pi^+n$, $\pi^0p$, and $\pi^+\pi^-p$, 
demonstrate that the extractions of these fundamental quantities are reliable as these different electroproduction 
channels have quite different backgrounds~\cite{Mo12}. Furthermore, this consistency also strongly suggests that the 
reaction models~\cite{Az09,Mo09,Mo12} developed for the description of these channels will provide a reliable 
evaluation of the $\gamma_vNN^*$ electrocouplings for analyzing either single- or charged double-pion 
electroproduction data. These models then make it possible to determine the electrocouplings for almost all 
well-established resonances that decay preferentially to the $N\pi$ and/or $N\pi\pi$ final states. The information 
on the $\gamma_vNN^*$ electrocouplings available from the exclusive charged double-pion electroproduction off the
proton is still rather limited and will be extended by the results of this paper.

In this paper we present the results on the electrocouplings of the $N(1440)1/2^+$, $N(1520)3/2^-$, and 
$\Delta(1620)1/2^-$ resonances at the photon virtualities 0.5~GeV$^2$ $< Q^2 <$ 1.5~GeV$^2$, obtained from the 
analysis of the CLAS data on $\pi^+\pi^-p$ electroproduction off the proton~\cite{Ri03}. The analysis was carried 
out employing the JM reaction model~\cite{Mo09,Mo12}, which has been further developed to provide a framework for 
the determination of the $\gamma_vpN^*$ electrocouplings from a combined fit of unpolarized differential cross 
sections in a broader kinematic area of $W$ and $Q^2$ in comparison with that covered in our previous studies
\cite{Mo09,Mo12}. This paper extends the available information on the $N(1440)1/2^+$ and $N(1520)3/2^-$ 
electrocouplings from the charged double-pion exclusive electroproduction off the proton and provides the first 
results on the electrocouplings and the hadronic decay widths of the $\Delta(1620)1/2^-$ resonance to the 
$\pi \Delta$ and $\rho N$ final states. 

The paper is organized as follows. In Section~\ref{genjm} we describe the JM reaction model employed for the 
extraction of the resonance parameters and the fitted experimental data. A special fitting procedure that allowed us 
to account for the experimental data and the reaction model uncertainties is presented in Section~\ref{fit}. The 
results on the $\gamma_vpN^*$ electrocouplings and the hadronic decays of the $N(1440)1/2^+$, $N(1520)3/2^-$, and 
$\Delta(1620)1/2^-$ resonances to the $\pi \Delta$ and $\rho N$ final states extracted from the CLAS data~\cite{Ri03} 
are presented in Section~\ref{nstarelectrocoupl}. Insights into the non-perturbative strong interaction mechanisms 
offered by our results and their impact on hadron structure theory are discussed in Section~\ref{impact} with 
summary and outlook in Section~\ref{concl}.

\section{Analysis Tools for Evaluation of the $\gamma_vpN^*$ Resonance Electrocouplings and Hadronic Decay Widths}
\label{genjm}

The $\gamma_vpN^*$ electrocouplings and hadronic decay widths of the $N(1440)1/2^+$, $N(1520)3/2^-$, and 
$\Delta(1620)1/2^-$ resonances to the $\pi \Delta$ and $\rho N$ final states were extracted from the fit of the 
CLAS charged double-pion electroproduction data~\cite{Ri03} at $W$ from 1.41~GeV to 1.66~GeV in three $Q^2$-bins 
centered at $Q^2$=0.65~GeV$^2$, 0.95~GeV$^2$, and 1.3~GeV$^2$. The JM meson-baryon model~\cite{Mo09,Mo12} was 
employed for the description of the measured observables in the $\gamma_v p \to  \pi^+\pi^-p$ exclusive channel. 
This model was successfully used in our previous extraction of the $N(1440)1/2^+$ and $N(1520)3/2^-$ resonance 
electrocouplings at smaller $Q^2 < 0.6$~GeV$^2$~\cite{Mo12} from the CLAS charged double-pion electroproduction 
data~\cite{Fe09} at $W < 1.57$~GeV. In our current analysis of the CLAS $\pi^+\pi^-p$ electroproduction data
\cite{Ri03}, the JM model was further developed in order to provide a data description in a wider area of $W$ 
from 1.40~GeV to 1.82~GeV and at photon virtualities $Q^2$ from 0.5~GeV$^2$ to 1.5~GeV$^2$. In this section we 
describe the differential cross sections we fit for the resonance parameter extraction. We also present the basic 
features of the JM model relevant for the extraction of the $\gamma_vpN^*$ electrocouplings from the data
\cite{Ri03}, focusing on the model updates needed to achieve a good description of the measured differential cross 
sections. 

\begin{figure}[htbp]
\begin{center}
\includegraphics[width=8cm]{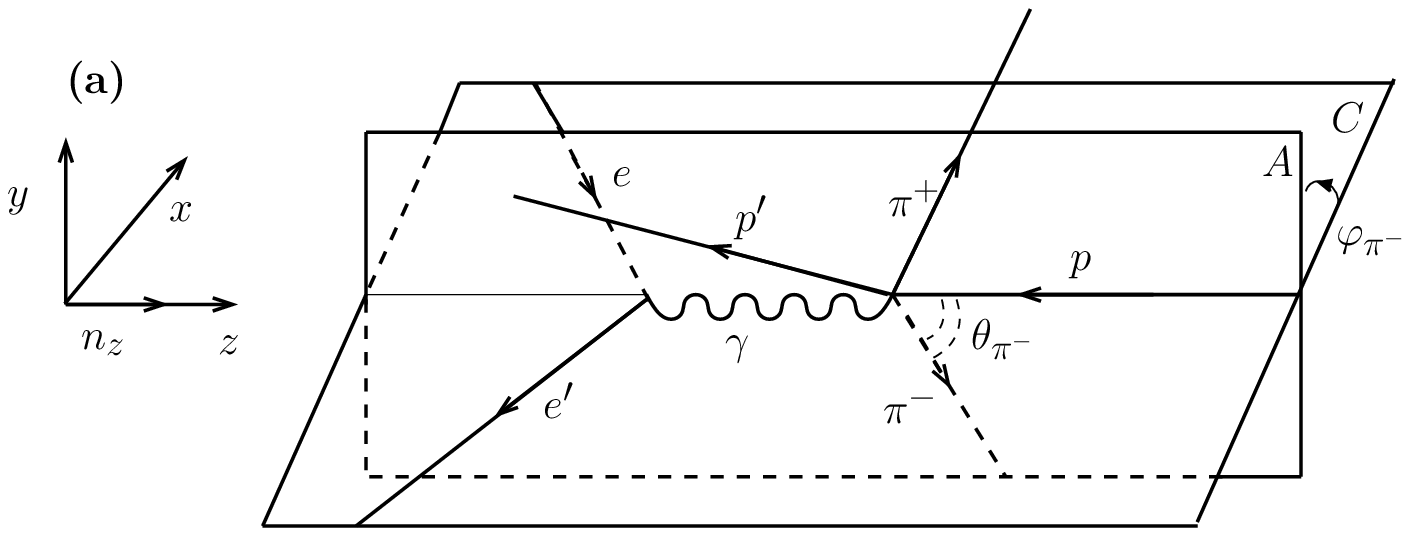}
\includegraphics[width=8cm]{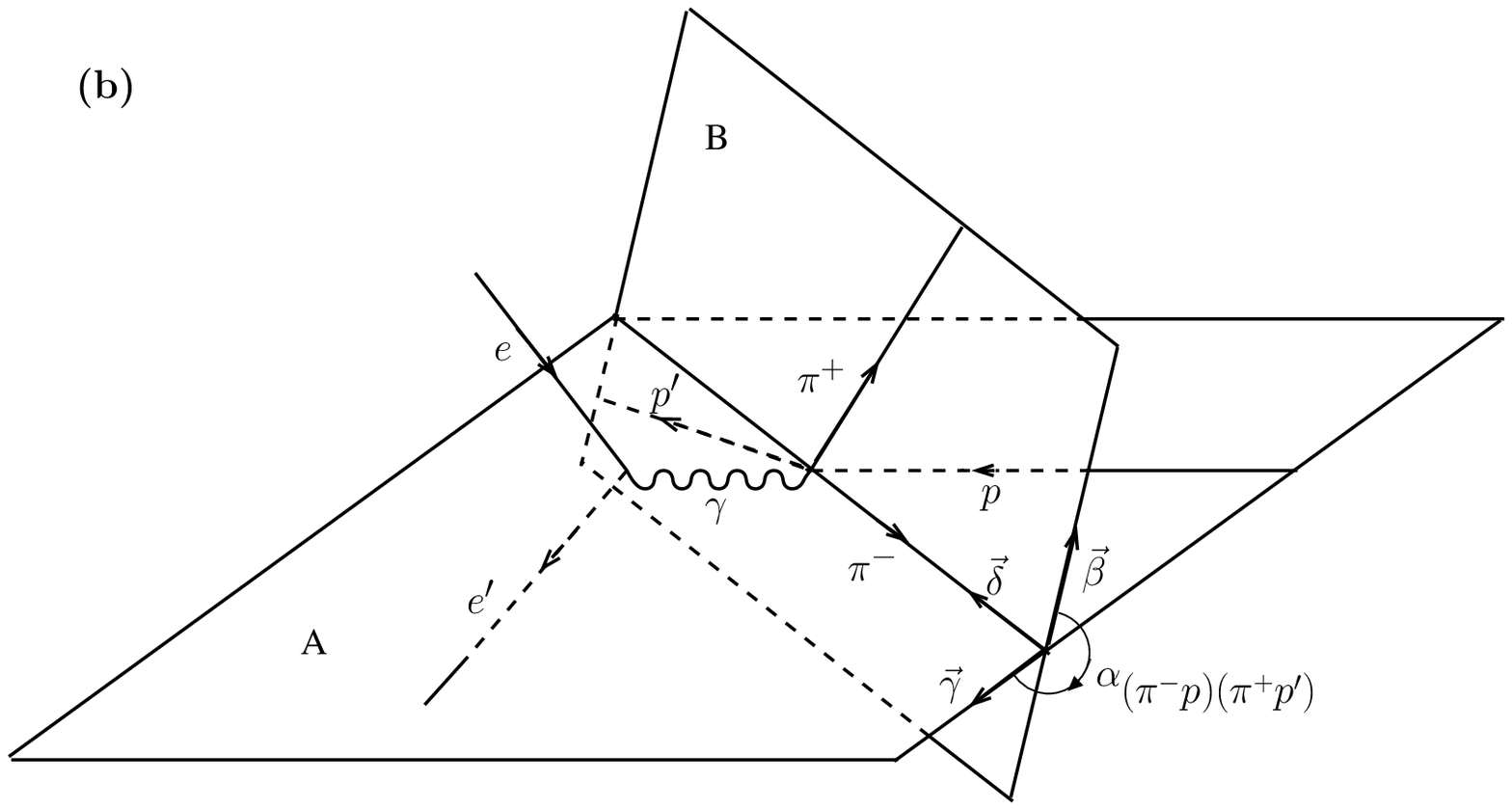}
\caption{Kinematic variables for the description of $e p \to e' p' \pi^+ \pi^-$ in the CM frame of 
the final-state hadrons corresponding to the explicit assignment presented in Section~\ref{kinxsect}. Panel (a) 
shows the $\pi^-$ spherical angles $\theta_{\pi^-}$ and $\varphi_{\pi^-}$. Plane C represents the electron 
scattering plane. Plane A is defined by the 3-momenta of the initial state proton and the final state $\pi^-$.  
Panel (b) shows the angle $\alpha_{[\pi^-p][\pi^+p']}$  between the two defined hadronic planes A and B or the 
plane B rotation angle around the axis aligned along the 3-momentum of the final $\pi^-$. Plane B is defined by 
the 3-momenta of the final state $\pi^+$ and $p'$. The unit vectors $\overline{\gamma}$ and $\overline{\beta}$ are 
normal to the $\pi^-$ three-momentum in the planes A and B, respectively.} 
\label{kinematic}
\end{center}
\end{figure}

\subsection{Differential Cross Sections and Kinematic Variables}
\label{kinxsect}

At a given invariant mass $W$ and photon virtuality $Q^2$, the $\gamma_v p \to \pi^+\pi^-p$ reaction
can be fully described as a five-fold differential cross section $d^5\sigma/d^5\tau$, where $d^5\tau$ is 
the differential of the five independent variables in the center-of-mass (CM) frame of the final
$\pi^+ \pi^- p$ state. There are many possible choices~\cite{Byc} of the five independent variables. After 
defining $M_{\pi^+p}$, $M_{\pi^-p}$, and $M_{\pi^+\pi^-}$ as invariant mass variables of the three possible 
two-particle pairs in the $\pi^+\pi^-p$ system, we adopt here the following assignment for the computation of the
five-fold differential cross section:\\
  \\
$d^5\tau=dM_{\pi^+p}dM_{\pi^+\pi^-}d\Omega_{\pi^-}d\alpha_{[\pi^-p][\pi^+p']}$, where $\Omega_{\pi^-}$ 
($\theta_{\pi^-}$, $\varphi_{\pi^-}$) are the final state $\pi^-$ spherical angles with respect to the 
direction of the virtual photon with the $\varphi_{\pi^-}$ defined as the angle between the hadronic plane A and 
the electron scattering plane C, see Fig~\ref{kinematic} (a), and $\alpha_{[\pi^-p][\pi^+p']}$ is the rotation 
angle of the plane B defined by the momenta of the final state $\pi^+p'$ around the axis defined by the final 
state $\pi^-$ momentum, see Fig.~\ref{kinematic} (b).\\ 
  \\
All frame-dependent variables are defined in the final hadron CM frame. The relations between the momenta of the 
final-state hadrons and the aforementioned five variables can be found in Ref.~\cite{Fe09}.

\begin{figure*}[htp]
\begin{center}
\includegraphics[width=11.5cm]{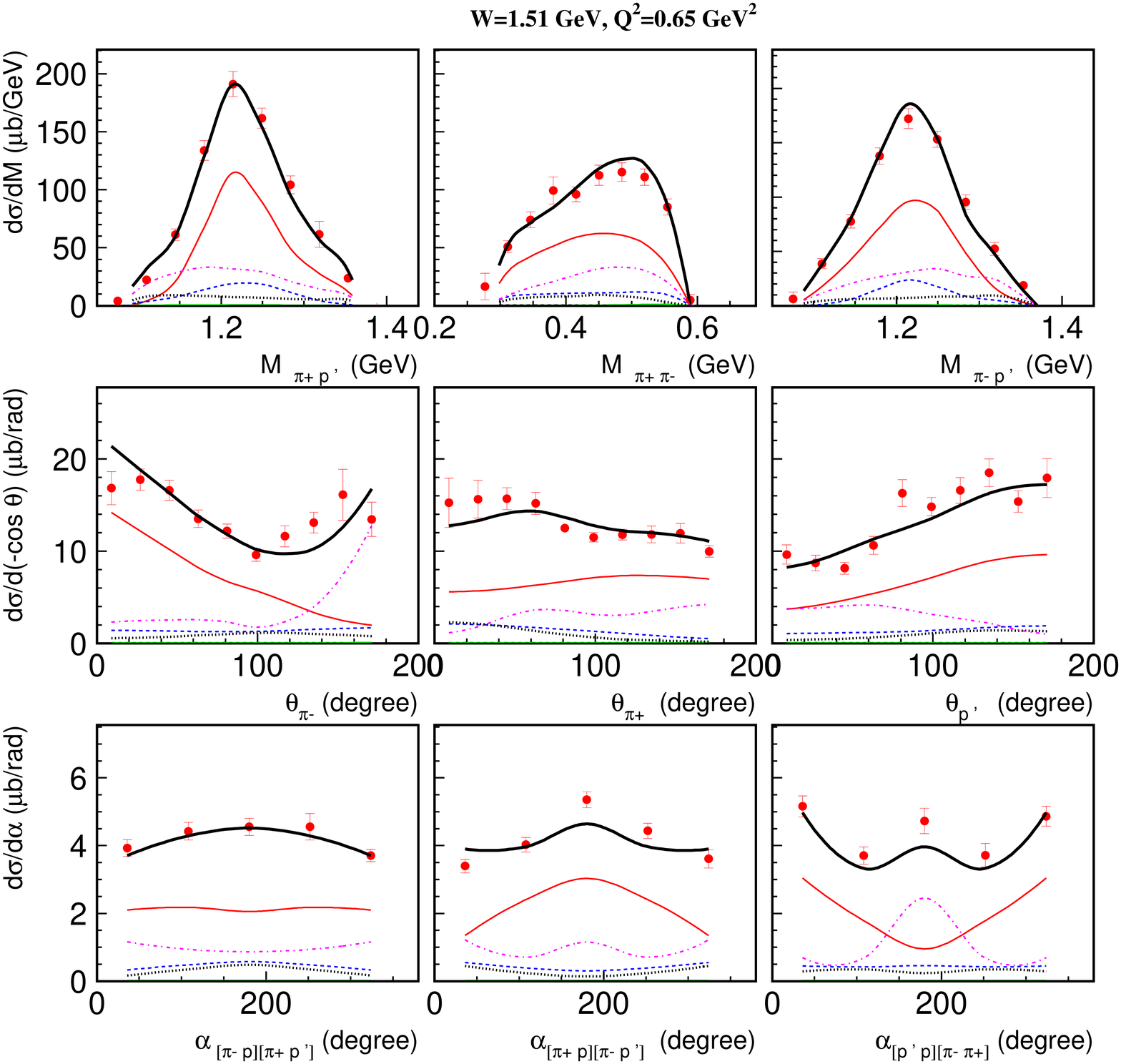}
\includegraphics[width=11.5cm]{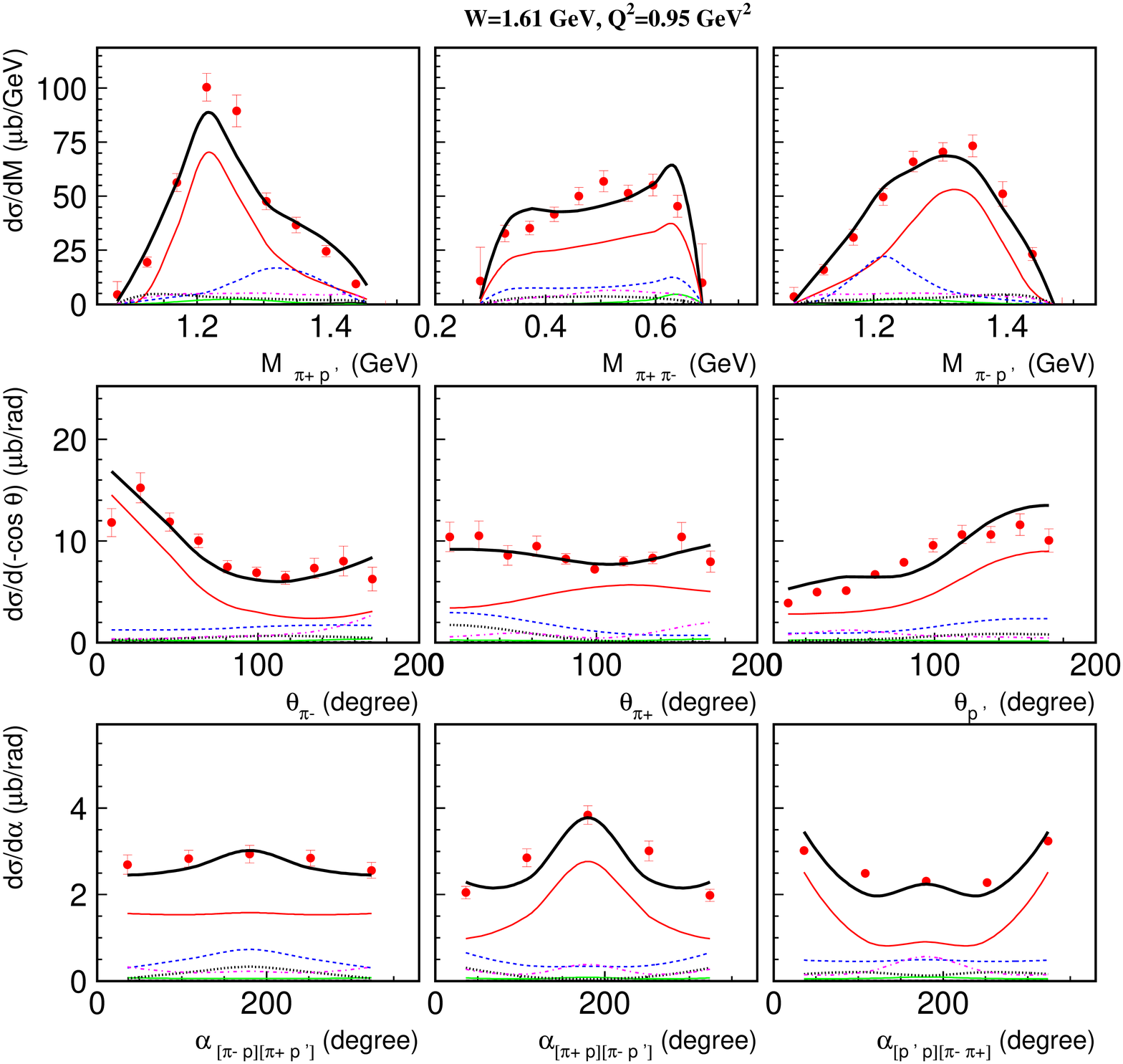}
\vspace{-0.1cm}
\caption{(Color Online) Description of the CLAS $ep \to e'p'\pi^+\pi^-$ data~\cite{Ri03} within the 
framework of the JM model~\cite{Mo09,Mo12} after implementation of the phases for the $2\pi$ direct production 
mechanisms discussed in Section~\ref{pipipmech} at $W$ = 1.51~GeV, $Q^2$=0.65~GeV$^2$ (top) and at $W$ = 1.61~GeV, 
$Q^2$=0.95~GeV$^2$ (bottom). Full model results are shown by the black thick solid lines together with the 
contributions from the isobar channels $\pi^-\Delta^{++}$ (thin red lines), $\pi^+ \Delta^0$ (blue dash-dotted 
lines), $\pi^+ D^0_{13}(1520)$ (black dotted lines), and the $2\pi$ direct production mechanisms (magenta 
dash-dotted lines). The contributions from other mechanisms described in Section~\ref{pipipmech} are comparable 
with the data uncertainties and are not shown in the plot.} 
\label{isochan}
\end{center}
\end{figure*}

\begin{figure*}[htp]
\begin{center}
\includegraphics[width=14.0cm]{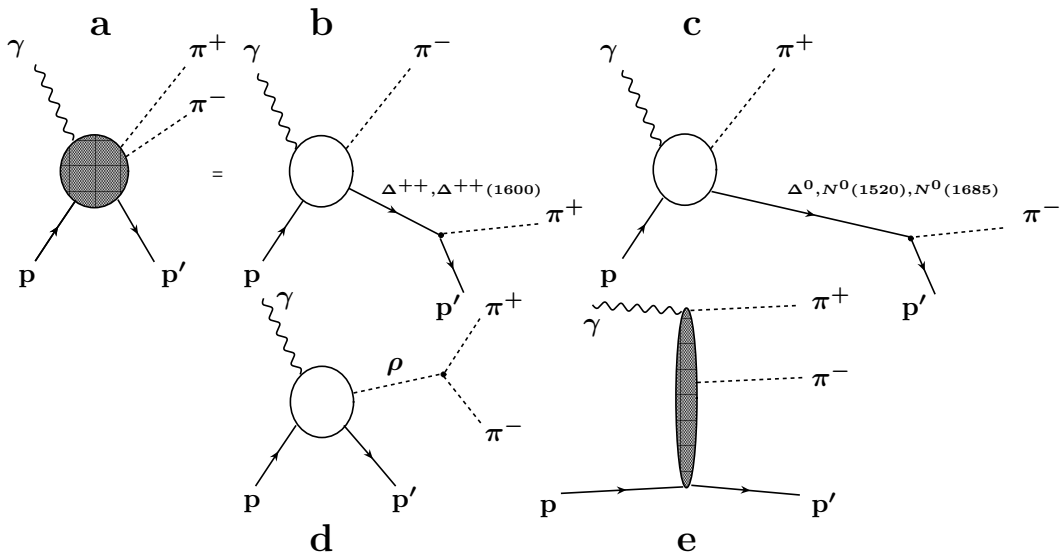}
\vspace{-0.1cm}
\caption{The $ep\to e'p' \pi^+\pi^-$ electroproduction mechanisms incorporated into the JM model
\cite{Mo09,Mo12}: a) full amplitude; b) $\pi^- \Delta^{++}$ and $\pi^-\Delta^{++}(1600)\frac{3}{2}^+$ isobar 
channels; c) $\pi^+ \Delta^0$, $\pi^+ N^0(1520)\frac{3}{2}^-$, and $\pi^+ N^0(1685)\frac{5}{2}^+$ isobar 
channels; d) $\rho p$ meson-baryon channel; e) the 2$\pi$ direct electroproduction mechanisms.}
\label{jmmech}
\end{center}
\end{figure*}

\begin{table}
\begin{center}
\begin{tabular}{|c|c|} \hline
                          & 0.5-0.8    \\
                          & 0.65 central value \\ \cline{2-2}
                          & 0.8-1.1   \\
$Q^2$ Intervals, GeV$^2$  & 0.95 central value \\ \cline{2-2}
                          & 1.1-1.5  \\
                          & 1.3 central value  \\ \hline
$W$ Intervals, GeV        & 1.41-1.71 \\
covered in each $Q^2$ bin & 13 bins   \\ \hline
\end{tabular}
\caption{Kinematic area covered in the fit of the CLAS $\pi^+\pi^-p$ electroproduction cross 
sections~\cite{Ri03} for the extraction of the resonance parameters.}
\label{wq2bins} 
\end{center}
\end{table}

\begin{table}
\begin{center}
\begin{tabular}{|c|c|c|}
\hline
One-fold differential & Interval  & Number of   \\
cross section         & Covered   & Bins        \\ \hline
$\frac{d\sigma}{dM_{\pi^+p}}$ ($\mu$b/GeV)         & $M_{\pi^+p_{min}}$-$M_{\pi^+p_{max}}$ & 10  \\
$\frac{d\sigma}{dM_{\pi^+\pi^-}}$ ($\mu$b/GeV)     & $M_{\pi^+\pi^-_{min}}$-$M_{\pi^+\pi^-_{max}}$ & 10  \\
$\frac{d\sigma}{dM_{\pi^-p}}$ ($\mu$b/GeV)         & $M_{\pi^-p_{min}}$-$M_{\pi^-p_{max}}$ & 10  \\
$\frac{d\sigma}{d(-cos(\theta_{\pi^-}))}$ ($\mu$b/rad)      & 0-180$^\circ$ & 10  \\
$\frac{d\sigma}{d(-cos(\theta_{\pi^+}))}$ ($\mu$b/rad)      & 0-180$^\circ$ & 10  \\
$\frac{d\sigma}{d(-cos(\theta_{p'}))}$ ($\mu$b/rad)        & 0-180$^\circ$ & 10  \\
$d\sigma/d\alpha_{[\pi^-p][\pi^+p']}$ ($\mu$b/rad) & 0-360$^\circ$ & 5\\
$d\sigma/d\alpha_{[\pi^+p][\pi^-p']}$ ($\mu$b/rad) & 0-360$^\circ$ & 5\\
$d\sigma/d\alpha_{[\pi^+\pi^-][p p']}$ ($\mu$b/rad)& 0-360$^\circ$ & 5\\ \hline
\end{tabular}
\caption{List of the fit one-fold differential cross sections measured with CLAS~\cite{Ri03} and the
binning over the kinematic variables. $M_{min_{i,j}}=M_{i}+M_{j}$, $M_{max_{i,j}}=W-M_{k}$, where 
$M_{i,j}$ and $M_k$ are the invariant masses of the final state hadron pair $i,j$, and the mass of the 
third final state hadron $k$, respectively.}
\label{1diffbins}
\end{center}
\end{table}

The $\pi^+\pi^-p$ electroproduction data have been collected in the bins of a seven-dimensional space. As 
mentioned above, five variables are needed to fully describe the final hadron kinematics, while to describe 
the initial state kinematics, two others variables, $W$ and $Q^2$, are required. The huge number of seven 
dimensional bins over the reaction phase space ($\approx$ 500,000 bins) does not allow us to use the
correlated multi-fold differential cross sections in the analysis of the data, where the statistics decrease 
drastically with increasing $Q^2$. More than half of the seven-dimensional phase-space bins of the final state
hadrons are not populated due to statistical limitations. This is a serious obstacle for any analysis method 
that employs information on the behavior of multi-fold differential cross sections. We therefore use the 
following one-fold differential cross sections in each bin of $W$ and $Q^2$ covered by the data:
\begin{itemize}
\item invariant mass distributions for the three pairs of the final state particles 
$d\sigma/dM_{\pi^+\pi^-} $, $d\sigma/dM_{\pi^+ p}$, and 
$d\sigma/dM_{\pi^- p}$;
\item angular distributions for the spherical angles of the three final state particles 
$d\sigma/d(-\cos\theta_{\pi^-})$, $d\sigma/d(-\cos\theta_{\pi^+})$, and $d\sigma/d(-\cos\theta_{p'})$ 
in the CM frame;
\item angular distributions for the three $\alpha$-angles determined in the CM frame: 
$d\sigma/d\alpha_{[\pi^-p][\pi^+p']}$,  $d\sigma/d\alpha_{[\pi^+p][\pi^-p']}$, and
$d\sigma/d\alpha_{[\pi^+\pi^-][p p']}$. The $d\sigma/d\alpha_{[\pi^+p][\pi^-p']}$ and
$d\sigma/d\alpha_{[\pi^+\pi^-][p p']}$ differential cross sections are defined analogously to 
$d\sigma/d\alpha_{[\pi^-p][\pi^+p']}$ describe above. More details on these observables can be found in 
Refs.~\cite{Fe09,Mo12}. 
\end{itemize} 
The one-fold differential cross sections were obtained by integrating the five-fold differential cross 
sections over the other four relevant kinematic variables of $d^5\tau$. However, the angular 
distributions for the spherical angles of the final state $\pi^+$ and $p$, as well as for the rotation angles 
around the axes along the momenta of these final state hadrons, cannot be obtained with the aforementioned 
$d^5\tau$, since this differential does not depend on these variables. Two other sets of $d^5\tau^{'}$ and 
$d^5\tau^{''}$  differentials are required, which contain $d\Omega_{\pi^+}d\alpha_{[\pi^+p][\pi^-p']}$ and 
$d\Omega_{p'}d\alpha_{[pp'][\pi^+\pi^-]}$, respectively, as described in Refs.~\cite{Fe09,Mo09}. The five-fold 
differential cross sections evaluated over the other two $d^5\tau^{'}$ and $d^5\tau^{''}$ differentials were 
computed from the five-fold differential cross section over the $d^5\tau$ differential detailed above by means 
of cross section interpolation. For each kinematic point in the five-dimensional phase space determined by the 
variables of the other two $d^5\tau^{'}$ and $d^5\tau^{''}$ differentials, the four-momenta of the three 
final state hadrons were computed, and from these values the five variables of the $d^5\tau$ were determined. 
The $d^5\sigma/d^5\tau$ cross sections were interpolated into this five-dimensional kinematic point.
All details related to the evaluation of the nine one-fold differential cross sections from the CLAS data 
on $\pi^+\pi^-p$ electroproduction off the proton can be found in Ref.~\cite{Fe09}. 

An example of the data analyzed in two particular bins of $W$ and $Q^2$ is shown in Fig.~\ref{isochan}. The 
kinematic area covered in our analysis and the data binning are summarized in Tables~\ref{wq2bins}
and \ref{1diffbins}.

\subsection{The Reaction Model for Extraction of the Resonance Parameters}
\label{pipipmech}

A phenomenological analysis of the CLAS $\pi^+\pi^- p$ electroproduction data~\cite{Ri03} was carried out for 
$W < 1.82$~GeV and at photon virtualities 0.5~GeV$^2$ $< Q^2 <$ 1.5~GeV$^2$. This work allows us to establish 
all essential mechanisms that contribute to the measured cross sections. The peaks in the invariant mass 
distributions provide evidence for the presence of the channels arising from 
$\gamma_v p \to Meson+Baryon \to \pi^+\pi^- p$ having an unstable baryon or meson in the intermediate state. 
Pronounced dependencies in the angular distributions further allow us to establish the relevant $t$-, $u$-, and 
$s$-channel exchanges. The mechanisms without pronounced kinematic dependencies are identified through 
examination of various differential cross sections, with their presence emerging from the correlation patterns.  
The phenomenological reaction model JM~\cite{Mo09,Mo12,Ri00,Az05} was developed with the primary objective to 
determine the $\gamma_vpN^*$ electrocouplings and the corresponding $\pi \Delta$ and $\rho N$ partial hadronic 
decay widths from fitting all measured observables in the $\pi^+\pi^- p$ electroproduction channel. The 
relationships between the $\pi^+\pi^-p$ electroproduction cross sections and the three-body production 
amplitudes employed in the JM model are given in Appendix D of Ref.~\cite{Mo09}. 

The amplitudes of  the $\gamma_v p \to \pi^+\pi^- p$ reaction are described in the JM model as a superposition 
of the $\pi^-\Delta^{++}$, $\pi^+\Delta^0$, $\rho p$, $\pi^+ D_{13}^0(1520)$, $\pi^+ F_{15}^0(1685)$, and
$\pi^- \Delta^{++}(1600)$ sub-channels with subsequent decays of the unstable hadrons to the final $\pi^+\pi^-p$ 
state, and additional direct 2$\pi$ production mechanisms, where the final $\pi^+\pi^- p$ state comes about 
without going through the intermediate process of forming unstable hadron states. The mechanisms incorporated 
into the JM model are shown in Fig.~\ref{jmmech}. 

The JM model incorporates contributions from all well-established $N^*$ states with listed in Table~\ref{nstlist} 
considering the resonant contributions only to $\pi \Delta$ and $\rho p$ sub-channels. We also have included the 
$3/2^+(1720)$ candidate state, whose existence is suggested in the analysis~\cite{Ri03} of the CLAS 
$\pi^+\pi^- p$ electroproduction data. In the versions of the JM model beginning in 2012~\cite{Mo12}, the 
resonant amplitudes are described by a unitarized Breit-Wigner ansatz as proposed in Ref.~\cite{Ait72}; the
model was modified to make it fully consistent with a relativistic Breit-Wigner parameterization of each individual 
$N^*$ state contributions in the JM model ~\cite{Ri00} that also accounts for the energy-dependent resonance hadronic 
decay widths. A unitarized Breit-Wigner ansatz accounts for the transition between the same and different resonances 
in the dressed resonant propagator, which makes the resonant amplitudes consistent with restrictions imposed by a 
general unitarity condition~\cite{Ait78,Lee08}. Quantum number conservation in the strong interaction allows 
for transitions between the pairs of $N^*$ states $N(1520)3/2^- \leftrightarrow N(1700)3/2^-$, 
$N(1535)1/2^- \leftrightarrow N(1650)1/2^-$, and $3/2^+(1720) \leftrightarrow N(1720)3/2^+$ incorporated into the 
JM model. We found that the use of the unitarized Breit-Wigner ansatz has a minor influence on the $\gamma_vNN^*$ 
electrocouplings, but it may substantially affect the $N^*$ hadronic decay widths determined from fits to the CLAS 
data.  

The non-resonant contributions to the $\pi \Delta$ sub-channels incorporate a minimal set of current-conserving 
Born terms~\cite{Mo09,Ri00}. They consist of $t$-channel pion exchange, $s$-channel nucleon exchange, $u$-channel 
$\Delta$ exchange, and contact terms. Non-resonant Born terms were reggeized and current conservation was 
preserved as proposed in Refs.~\cite{Gu97hy,Gu97by}. The initial- and final-state interactions in $\pi \Delta$ 
electroproduction are treated in an absorptive approximation, with the absorptive coefficients estimated from the 
data from $\pi N$ scattering~\cite{Ri00}. Non-resonant contributions to the $\pi \Delta$ sub-channels further 
include additional contact terms that have different Lorentz-invariant structures with respect to the contact 
terms in the sets of Born terms. These extra contact terms effectively account for non-resonant processes in the 
$\pi \Delta$ sub-channels beyond the Born terms, as well as for the final-state interaction effects that are 
beyond those taken into account by the absorptive approximation. Parameterizations of the extra contact terms in 
the $\pi \Delta$ sub-channels are given in Ref.~\cite{Mo09}. A phenomenological treatment of the initial and 
final state interactions~\cite{Ri00} along with the extra-contact-terms~\cite{Mo09} in the $\pi \Delta$ sub-channels
determined from fits to the data are important in order to account for the constraints imposed by unitarity on the 
non-resonant amplitudes of these sub-channels.

\begin{figure*}[htp]
\begin{center}
\includegraphics[width=7.0cm]{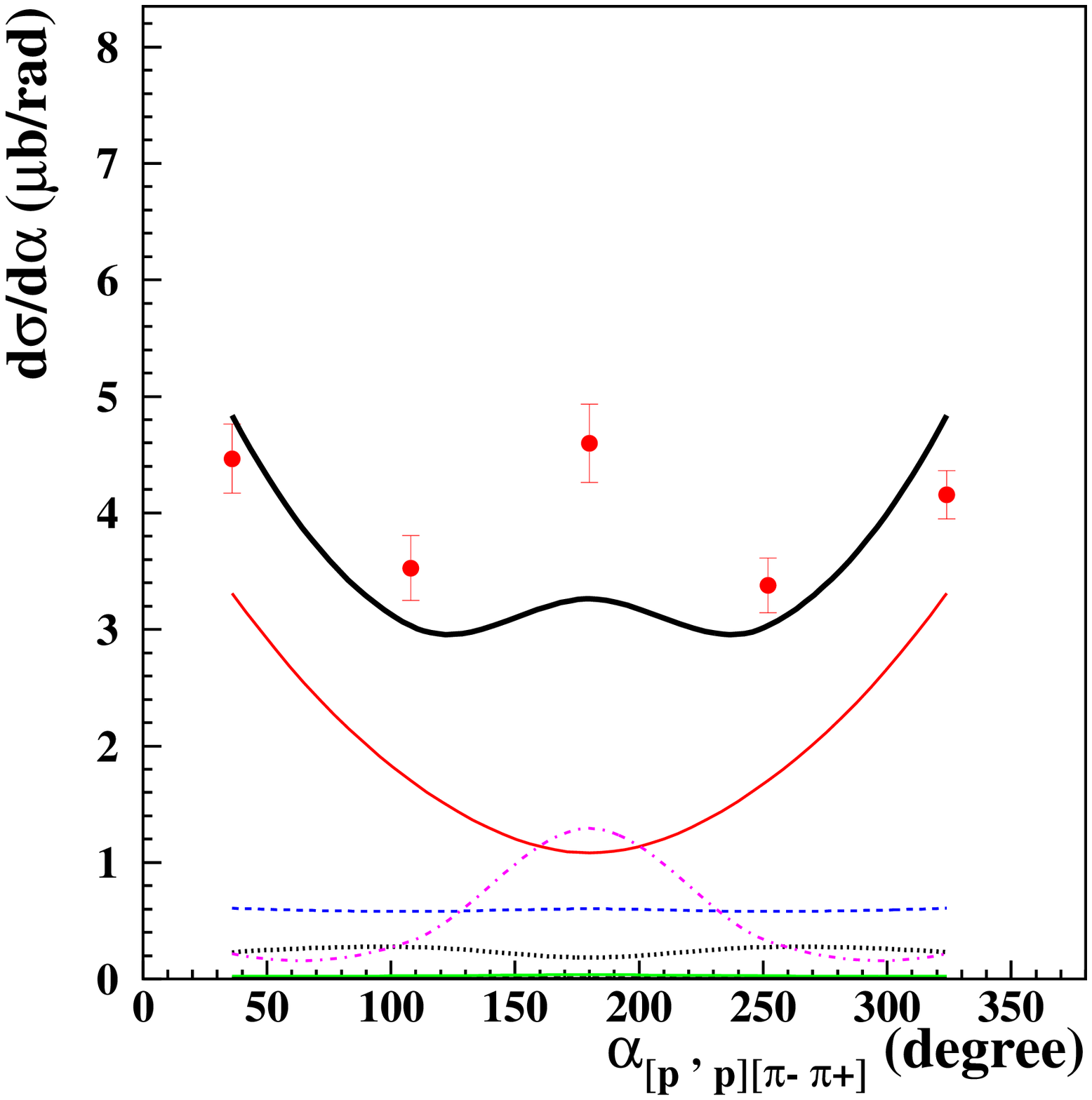}
\includegraphics[width=7.0cm]{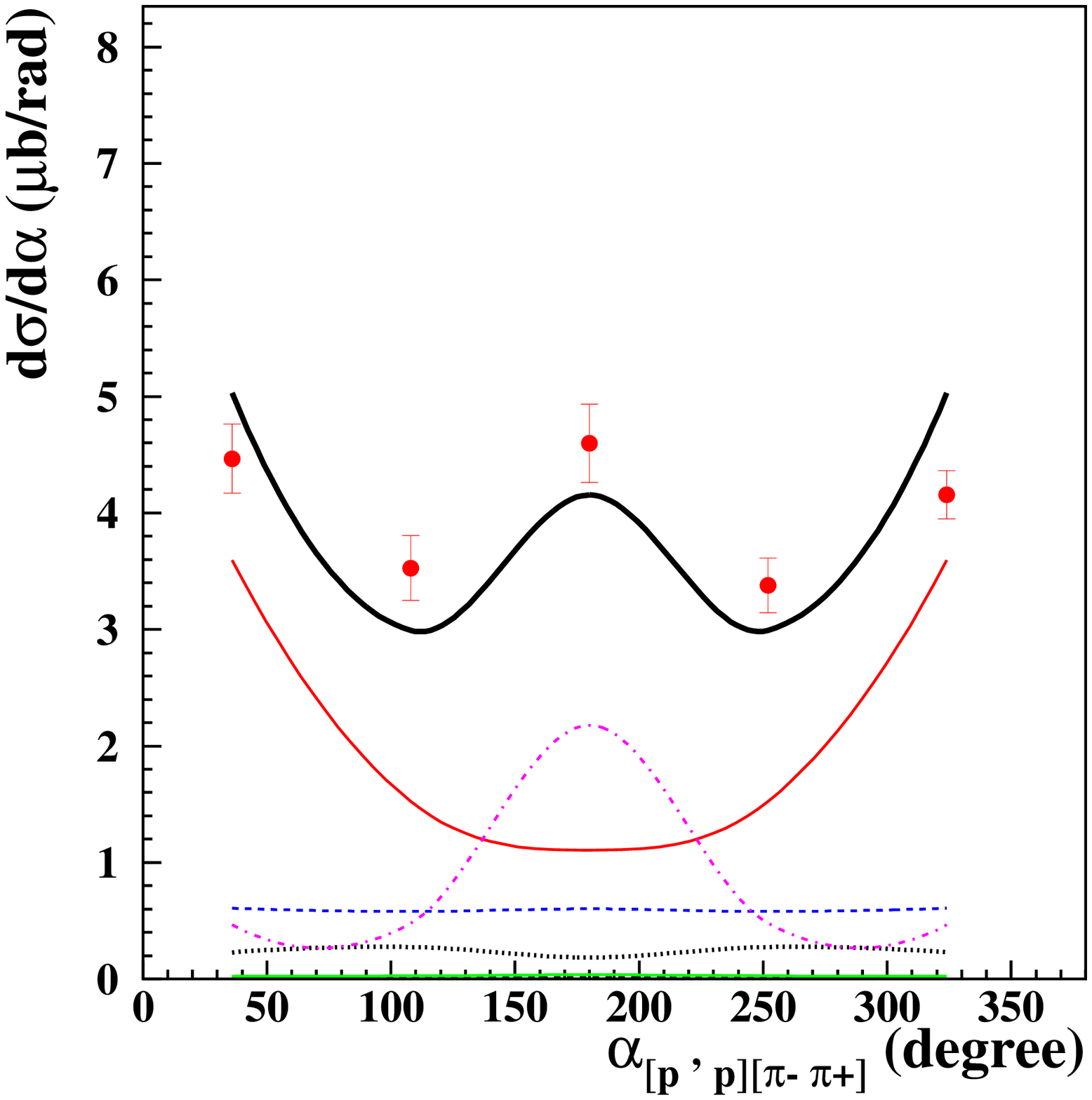}\\
\includegraphics[width=7.0cm]{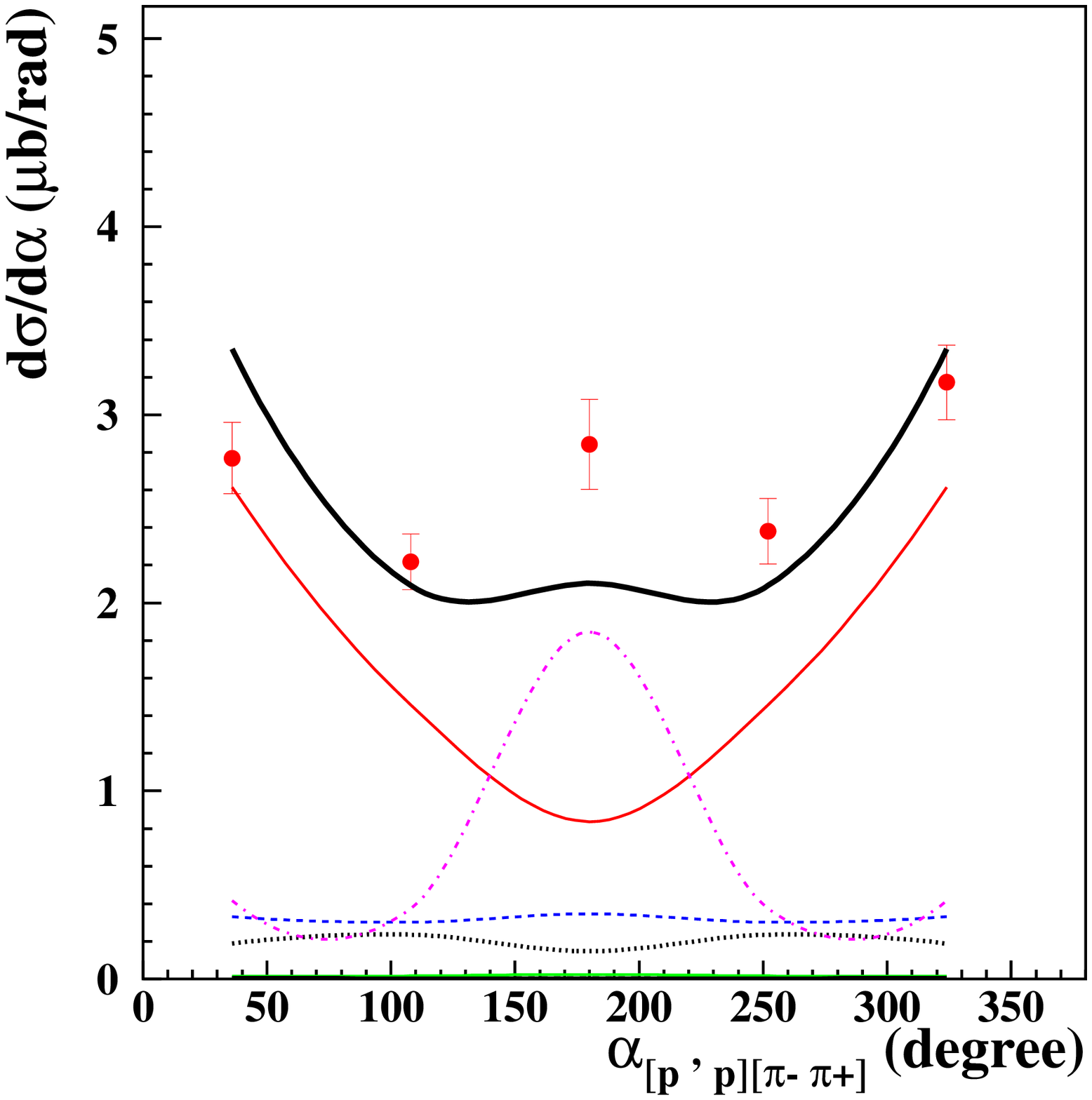}
\includegraphics[width=7.0cm]{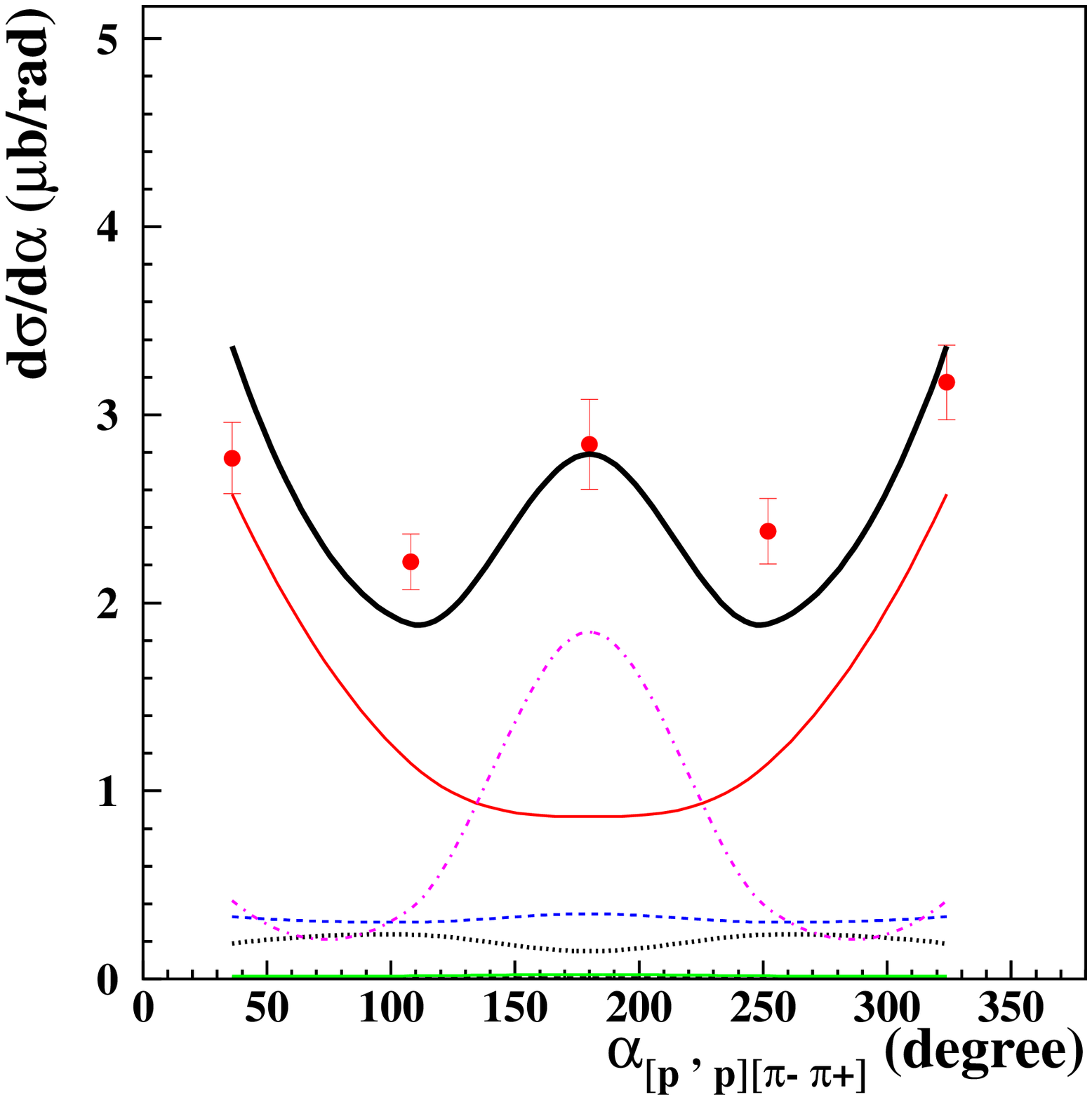}
\vspace{-0.1cm}
\caption{(Color Online) Manifestation of the direct 2$\pi$ electroproduction mechanism relative phases in the 
CLAS data~\cite{Ri03} on the angular distributions over the angle $\alpha_{[\pi^+\pi^-][p p']}$. The JM model 
results with the relative phases equal to zero are shown in the left column, while the computed cross sections with 
phases based on fits to the CLAS data~\cite{Ri03} are shown in the right column. The sample plots shown are for
$W$=1.56~GeV, $Q^2$=0.65~GeV$^2$ (top row) and $W$=1.54~GeV, $Q^2$=0.95~GeV$^2$ (bottom row). The curves for
the different contributing meson-baryon channels are the same as those shown in Fig.~\ref{isochan}.}
\label{pidirphase}
\end{center}
\end{figure*}

\begin{figure*}[htp]
\begin{center}
\includegraphics[width=5.5cm]{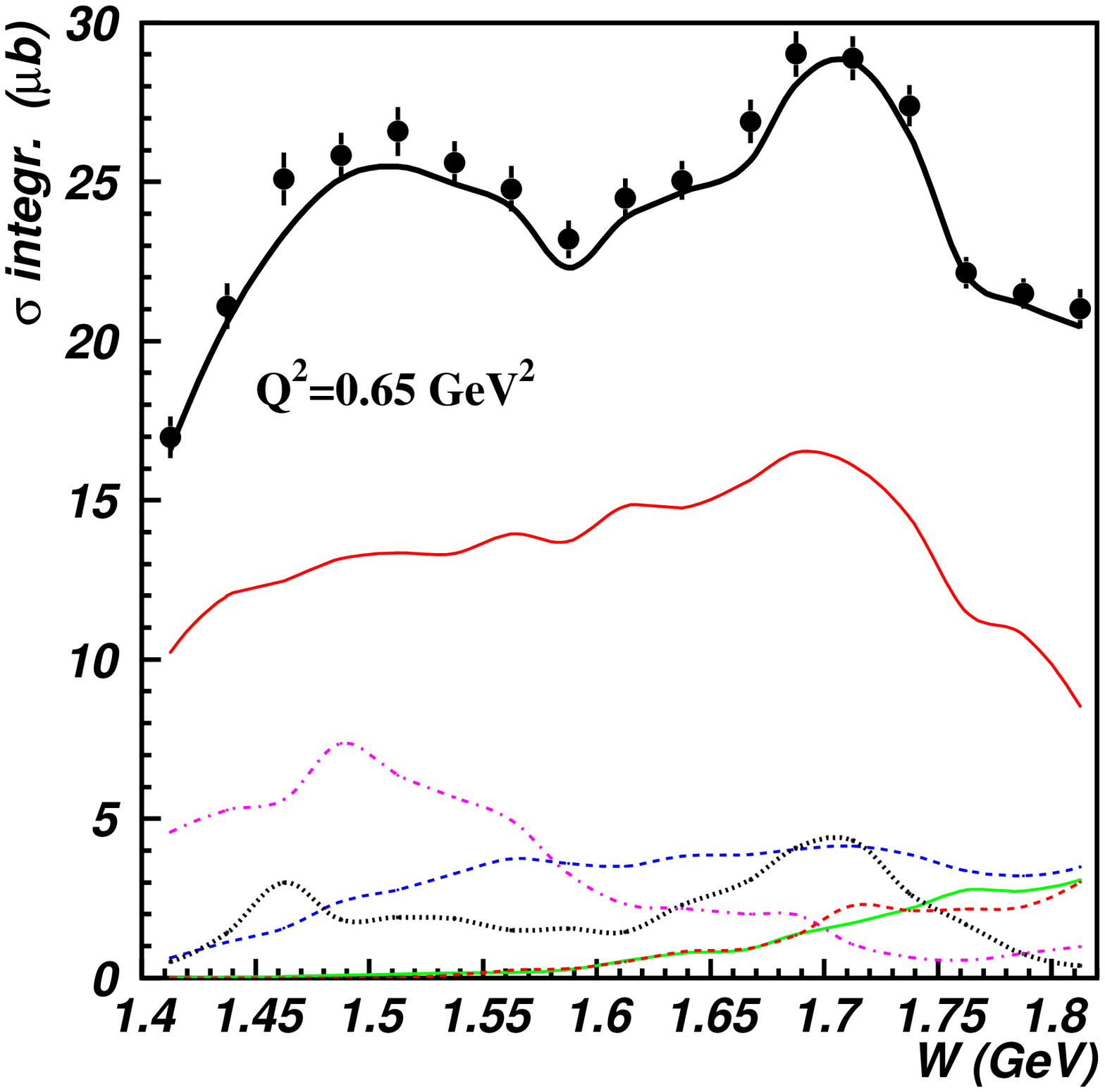}
\includegraphics[width=5.5cm]{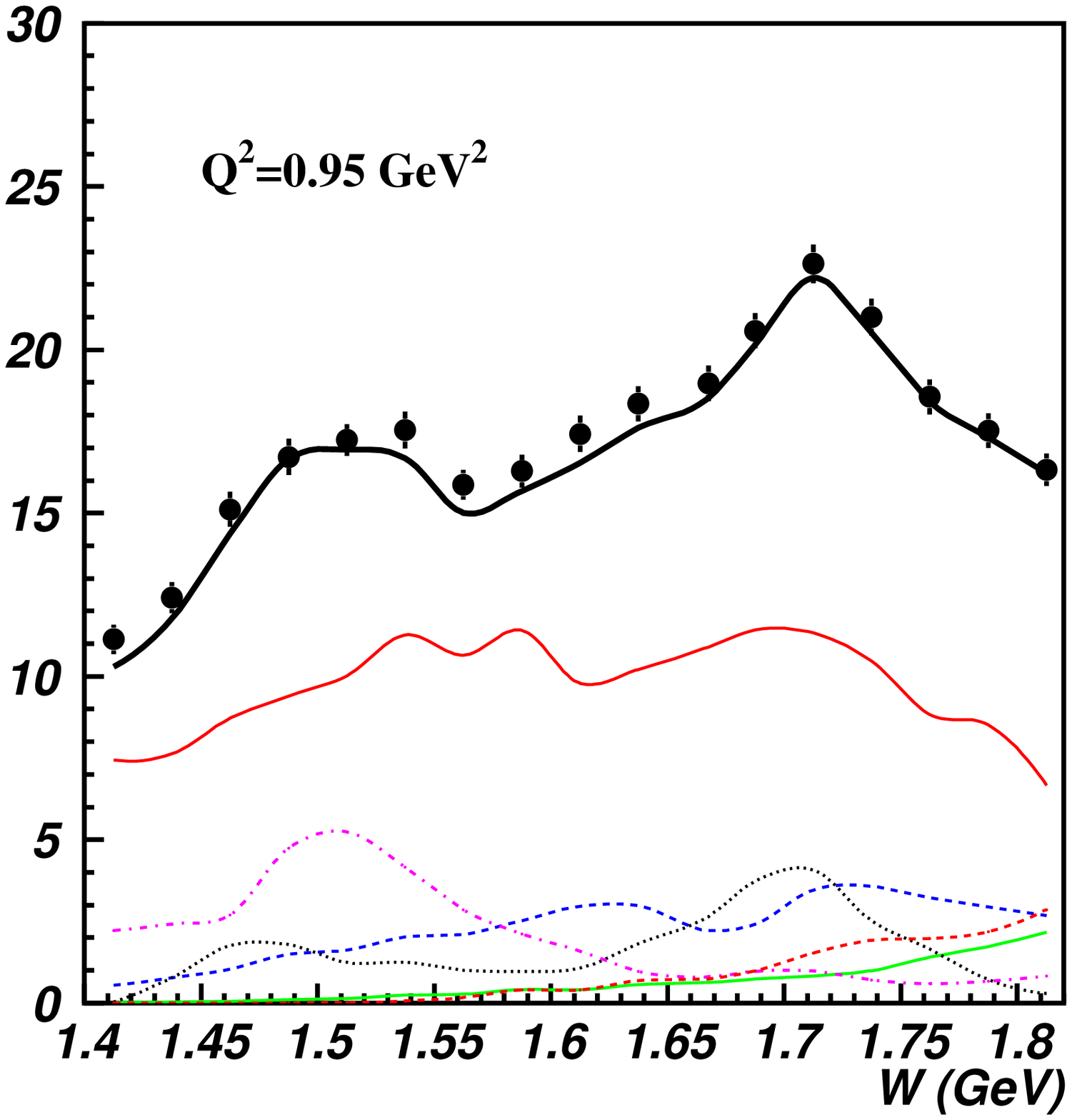}
\includegraphics[width=5.5cm]{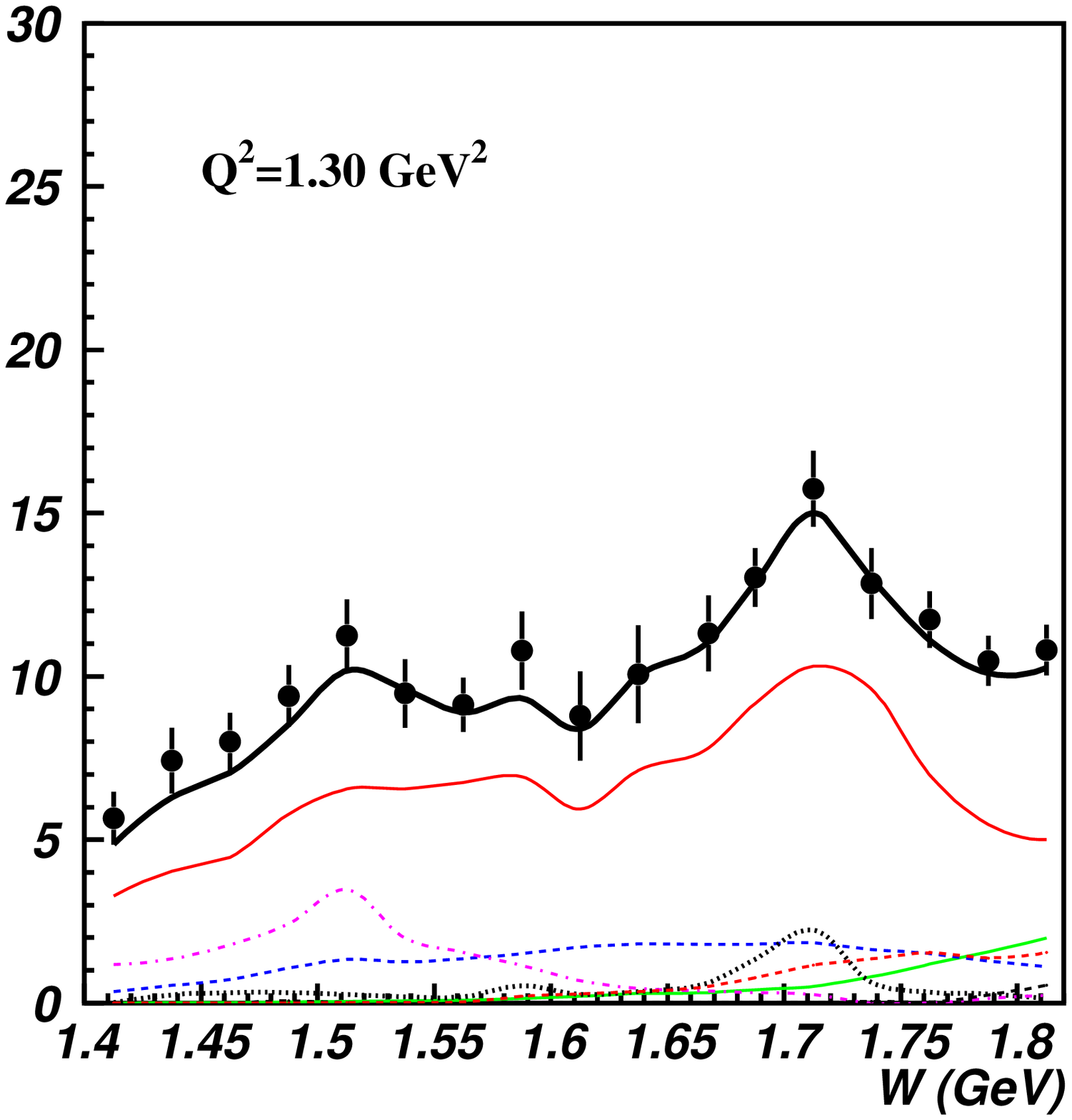}
\vspace{-0.1cm}
\caption{(Color Online) Description of the fully integrated $\pi^+\pi^-p$ electroproduction cross sections 
achieved within the framework of the updated JM model discussed in Section~\ref{pipipmech} together with the 
cross sections for the various contributing mechanisms: full cross section (black solid), $\pi^-\Delta^{++}$ 
(red thin solid), $\rho p$ (green thin solid), $\pi^+\Delta^0$ (blue thin dashed), $\pi^+ N^0(1520)3/2^-$ (black 
dotted), direct 2$\pi$ mechanisms (magenta  thin dot-dashed), and $\pi^+ N^0(1685)5/2^+$ (red thin dashed). The data 
fits were carried out at $W < 1.71$~GeV.}  
\label{integsec}
\end{center}
\end{figure*}

The contributions from the $\rho p$ meson-baryon channel are quite small in the range of $W < 1.71$~GeV where 
the resonance parameters presented in this paper are determined. However, reliable accounting of this channel 
is important for ascertaining the electrocouplings and the corresponding hadronic parameters of the resonances 
in the aforementioned range of $W$. Non-resonant amplitudes in the $\rho p$ sub-channel are described within the 
framework of a diffractive approximation, which also takes into account the effects caused by $\rho$-line 
shrinkage~\cite{Bu07}. The latter effects play a significant role in near-threshold and sub-threshold 
$\rho$-meson production for $W < 1.71$~GeV. The previous analyses of the CLAS data ~\cite{Fe09,Ri03} have 
revealed the presence of the $\rho p$ sub-channel contributions for $W > 1.5$~GeV.

\begin{table}
\begin{center}
\begin{tabular}{|c|c|c|c|c|c|} \hline
$N^*$ states & Mass,\     & Total           & BF                 & BF             & $N^*$ electro-  \\
incorporated &  GeV       & decay           & ${\pi \Delta}$, \% & ${\rho p}$, \% & coupling  \\
into the     &            & width           &                    &                & variation  \\
data fit     &            & $\Gamma_{tot}$, &                    &                &  in the fit \\
             &            &        GeV      &                    &                &             \\ \hline
$N(1440)1/2^+$ & var & var & var & var & \cite{Az051} var \\
$N(1520)3/2^-$ & var & var & var & var & \cite{Az051} var \\
$N(1535)1/2^-$ & var & var & var & var & \cite{Az09} fix \\
$\Delta(1620)1/2^-$ & var & var &var & var & \cite{Az05,Mo05nstar} var\\
$N(1650)1/2^-$ & var & var & var & var & \cite{Bu03} var \\
$N(1680)5/2^+$ & 1.68 & 0.12 & 12 & 5.5 & \cite{Az05,Mo05nstar} var \\
$N(1700)3/2^-$ & 1.74 & 0.19 & 53 & 45 & \cite{Az05,Mo05nstar} fix \\
$\Delta(1700)3/2^-$ & 1.70 & 0.26 & 89 & 2 & \cite{Az05,Mo05nstar} fix  \\
$3/2^+(1720)$ & 1.72 & 0.09 & 55 & 7 & \cite{Mo14} fix \\
$N(1720)3/2^+$ & 1.73 & 0.11 & 47 & 36 & \cite{Mo14} fix \\ \hline
\end{tabular}
\caption{List of resonances invoked in the $\pi^+\pi^-p$ fit and their parameters: total decay widths 
$\Gamma_{tot}$ and branching fractions (BF) to the $\pi \Delta$ and $\rho N$ final states. The quoted values 
for the hadronic parameters are taken from earlier fits~\cite{Az05,Mo14} to the CLAS $\pi^+\pi^-p$ data
\cite{Ri03}. The quantities labeled as $var$ correspond to the variable parameters fit to the CLAS 
$\pi^+\pi^-p$ data~\cite{Ri03} within the framework of the current JM model version. Start values for 
the resonance electrocouplings are taken from the references listed in the last column and extrapolated to 
the $Q^2$ area covered by the CLAS experiment~\cite{Ri03}. $3/2^+(1720)$ represents the candidate $N^*$ state 
with the signal reported in a previous analysis of CLAS data~\cite{Ri03}.}
\label{nstlist} 
\end{center}
\end{table}

The $\pi^+ N^0(1520)3/2^-$, $\pi^+ N^0(1685)5/2^+$, and $\pi^- \Delta(1600)3/2^+$ sub-channels are described 
in the JM model by non-resonant contributions only. The amplitudes of the $\pi^+ N^0(1520)3/2^-$ sub-channel 
were derived from the non-resonant Born terms in the $\pi \Delta$ sub-channels by implementing an additional 
$\gamma_5$-matrix  that accounts for the opposite parities of $\Delta(1232)3/2^+$ and $ N(1520)3/2^-$~\cite{Mo05nstar}. 
The magnitudes of the $\pi^+ N^0(1520)3/2^-$ production amplitudes were independently fit to the data for each 
bin in $W$ and $Q^2$. The contributions from the $\pi^+ N^0(1520)3/2^-$ sub-channel should be taken into account 
for $W > 1.5$~GeV.

The $\pi^+ N^0(1685)5/2^+$ and $\pi^- \Delta^{++}(1600)3/2^+$ sub-channel contributions are seen in the data
\cite{Ri03} at $W > 1.6$~GeV. These contributions are almost negligible at smaller $W$. The effective contact 
terms were employed in the JM model for parameterization of these sub-channel amplitudes~\cite{Mo05nstar}. The
magnitudes of the $\pi^+ N^0(1685)5/2^+$ and $\pi^- \Delta^{++}(1600)3/2^+$ sub-channel amplitudes were fit to 
the data for each bin in $W$ and $Q^2$.

In general, unitarity requires the presence of so-called $2\pi$ direct production mechanisms in the $\pi^+\pi^- p$ 
electroproduction amplitudes, where the final $\pi^+\pi^- p$ state is created without going through the intermediate 
step of forming unstable hadron states~\cite{Ait78}. These $2\pi$ direct production processes 
are beyond the aforementioned contributions from the two-body sub-channels. 2$\pi$ direct production amplitudes 
were established for the first time in the analysis of the CLAS $\pi^+\pi^-p$ electroproduction data~\cite{Mo09}. 
They are described in the JM model by a sequence of two exchanges in the $t$ and/or $u$ channels by unspecified 
particles that may come from two Regge trajectories. The amplitudes of the $2\pi$ direct production mechanisms are 
parameterized by a Lorentz-invariant contraction between spin-tensors of the initial and final-state particles, 
while two exponential propagators describe the above-mentioned exchanges by unspecified particles. The magnitudes 
of these amplitudes are fit to the data for each bin in $W$ and $Q^2$. The contributions from the $2\pi$ direct  
production mechanisms are maximal and substantial ($\approx$ 30\%) for $W < 1.5$~GeV and they decrease with 
increasing $W$, contributing less than 10\% for $W > 1.7$~GeV. However, even in this kinematic regime, $2\pi$ direct 
 production mechanisms can be seen in the $\pi^+\pi^-p$ electroproduction cross sections due to an 
interference of the amplitudes with the two-body sub-channels. Explicit expressions for the above-mentioned 
2$\pi$ direct production amplitudes can be found in Appendices A-C of Ref.~\cite{Mo09}. We are planning to 
explore the possibility to replace this phenomenological ansatz by the $B_{5}$ Veneziano model that was employed 
successfully in the studies of charged double-kaon photoproduction \cite{Jpac}. 

The studies of the final state hadron angular distributions over $\alpha_{i}~(i=[\pi^-p][\pi^+p'],[\pi^+p][\pi^-p'], 
[\pi{+}\pi^-][p p']$) conclusively demonstrated the need to implement the relative phases for all $2\pi$ direct 
production mechanisms included in the JM model. Figure~\ref{pidirphase} shows the comparison of the measured data
\cite{Ri03} to the differential cross sections $d\sigma/d\alpha_{[\pi^+\pi^-][p p']}$ computed within the 
framework of the JM model for values of the relative phases of the $2\pi$ direct production mechanisms fit to 
the data and for values of these phases equal to zero. The computed cross sections, assuming zero phases for all 
$2\pi$ direct production amplitudes, underestimate the measured $d\sigma/d\alpha_{[\pi^+\pi^-][p p']}$ cross 
sections at $\alpha_{[\pi^+\pi^-][p p']}$ around 180$^\circ$ (left panel in Fig.~\ref{pidirphase}). This is a 
consequence of destructive interference of these contributions with the amplitudes of other relevant processes at 
$Q^2$=0.95 GeV$^2$ and insufficient constructive interference at $Q^2$=0.65 GeV$^2$. Fits to the data phases of 
$2\pi$ direct production mechanisms change the interference pattern and allow us to 
improve the description of the $d\sigma/d\alpha_{[\pi^+\pi^-][p p']}$ angular distributions at $W > 1.48$~GeV in 
all three $Q^2$-bins under study, while retaining the same or even better quality of description of the other 
eight one-fold differential cross sections. Examples of the achieved improvements implementing the relative 
phases of the $2\pi$ direct production mechanisms are shown in the right column of Fig.~\ref{pidirphase}. 


The JM model provides a reasonable description of the nine $\pi^+\pi^- p$ one-fold differential cross sections for
$W  < 1.8$~GeV and $Q^2 < 1.5$~GeV$^2$. As a typical example, the nine one-fold differential cross 
sections and their corresponding descriptions for $W = 1.51$~GeV and $Q^2$ = 0.65~GeV$^2$ and for $W = 1.61$~GeV 
and $Q^2$ = 0.95~GeV$^2$ are shown in Fig.~\ref{isochan}, together with the contributions from each of the
individual mechanisms incorporated into the JM model. Any contributing mechanism will be manifested by
substantially different shapes in the cross sections for the observables, all of which are highly correlated 
through the underlying reaction dynamics. The simultaneous description of all the nine one-fold differential 
cross sections allows for identifying the essential mechanisms contributing to the $\pi^+\pi^-p$ 
electroproduction off the proton. 

Descriptions of the fully integrated $\pi^+\pi^-p$ electroproduction cross sections are shown in 
Fig.~\ref{integsec} together with the contributions from the meson-baryon mechanisms of the JM model. The 
major part of the $\pi^+\pi^-p$ electroproduction off the proton at $W < 1.6$~GeV is due to contributions 
from the two $\pi \Delta$ isobar channels, $\pi^- \Delta^{++}$ and $\pi^+ \Delta^0$. The $\Delta^{++}$(1232) 
resonance is clearly seen in all $\pi^+ p$ mass distributions for $W > 1.4$~GeV, while contributions from the 
$\pi^+ \Delta^0$ isobar channel are needed to better describe the data in the low mass regions of the $\pi^- p$
 mass distributions. The strength of the $\pi^- \Delta^{++}$ isobar channel observed in the data~\cite{Ri03,Fe09} 
is approximately nine times larger than that of $\pi^+ \Delta^0$~\cite{Mo09} due to isospin invariance. The CLAS 
data~\cite{Ri03} demonstrated sub-leading but still important contributions from the $\pi^+D_{13}^0(1520)$ 
meson-baryon channel. This contribution is needed for a description of the $\pi^+$ CM-angular 
distributions at forward angles and allows us to better describe the $\pi^-p$ invariant mass distributions as $W$
increases (see Fig.~\ref{isochan}). The contributions from $2\pi$ direct production mechanisms shown in 
Fig.~\ref{integsec} were obtained with the phases of these mechanisms derived from the CLAS data~\cite{Ri03}. 
These contributions are substantial, up to 30\% at $W < 1.5$~GeV. They decrease sharply as $W$ increases. Direct 
$2\pi$ production mechanisms become minor at $W > 1.7$~GeV, but they still should be taken into account because 
of their interference with larger amplitudes of other contributing mechanisms. $2\pi$ direct production mechanisms 
are kinematically allowed in the entire range of $W$, while meson-baryon channels with final mesons/baryons 
heavier than the pion/nucleon can be open at $W$ larger than the respective threshold values. This may explain the 
biggest contributions from $2\pi$ direct production mechanisms at small $W < 1.5$~GeV. The $\pi^+\pi^-p$ final 
state interaction with all open meson-baryon channels may be a plausible explanation for the sharp decrease of 
these mechanism contributions at $W > 1.5$~GeV, see Fig~\ref{integsec}. A quantitative description of this 
pronounced effect in the behavior of the $2\pi$ direct production mechanisms represents a challenging task for the 
global multi-channel analysis of exclusive meson electroproduction within the framework of the coupled channel 
approaches under development by the Argonne-Osaka group~\cite{Lee10,Lee13a}.

Accounting for the restrictions imposed by unitarity on the $\pi^+\pi^-p$ electroproduction amplitudes 
represents an important requirement for reliable extraction of the resonance parameters. However, a rigorous
implementation of unitarity for this three-body exclusive channel is still far from the reach of reaction theory 
and is outside the scope of this paper. To our knowledge, none of the available models is capable of providing 
fully unitarized amplitudes to fit the $\pi^+\pi^-p$ data to determine the electroproduction amplitudes. A very 
promising step in this direction was made by the Argonne-Osaka group~\cite{Lee13,Kam09}. Nevertheless, their 
approach is still under development. In this paper we employ a strategy that allows us to account phenomenologically 
for unitarity constraints in extracting the resonance parameters. As was mentioned above, we incorporated several 
features in the JM model in order to account for the unitarity restrictions on the resonant/non-resonant 
$\pi^+\pi^-p$ electroproduction amplitudes: a) the unitarized Breit-Wigner ansatz for the resonant amplitudes, 
b) the phenomenological treatment of the initial and final state interactions and the inclusion of the 
extra contact terms for the non-resonant amplitudes of the $\pi \Delta$ sub-channels, and c) direct 2$\pi$ 
production mechanisms. A good description of the nine one-fold differential cross sections in the entire 
kinematic area of $W$ and $Q^2$ analyzed in this paper strongly suggests a reliable parameterization of the 
squared $\pi^+\pi^-p$ electroproduction amplitudes achieved for the CLAS data~\cite{Ri03} fit within the framework 
of the JM model updated as was described in earlier. The $\pi^+\pi^-p$ electroproduction amplitudes determined
from a fit to the data account for the restrictions imposed by unitarity on their magnitudes at the real energy 
axis because the measured differential cross sections should be consistent with the unitarity constraints. The 
resonant contributions to the differential cross sections were obtained from these amplitudes switching off the 
non-resonant parts. In Section~\ref{fit} we will discuss in detail the extraction of the resonant contributions to 
the differential cross sections. The resonant parameters were extracted from the resonant contributions to the
differential cross sections employing the unitarized Breit-Wigner ansatz for the resonant amplitudes. Therefore, 
the unitarity constraints on the resonant amplitudes were fully taken into account. The resonant parameters were 
obtained at the real energy axis at the resonant point $W=M_{N^*}$. The reliability of the resonance parameters 
obtained in this way is determined by credible isolation of the resonant contributions to the differential cross 
sections, which will be discussed in Section~\ref{fit}. 

\section{The CLAS Data Fit}
\label{fit}

The resonance parameters obtained in our work were determined in the simultaneous fit to the CLAS $\pi^+\pi^-p$ 
electroproduction differential cross sections ~\cite{Ri03} in the three $Q^2$-bins listed in Table~\ref{wq2bins}.
The $W$-area included in the fit is limited to $W < 1.71$~GeV. Currently the resonance content for the structure 
in the $W$-dependence of the fully integrated $\pi^+\pi^-p$ cross sections at $W \approx 1.7$~GeV~\cite{Ri03} is 
still under study~\cite{Mo15}. For this reason the resonance parameters for the states located at $W$ above 
1.64~GeV are outside of this paper scope. 

In order to provide a realistic evaluation of the resonance parameters, we abandoned the traditional least-squares 
fit, since the parameters extracted in such a fit correspond to a single presumed global minimum, while the 
experimental data description achieved with other local minima may be equally good within the data uncertainties. 
Furthermore, the traditional evaluation of the fit-parameter uncertainties, based on the error propagation matrix, 
cannot be used for the same reason.

\begin{table}
\begin{center}
\begin{tabular}{|c|c|} \hline
                                & Ranges covered  \\
                                & in variations of the  \\ 
Variable \, parameters          & start parameters,  \\
		                & \% from their values \\ \hline
Magnitude of the additional     &      \\
contact term amplitude in the   & 40.0 \\
$\pi^- \Delta^{++}$ sub-channel &      \\ \hline
Magnitude of the additional     &      \\
contact term amplitude in the   & 45.0 \\
$\pi^+ \Delta^0$ sub-channel    &      \\ \hline
Magnitude of the                &      \\
$\pi^+ N^0(1520)3/2^-$ amplitude& 45.0 \\ \hline
Magnitudes of the six           &      \\
2$\pi$ direct production        & 20.0-30.0 \\
amplitudes                      &      \\ \hline
\end{tabular}
\caption{Variable parameters of the non-resonant mechanisms incorporated into the JM model~\cite{Mo09,Mo12}. 
The ranges in the table correspond to the 3$\sigma$ areas around the start values of the parameters.}
\label{bckpar} 
\end{center}
\end{table}

The special data fit procedure described in Ref.~\cite{Mo12} was employed for extraction of the resonance 
parameters. It allows us to obtain not only the best fit, but also to establish bands of the computed cross 
sections that are compatible with the data within their uncertainties. In the fit we simultaneously vary the 
resonant and non-resonant parameters of the JM model given in Tables~\ref{nstlist} and~\ref{bckpar}, respectively. 
More details on the non-resonant mechanisms listed in Table~\ref{bckpar} can be found in Refs.~\cite{Mo09,Mo12}. 
These non-resonant mechanisms have an essential influence on the data description at $W < 1.71$~GeV. The values 
of the aforementioned non-resonant/resonant parameters were evaluated under simultaneous variation of:
\begin{itemize}
\item the magnitudes of additional contact-term amplitudes in the $\pi^- \Delta^{++}$ and $\pi^+ \Delta^0$ isobar 
channels (2 parameters per $Q^2$-bin);
\item the magnitudes of the $\pi^+ N^0(1520)3/2^-$ isobar channel (1 parameter per $Q^2$-bin);
\item the magnitudes of all direct 2$\pi$ production amplitudes (9 parameters per $Q^2$-bin including the phases 
described in the Section~\ref{pipipmech});
\item and the variable resonant parameters listed in Table~\ref{nstlist}. The CLAS $\pi^+\pi^-p$ data~\cite{Ri03} at 
$W < 1.71$~GeV are mostly sensitive to the variable electrocouplings of the $N(1440)1/2^+$, $N(1520)3/2^-$, 
$\Delta(1620)1/2^-$, and $N(1650)1/2^-$ states (9 resonance electrocouplings per $Q^2$-bin), as well as the $\pi \Delta$ 
and $\rho p$ hadronic decay widths of these four resonances and of the $N(1535)1/2^-$ state (12 parameters that 
remain the same in the entire kinematic area covered by the fit). 
\end{itemize}

All of the aforementioned parameters are sampled around their start values, employing unrestricted normal 
distributions. In this way we mostly explore the range of $\approx$ 3$\sigma$ around the start parameter values. 
The $W$-dependencies of the magnitudes of the amplitudes of all non-resonant contributions are determined by 
adjusting their values to reproduce the measured nine one-fold differential charged double-pion electroproduction 
cross sections~\cite{Ri03}. We apply multiplicative factors to the magnitudes of all non-resonant amplitudes. 
They remain the same in the entire $W$-interval covered by the fit within any $Q^2$-bin, but they depend on the
photon virtuality $Q^2$ and are fit to the data in each $Q^2$-bin independently. The multiplicative factors are 
varied around unity employing normal distributions with the $\sigma$ values in the ranges listed in 
Table~\ref{bckpar}. In this way we retain the smooth $W$-dependencies of the non-resonant contributions
established in the adjustment to the data and explore the possibility to improve the data description in the
simultaneous variation of the resonant and non-resonant parameters. 

\begin{table}
\begin{center}
\begin{tabular}{|c|c|c|c|} \hline
Resonances        &$Q^2_{cent.}$=0.65 &$Q^2_{cent.}$=0.95 &$Q^2_{cent.}$=1.30   \\
                  & GeV$^2$           & GeV$^2$           & GeV$^2$                    \\
\hline
$N(1440)1/2^+$ & 40  & 30 & 40 \\
$N(1520)3/2^-$ & 20 & 20 & 30 \\
$\Delta(1620)1/2^-$ & 40 & 40 & 40 \\
$N(1650)1/2^-$ & 40 & 40 & 50  \\ \hline
\end{tabular}
\caption{$\sigma$ parameters employed in the variation of the resonance electrocouplings in \% of their start 
values. The $\sigma$ parameters listed for the $\Delta(1620)1/2^-$ were applied as a variation of the 
$S_{1/2}(Q^2)$ electrocouplings only. The variation of the $A_{1/2}(Q^2)$ electrocouplings of this state is 
described in Section~\ref{nstarelectrocoupl}.}
\label{varelcoupl}  
\end{center}
\end{table}

In this fit we also vary the $\gamma_vpN^*$ electrocouplings and the $\pi \Delta$ and $\rho p$ hadronic 
partial decay widths of the $N(1440)1/2^+$, $N(1520)3/2^-$, and $\Delta(1620)1/2^-$ resonances around their 
start values. The start values of the $N(1440)1/2^+$ and $N(1520)3/2^-$ electrocouplings were determined by 
interpolating the results from the analyses~\cite{Az09,Mo12} of the CLAS data on $N\pi$ and $\pi^+\pi^-p$ 
electroproduction off the proton in the range 0.5~GeV$^2$ $< Q^2 <$ 1.5~GeV$^2$. The electrocouplings of the 
$N(1440)1/2^+$ and $N(1520)3/2^-$ resonances were varied employing normal distributions with the $\sigma$ 
parameters listed in Table~\ref{varelcoupl} in terms of \% of their start values. There were no restrictions on 
the minimum or maximum trial electrocoupling values, allowing us to mostly explore the area of $\approx$ 3$\sigma$ 
around their start values.

The $\pi^+\pi^-p$ electroproduction channel also has some sensitivity to the $N(1535)1/2^-$ state, which couples 
dominantly  to the $N\pi$ and $N\eta$ final states. The $N(1535)1/2^-$ electrocouplings were taken from the CLAS 
analysis of $N \pi$ electroproduction~\cite{Az09} and varied strictly inside the uncertainties reported in that
paper. 

The start values of the $\Delta(1620)1/2^-$ and $N(1650)1/2^-$ electrocouplings were taken from a preliminary 
analysis~\cite{Mo14}. In this study the resonance electrocouplings were adjusted to describe the nine 
$\pi^+\pi^-p$ one-fold differential cross sections~\cite{Ri03} in the $W$-interval from 1.41 to 1.80~GeV and at 
0.5~GeV$^2$ $< Q^2 <$ 1.5~GeV$^2$. However, the results~\cite{Mo14} do not allow us to draw unambiguous 
conclusions regarding the resonant content of the structure at $W$ $\approx$ 1.7~GeV. Therefore, we are using the 
results of this analysis as an estimate for the resonance electrocoupling start points to fit the charged 
double-pion electroproduction data~\cite{Ri03} for $W < 1.71$~GeV and 0.5~GeV$^2$ $< Q^2 <$ 1.5~GeV$^2$.

Since the resonance content of the structure at $W \approx$ 1.7~GeV is still under study, we present in this paper 
the fit results for the resonances with masses less than 1.64 GeV. In the extraction of these resonance parameters 
we also account for the contributions of the tails from several excited proton states in the third resonance region, 
$N(1685)5/2^+$, $N(1720)3/2^+$, and $\Delta(1700)3/2^-$, with their start electrocoupling values taken 
from the analyses of Refs.~\cite{Mo14,Mo12a} and varied within 15\% of their parameters. The 
$N(1685)5/2^+$ state decays mostly to the $N\pi$ final states. The electrocouplings of this state determined in 
the analyses of $\pi^+\pi^-p$ electroproduction~\cite{Mo14,Mo12a} are consistent with the results of independent 
analysis of $N\pi$ electroproduction~\cite{Tia11}. This suggests a reasonable evaluation of the aforementioned 
third resonance region state electrocouplings in the analyses~\cite{Mo14,Mo12a} of the $\pi^+\pi^-p$ electroproduction 
data as the start values for extraction of the resonance parameters for the states with masses less than 1.65~GeV. 
The contributions from the tails of the $N(1675)5/2^-$, $N(1700)1/2^+$, and $N(1700)3/2^-$ resonances were found 
to be negligible for $W < 1.64$~GeV.

\begin{table}
\begin{center}
\begin{tabular}{|c|c|c|} \hline
$N^*$ states        & Mass,     & Total decay width, $\Gamma_{tot}$, \\
                    &  MeV      &   MeV   \\ \hline
$N(1440)1/2^+$      & 1430-1480 & 200-450 \\
$N(1520)3/2^-$      & 1515-1530 & 100-150 \\
$N(1535)1/2^-$      & 1510-1560 & 100-200 \\
$\Delta(1620)1/2^-$ & 1600-1660 & 130-160 \\
$N(1650)1/2^-$      & 1640-1670 & 140-190 \\ \hline
\end{tabular}
\caption{The ranges of the resonance masses and total $N^*$ hadronic decay widths employed to constrain 
the variation of their partial hadronic decay widths to the $\pi \Delta$ and $\rho N$ final states in the 
fit of the CLAS $\pi^+\pi^-p$ electroproduction data~\cite{Ri03}.}
\label{hadrrange} 
\end{center}
\end{table}

The $\pi \Delta$ and $\rho p$ hadronic decay widths of the $N(1440)1/2^+$, $N(1520)3/2^-$, and $N(1535)1/2^-$ 
resonances were varied around their start values taken from previous analyses of the CLAS double-pion 
electroproduction data~\cite{Mo12}. For the $\Delta(1620)1/2^-$ state, the start values of these parameters 
were computed as the products of the $N^*$ total decay widths from Ref.~\cite{Rpp12} and the branching 
fractions for decays to the $\pi \Delta$ and $\rho N$ final states were taken from analyses of $\pi N \to \pi\pi N$ 
hadroproduction~\cite{Man92}. The ranges for the variations of the $\pi \Delta$ and $\rho p$ hadronic decay 
widths were restricted by the total $N^*$ decay widths and their uncertainties shown in Table~\ref{hadrrange}. 
The total $N^*$ decay widths were obtained by summing the partial widths over all decay channels. The partial 
hadronic decay widths to all final states other than $\pi \Delta$ and $\rho p$ were computed as the products of 
RPP~\cite{Rpp12} values of the $N^*$ total decay widths and branching fractions for decays to particular hadronic 
final states, which were taken from the analysis of Ref.~\cite{Man92}. We varied the $\pi \Delta$ and $\rho p$ 
hadronic decay widths of the $N(1440)1/2^+$, $N(1520)3/2^-$, $N(1535)1/2^-$, $\Delta(1620)1/2^-$, and 
$N(1650)1/2^-$ resonances simultaneously with their masses, keeping the hadronic $N^*$ parameters independent 
of $Q^2$. The $\pi \Delta$ and $\rho p$ hadronic decay widths of all other resonances obtained in the analyses of 
Refs.~\cite{Mo14,Mo12a} and noted in Table~\ref{nstlist} as ``fix" were kept unchanged.

\begin{table}
\begin{center}
\begin{tabular}{|c|c|}
\hline
$W$ intervals, & $\chi^2/d.p.$  intervals for selected\\
 GeV           & computed $\pi^+\pi^-p$ cross sections \\ \hline
1.41-1.51 & 2.12-2.30  \\
1.46-1.55 & 2.27-2.60  \\
1.51-1.61 & 2.55-2.85  \\
1.56-1.66 & 2.63-2.72  \\ 
1.61-1.71 & 2.49-2.68  \\ \hline
\end{tabular}
\caption{Quality of the fit of the CLAS data~\cite{Ri03} on $\pi^+\pi^-p$ electroproduction off the proton 
within the framework of the updated JM model described in Section~\ref{pipipmech}. The resonance parameters 
are determined from the data fit at $W$ $<$ 1.71 GeV.}
\label{fitqual} 
\end{center}
\end{table}

\begin{figure*}[htp]
\begin{center}
\includegraphics[width=11.5cm]{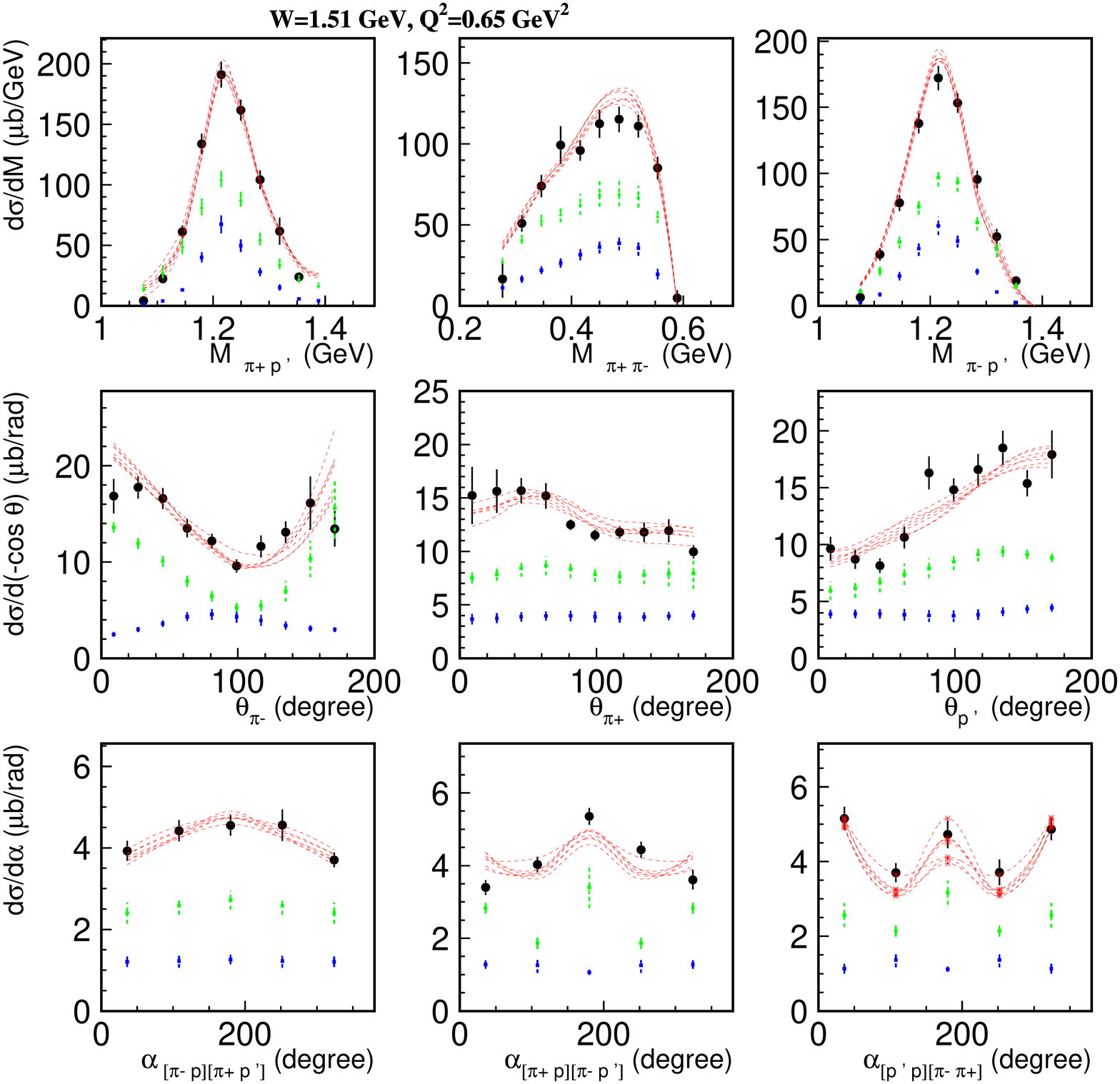}
\includegraphics[width=11.5cm]{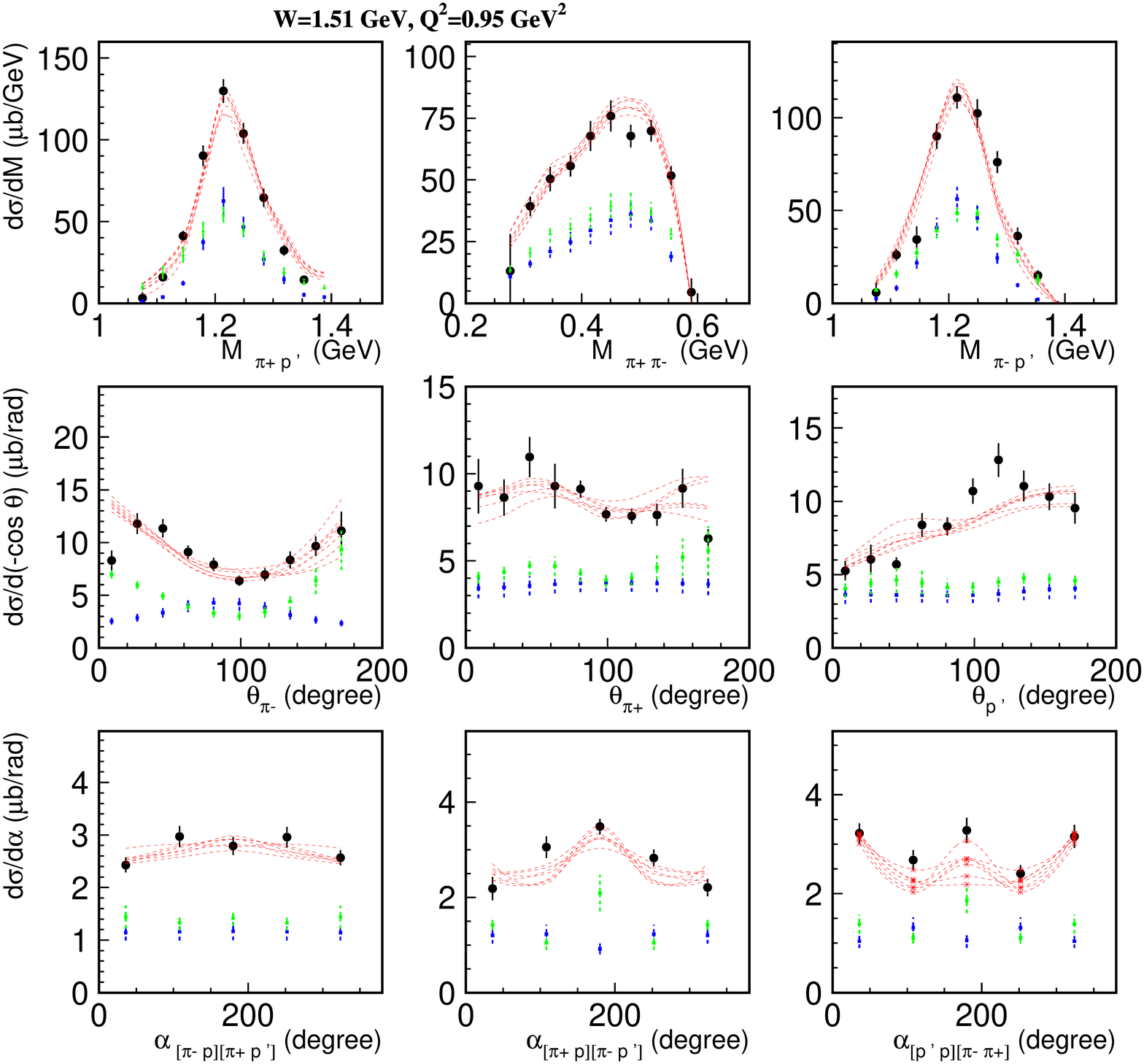}\\
\vspace{-0.1cm}
\caption{(Color Online) Examples of fits to the CLAS data~\cite{Ri03} on the nine one-fold differential 
$\pi^+\pi^-p$ electroproduction cross sections in particular bins of $W$ and $Q^2$ within the framework of the 
updated JM model described in Section~\ref{pipipmech}. The curves correspond to those fits with $\chi^2/d.p.$ 
within the intervals listed in Table~\ref{fitqual}. The resonant and non-resonant contributions determined 
from the data fit within the framework of the JM15 model are shown by blue triangles and green squares, 
respectively.} 
\label{fitsec}
\end{center}
\end{figure*}

\begin{figure*}[htp]
\begin{center}
\includegraphics[width=11.5cm]{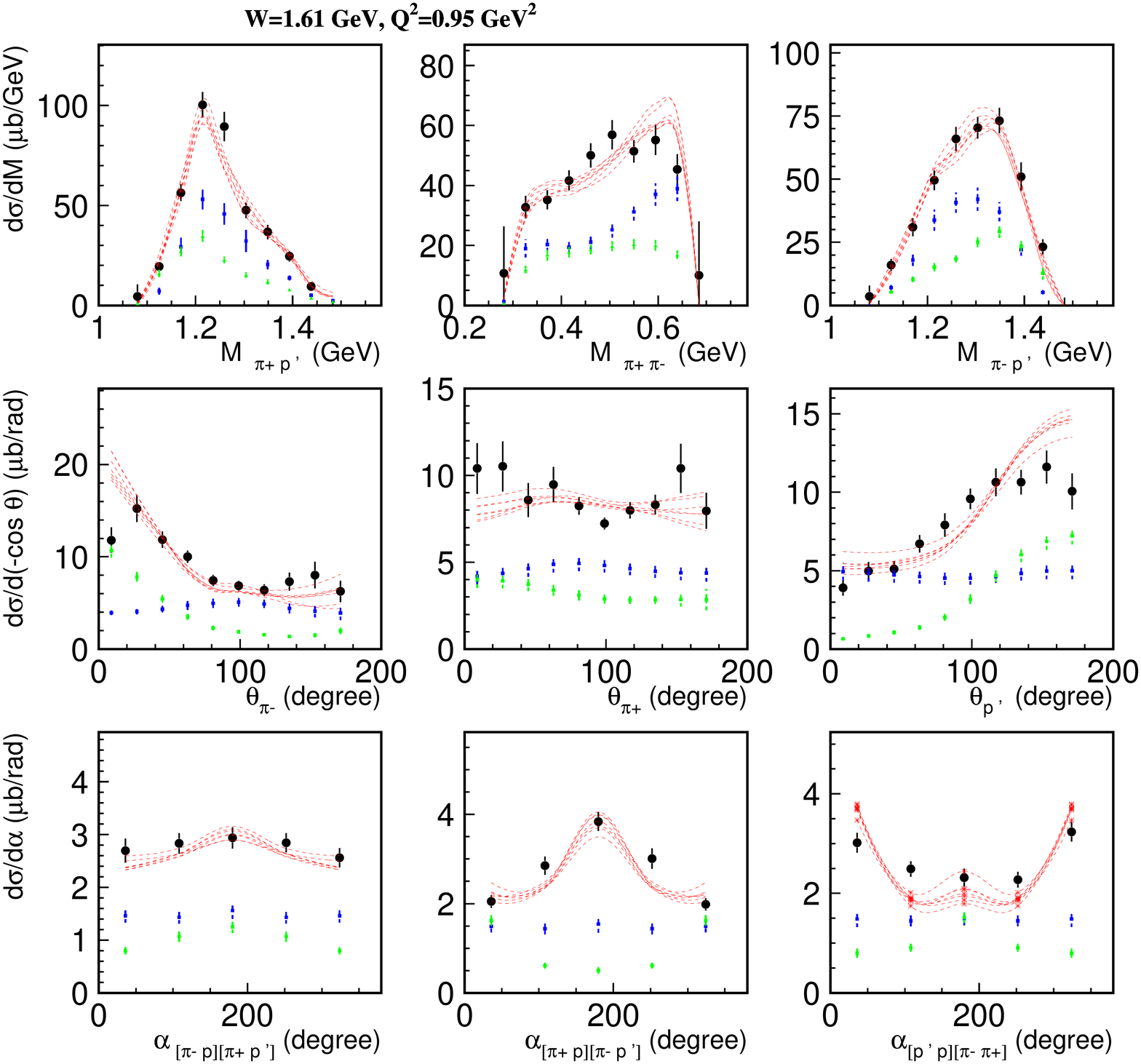}
\includegraphics[width=11.5cm]{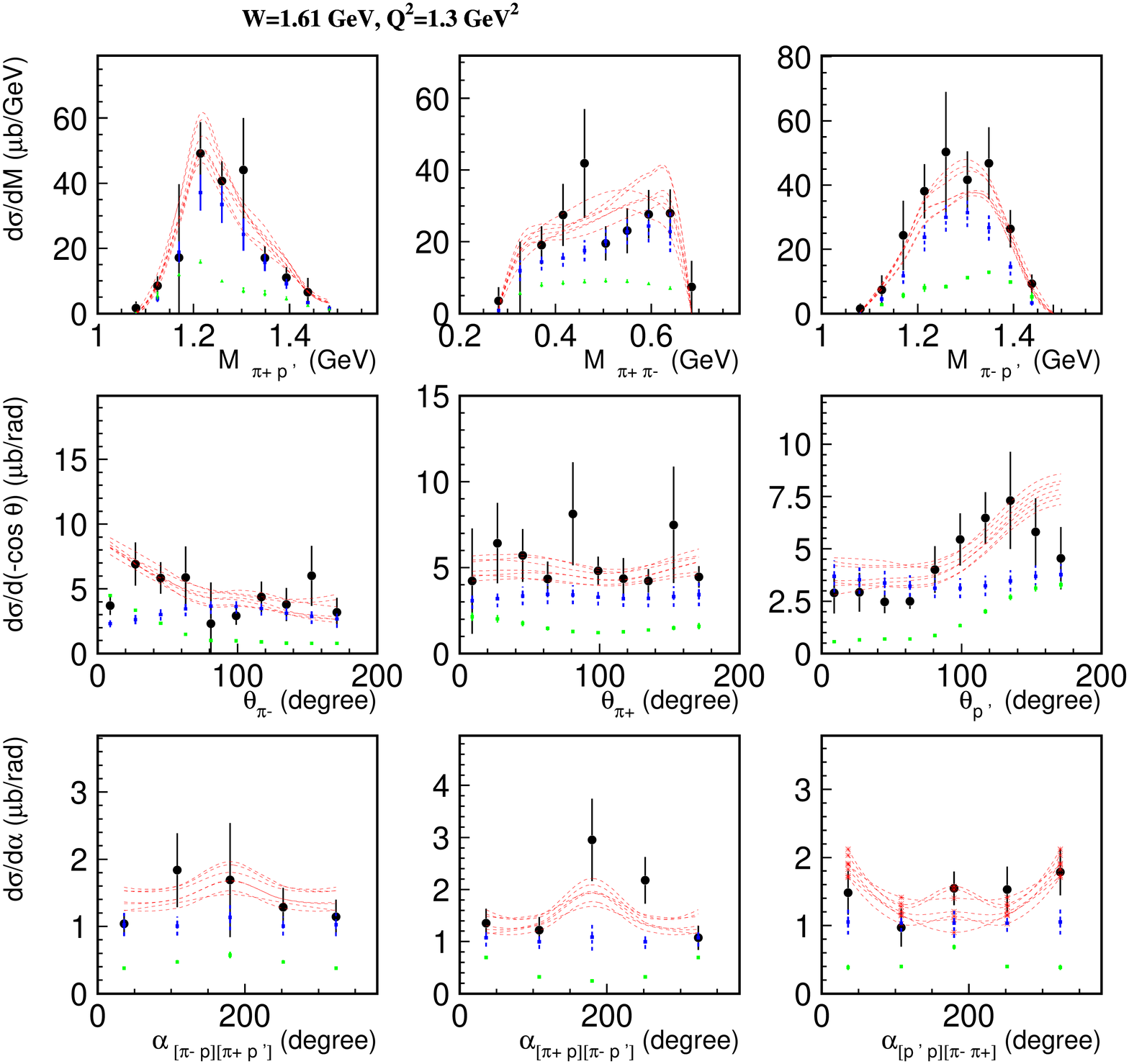}
\vspace{-0.1cm}
\caption{(Color Online) The same as in Fig.~\ref{fitsec}, but in other bins of $W$ and $Q^2$.} 
\label{fitsec1}
\end{center}
\end{figure*}

 For each trial set of the JM model resonant and non-resonant parameters we computed nine one-fold differential 
$\pi^+\pi^-p$ cross sections and the $\chi^2$ per data point values ($\chi^2$/$d.p.$). The $\chi^2$/$d.p.$
values were estimated in point-by-point comparisons between the measured and computed one-fold differential 
cross sections in all bins of $W$ from 1.41~GeV to 1.71~GeV and in the three $Q^2$-bins covered by the CLAS 
$\pi^+\pi^-p$ data~\cite{Ri03}. In the fit we selected the computed one-fold differential cross sections closest 
to the data with $\chi^2/d.p.$ less than a predetermined maximum value $\chi^2_{max}/d.p.$. The values of  
$\chi^2_{max}/d.p.$ were obtained by requiring that the computed cross sections with smaller $\chi^2/d.p.$ be 
within the data uncertainties for the majority of the data points, based on point-by-point comparisons between 
the measured and the computed cross sections, see examples in Fig.~\ref{fitsec},~\ref{fitsec1}. In this fit procedure we 
obtained the $\chi^2/d.p.$ intervals within which the computed cross sections described the data equally well 
within the data uncertainties.

We fit the CLAS data~\cite{Ri03} consisting of nine one-fold differential cross sections of the
$ep \to e'p'\pi^+\pi^-$ electroproduction reaction in all bins of $W$ and $Q^2$ in the kinematic regions of 
$W$: 1.41~GeV $< W <$ 1.51~GeV, 1.46~GeV $< W <$ 1.56~GeV, 1.51~GeV $< W <$ 1.61~GeV, 1.56~GeV $< W <$ 1.66~GeV,
 1.61~GeV $< W <$ 1.71~GeV and 0.5~GeV$^2$ $< Q^2 <$ 1.5~GeV$^2$ within the framework of the fit procedure 
described above. The five intervals of $W$ listed in Table~\ref{fitqual} were fit independently. Each of the 
aforementioned $W$ intervals contained 375 fit data points. The $\chi^2/d.p.$ intervals that correspond to an 
equally good data description within the data uncertainties are shown in Table~\ref{fitqual}. Their values 
demonstrate the quality of the CLAS $\pi^+\pi^-p$ data description achieved in the fits. Examining the 
description of the nine one-fold differential cross sections, we found that the $\chi^2/d.p.$ values were 
determined mostly by the deviations of only a few experimental data points from the computed fit cross sections. 
There were no discrepancies in describing the shapes of the differential cross sections, which would manifest 
themselves systematically in neighboring bins of $W$ and $Q^2$. Typical fit examples for $W$=1.51~GeV and 
neighboring $Q^2$ intervals centered at 0.65~GeV$^2$ and 0.95~GeV$^2$, as well as for $W$=1.61~GeV and $Q^2$ 
intervals centered at 0.95~GeV$^2$ and 1.30~GeV$^2$, are shown in Fig.~\ref{fitsec},~\ref{fitsec1}.
 
Since only statistical data uncertainties were used in the computation of the $\chi^2/d.p.$ values listed in 
Table~\ref{fitqual}, we concluded that a reasonable data description was achieved. The $\chi^2/d.p.$ values of 
our fits are comparable with those obtained in the fit of the CLAS $N\pi$ and $\pi^+\pi^-p$ electroproduction 
data published in Refs.~\cite{Az09,Park15} and in Ref.~\cite{Mo12}, respectively.

For each computed cross section point the resonant/non-resonant contributions were estimated switching off
the non-resonant/resonant amplitudes, respectively. The determined resonant/non-resonant contributions to the 
nine one-fold differential cross sections are shown in Fig.~\ref{fitsec},~\ref{fitsec1}. The results suggest the unambiguous 
and credible separation between the resonant/non-resonant contributions achieved fitting the CLAS data~\cite{Ri03} 
within the framework of the JM model. The determined resonant/non-resonant contributions are located within well 
defined ranges (see Fig.~\ref{fitsec},~\ref{fitsec1}) and show no evidence for separation ambiguities, which would manifest
themselves as substantial differences between the particular computed resonant/non-resonant cross sections and the 
ranges determined for the resonant/non-resonant contributions as shown in Fig.~\ref{fitsec},~\ref{fitsec1}. Such features in 
the behavior of the resonant/non-resonant contributions remain unseen in the entire area of $W$ and $Q^2$ covered 
by our analysis. Furthermore, the uncertainties of the resonant/non-resonant contributions are comparable with the 
uncertainties of the measured cross sections, demonstrating again unambiguous resonant/non-resonant separation of 
a good accuracy. The credible isolation of the resonant contributions makes it possible to determine the resonance 
parameters from the resonant contributions employing for their description the amplitudes of the unitarized 
Breit-Wigner ansatz that fully accounts for the unitarity restrictions on the resonant amplitudes. 

\begin{figure*}[htp]
\begin{center}
\includegraphics[width=8.5cm]{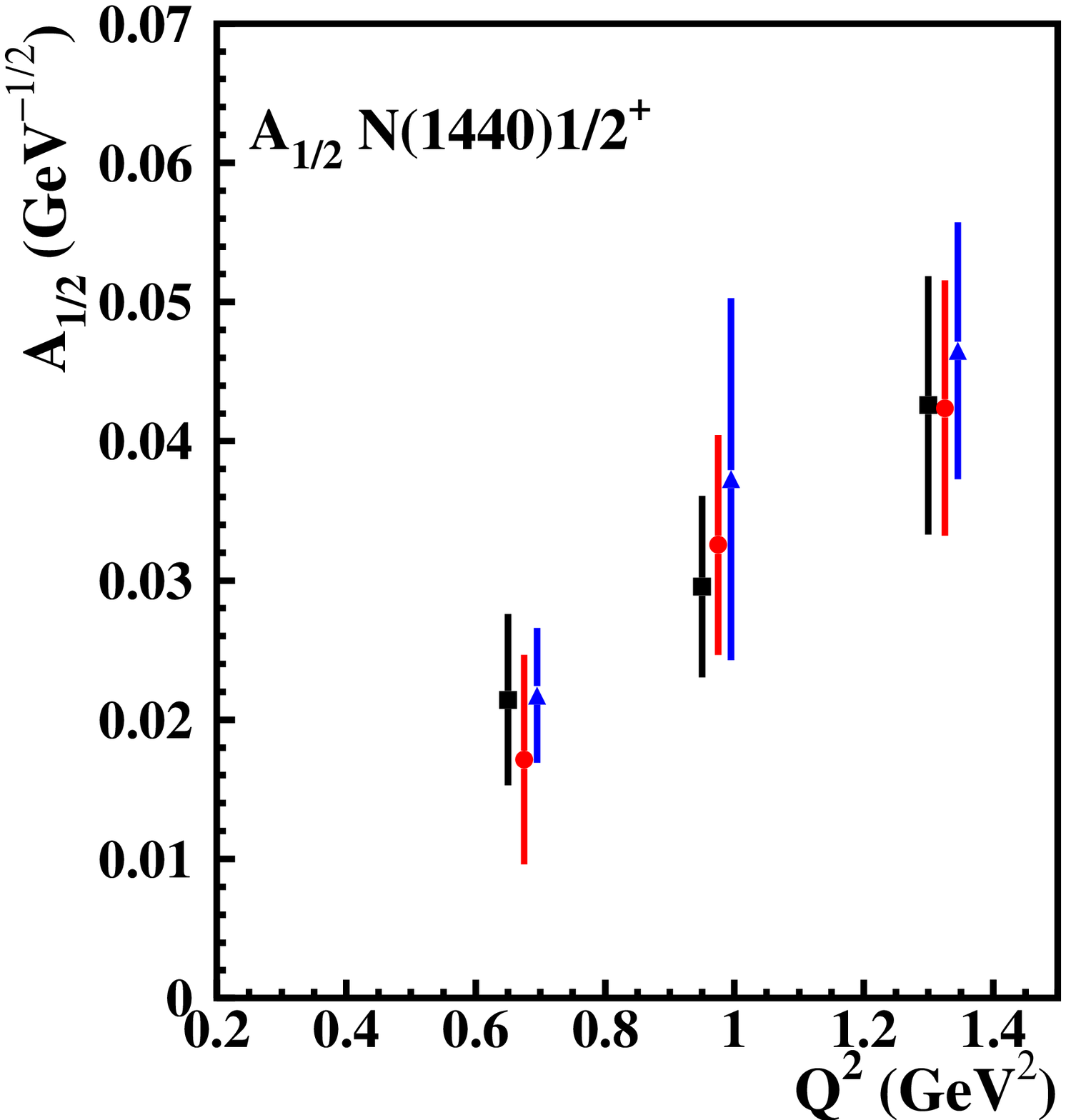}
\includegraphics[width=8.5cm]{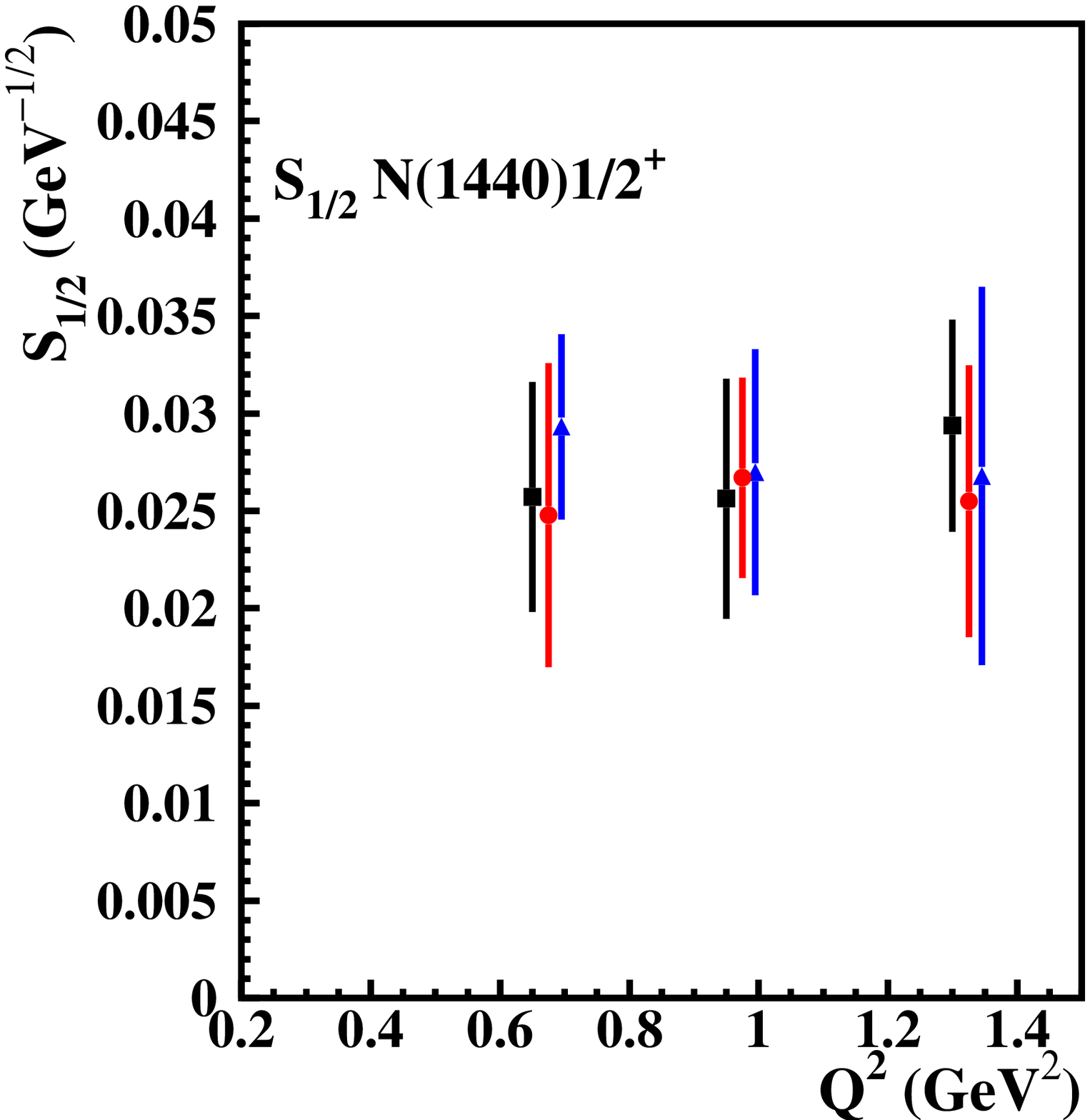}
\vspace{-0.1cm}
\caption{(Color Online) Electrocouplings of the $N(1440)1/2^+$ resonance determined from analysis of the 
CLAS $\pi^+\pi^-p$ electroproduction data~\cite{Ri03} carried out independently in three intervals of $W$: 
1.41~GeV $<$ $W$ $<$ 1.51~GeV (black squares), 1.46~GeV $<$ $W$ $<$ 1.56~GeV (red circles), and 1.51~GeV 
$<$ $W$ $<$ 1.61~GeV (blue triangles).}  
\label{p11comp}
\end{center}
\end{figure*}

\begin{figure*}[htp]
\begin{center}
\includegraphics[width=5.5cm]{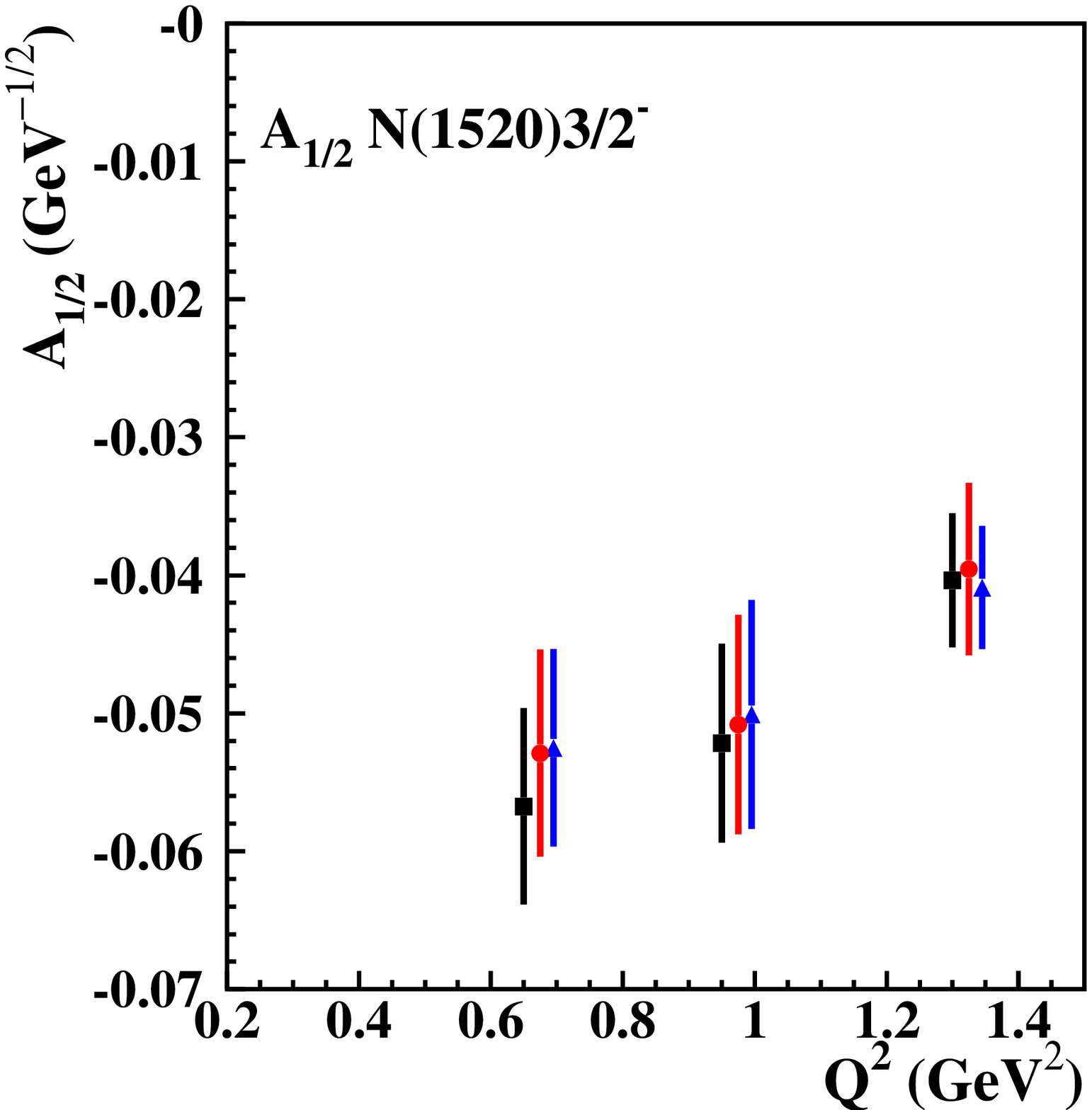}
\includegraphics[width=5.5cm]{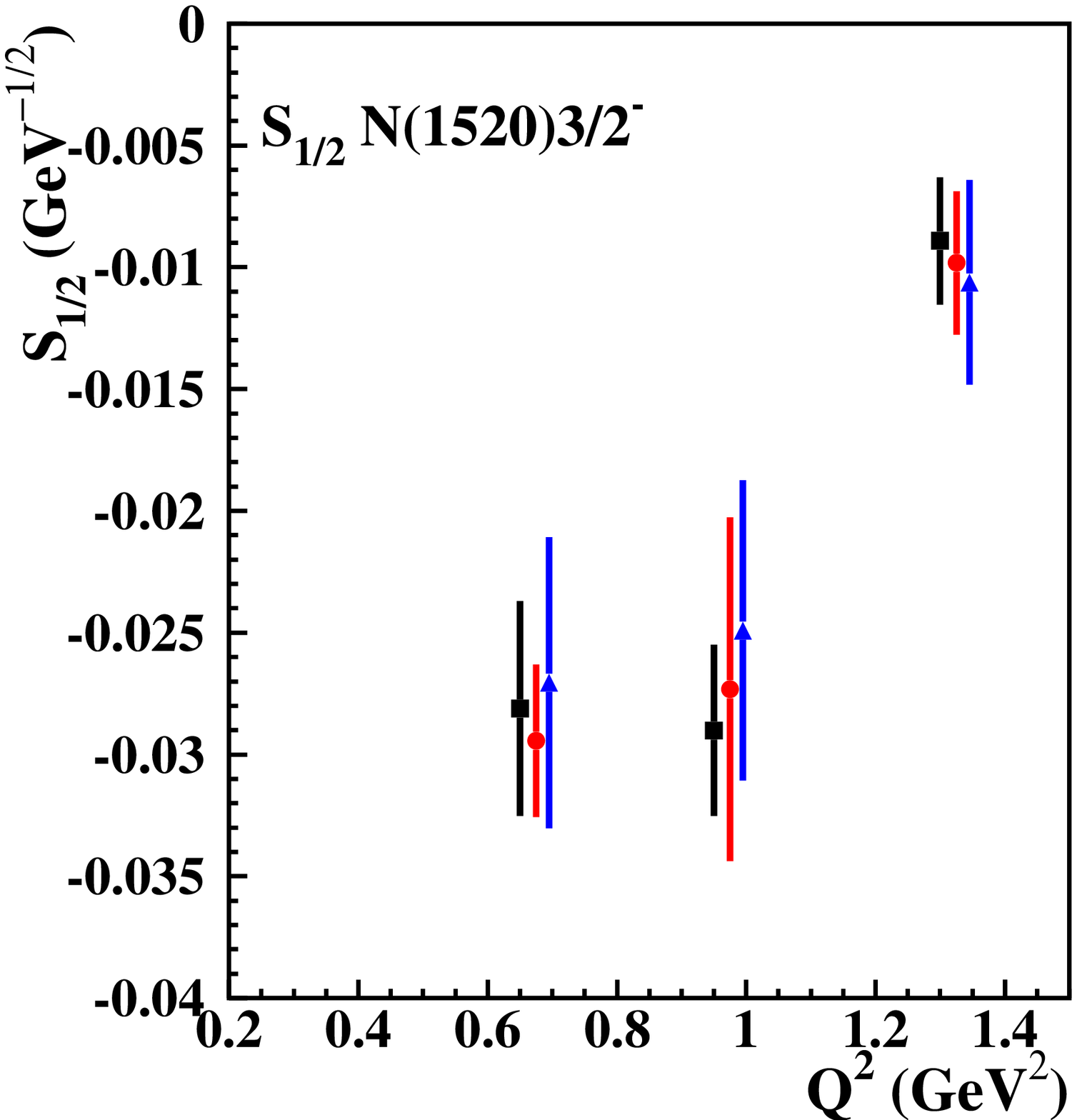}
\includegraphics[width=5.5cm]{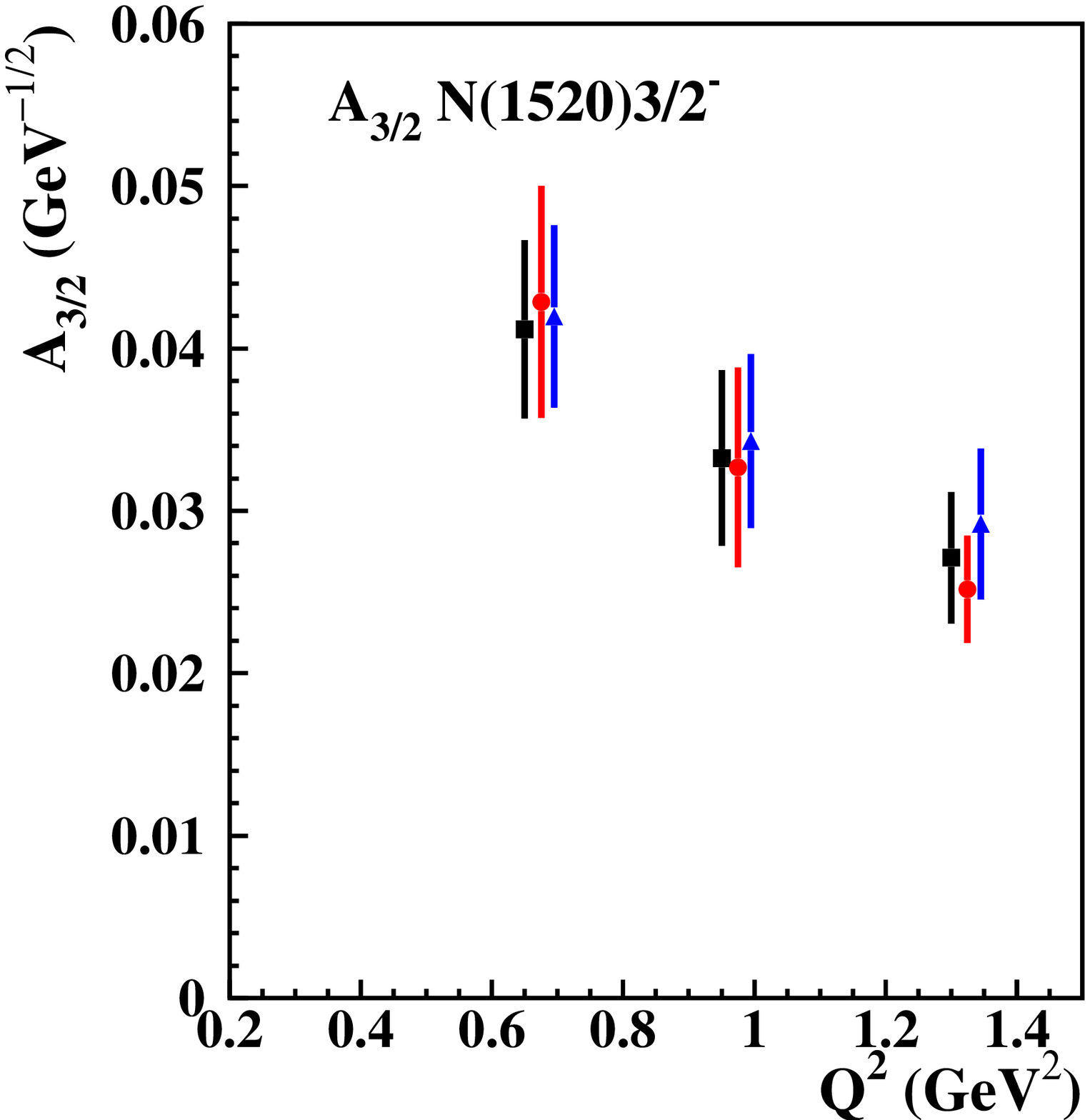}
\vspace{-0.1cm}
\caption{(Color Online) Electrocouplings of the $N(1520)3/2^-$ resonance determined from analysis of the 
CLAS $\pi^+\pi^-p$ electroproduction data~\cite{Ri03} carried out independently in three intervals of $W$: 
1.41~GeV $<$ $W$ $<$ 1.51~GeV (black squares), 1.46~GeV $<$ $W$ $<$ 1.56~GeV (red circles), and 1.51~GeV 
$<$ $W$ $<$ 1.61~GeV (blue triangles).}
\label{d13comp}
\end{center}
\end{figure*}

\begin{figure*}[htp]
\begin{center}
\includegraphics[width=8.5cm]{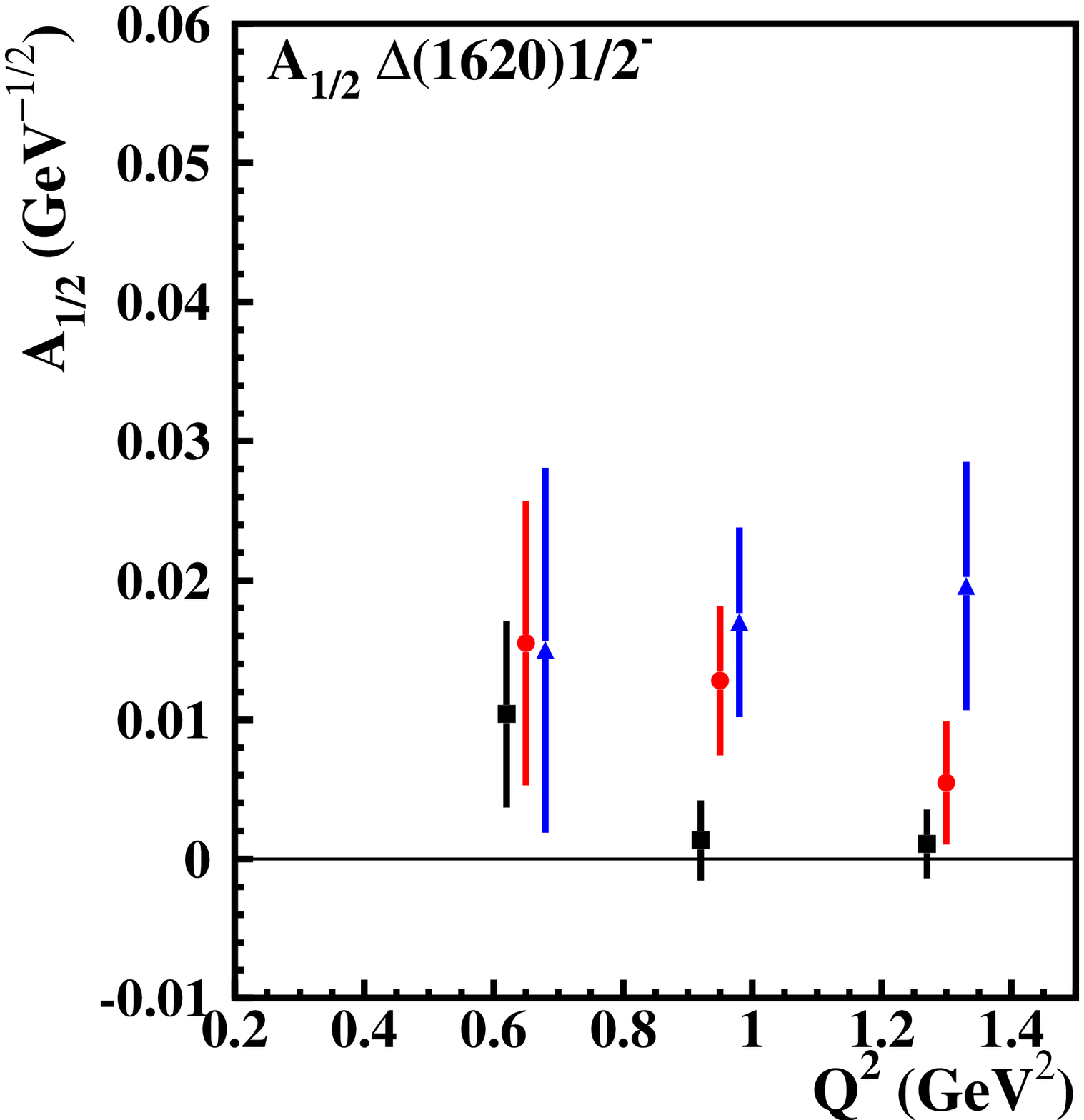}
\includegraphics[width=8.5cm]{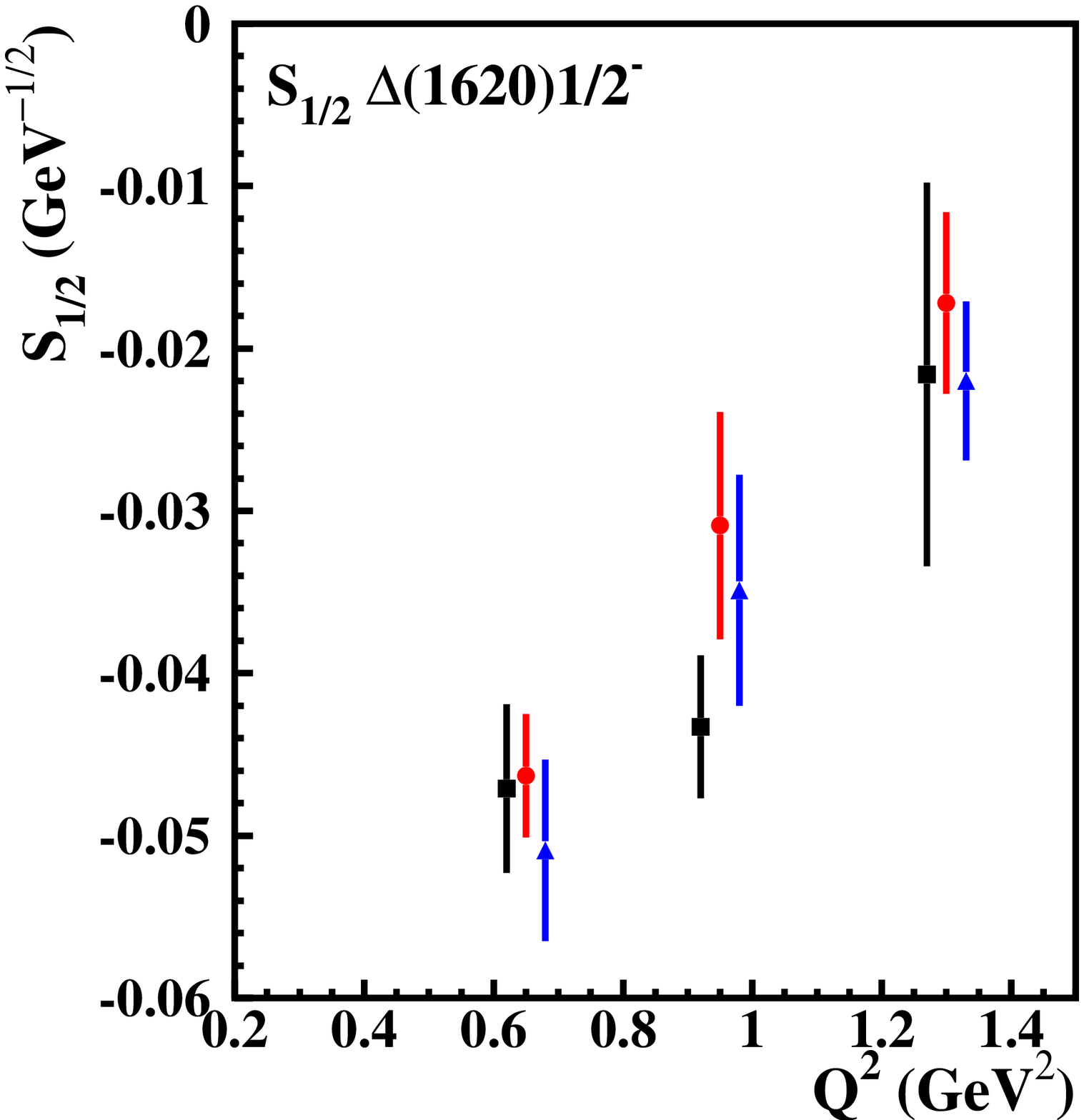}
\vspace{-0.1cm}
\caption{(Color Online) Electrocouplings of the $\Delta(1620)1/2^-$ resonance determined from analysis of 
the CLAS $\pi^+\pi^-p$ electroproduction data~\cite{Ri03} carried out independently in three intervals of 
$W$: 1.51~GeV $<$ $W$ $<$ 1.61~GeV (black squares), 1.56~GeV $<$ $W$ $<$ 1.66~GeV (red circles), and 
1.61~GeV $<$ $W$ $<$ 1.71~GeV (blue triangles).}
\label{s31comp}
\end{center}
\end{figure*}

The resonance parameters obtained from each of these equally good fits were averaged and their mean values were 
taken as the resonance parameters extracted from the data. The dispersions in these parameters were taken as 
the uncertainties. The resonance electrocoupling uncertainties obtained in this manner are shown in 
Figs.~\ref{p11comp}, \ref{d13comp}, \ref{s31comp}. Our fitting procedure allowed us to obtain more realistic 
uncertainties that take into account both statistical uncertainties in the data and systematic uncertainties 
imposed by the use of the JM reaction model. Furthermore, we consistently account for the correlations between 
variations of the non-resonant and resonant contributions while extracting the resonance parameters. 

\section{Evaluation of the $\gamma_vpN^*$ Resonance Electrocouplings and Hadronic Decay Widths to the 
$\pi \Delta$ and $\rho N$ Final States}
\label{nstarelectrocoupl}

The procedure described in Section~\ref{fit} allowed us to determine the $\gamma_vpN^*$ electrocouplings of 
the $N(1440)1/2^+$, $N(1520)3/2^-$, and $\Delta(1620)1/2^-$ resonances and their uncertainties. Our analysis 
extended the information on the electrocouplings of the $N(1440)1/2^+$ and  $N(1520)3/2^-$  states providing the 
first results in the range of photon virtualities 0.5~GeV$^2$ $< Q^2 <$ 1.5~GeV$^2$ from the CLAS data. The 
$\Delta(1620)1/2^-$ resonance decays preferentially to the $N\pi\pi$ final state, making the charged double-pion 
electroproduction channel the major source of information on the electrocouplings of this state. Our 
studies provide the first results on the electrocouplings and hadronic decays of this resonance to the 
$\pi \Delta$ and $\rho p$ final states from analysis of exclusive charged double-pion electroproduction.

A special approach was developed for the evaluation of the $\Delta(1620)1/2^-$ electrocouplings. The analysis 
of the CLAS $\pi^+\pi^-p$ electroproduction data revealed that the $A_{1/2}$ electrocoupling of this resonance 
was much smaller than the $S_{1/2}$ for 0.5~GeV$^2$ $< Q^2 <$ 1.5~GeV$^2$~\cite{Mo14}. The $A_{1/2}$ variations 
computed as a percentage of the start value became too small. For realistic uncertainty estimates we varied 
$A_{1/2}$ in a much wider range that made its tested absolute values comparable with those for the $S_{1/2}$ 
electrocoupling. We fit the CLAS data~\cite{Ri03} on $\pi^+\pi^-p$ electroproduction by varying $A_{1/2}$, as 
described above, keeping the variation of all other resonant and non-resonant parameters as described in 
Section~\ref{fit}.

In order to compare our results on the $N(1440)1/2^+$ and $N(1520)3/2^-$ electrocouplings in the $\pi^+\pi^-p$ 
electroproduction channel with their values from the analysis of $N\pi$ electroproduction, we must use in both 
of the exclusive electroproduction channels common branching fractions for the decays of these resonances to 
the $N\pi$ and $N\pi\pi$ final states. According to the RPP~\cite{Rpp12}, the sum of the branching fractions 
into the $N\pi$ and $N\pi\pi$ final states accounts for almost 100\% of the total decay widths of the 
$N(1440)1/2^+$ and $N(1520)3/2^-$ states. Since the $N\pi$ exclusive electroproduction channels are most 
sensitive to contributions from the $N(1440)1/2^+$ and $N(1520)3/2^-$ resonances, we re-evaluated the branching 
fraction for decay to the $N\pi\pi$ final states $BF(N\pi\pi)_{corr}$ as:
\begin{equation}
\label{bnpipi}
BF(N\pi\pi)_{corr}=1-BF(N\pi).
\end{equation} 
For these resonance decays to the $N\pi\pi$ final states it turns out that the estimated branching fractions 
$BF(N\pi\pi)_{corr}$ from Eq.(\ref{bnpipi}) are slightly ($<$10\%) different with respect to those obtained 
from the $\pi^+\pi^-p$ fit ($BF(N\pi\pi)_{0}$). Therefore, we multiplied the $\pi \Delta$ and $\rho p$ hadronic 
decay widths of the $N(1440)1/2^+$ and $N(1520)3/2^-$ from the $\pi^+\pi^-p$ fit by the ratio 
$\frac{BF(N\pi\pi)_{corr}}{BF(N\pi\pi)_{0}}$. The $N(1440)1/2^+$ and $N(1520)3/2^-$ electrocouplings obtained 
in our analysis were then multiplied by the correction factors 
\begin{equation}
\label{bnpipi1}
C_{hd}=\sqrt{\frac{BF(N\pi\pi)_{0}}{BF(N\pi\pi)_{corr}}}
\end{equation} 
in order to keep the resonant parts and the computed differential $\pi^+\pi^-p$ cross sections unchanged 
under the re-scaling of the resonance hadronic decay parameters described above.

The electrocouplings of the $N(1440)1/2^+$, $N(1520)3/2^-$, and $\Delta(1620)1/2^-$ resonances were determined 
in our analysis for 0.5~GeV$^2$ $< Q^2 <$ 1.5~GeV$^2$, where there is still no data on observables of other 
exclusive meson electroproduction channels measured with CLAS. We have developed special procedures to test 
the reliability of the resonance $\gamma_vpN^*$ electrocouplings and their $\pi\Delta$ and $\rho p$ partial 
hadronic decay widths extracted from the charged double pion electroproduction data only. In order to check the 
reliability of the extracted $\gamma_vpN^*$ electrocouplings, we carried out the extraction of the resonance 
parameters of all of the aforementioned resonances independently, fitting the CLAS $\pi^+\pi^-p$ electroproduction 
data~\cite{Ri03} in the five overlapping intervals of $W$ given in Table~\ref{fitqual} covering in each fit the 
three $Q^2$-bins centered at 0.65~GeV$^2$, 0.95~GeV$^2$, and 1.30~GeV$^2$. The non-resonant amplitudes in each of 
the aforementioned $W$-intervals are different, while the resonance parameters should remain the same as they are 
determined from the data fit in different $W$-intervals. The $N(1440)1/2^+$, $N(1520)3/2^-$, and $\Delta(1620)1/2^-$ 
state electrocouplings extracted in the fit of the $\pi^+\pi^-p$ CLAS data~\cite{Ri03} in the different $W$-intervals 
are shown in Figs.~\ref{p11comp}, \ref{d13comp}, and \ref{s31comp}. 

The values of the $N(1440)1/2^+$ and $N(1520)3/2^-$ electrocouplings, as well as the $S_{1/2}$ electrocoupling
of the $\Delta(1620)1/2^-$, obtained from independent analyses of the different $W$-intervals, are consistent 
within the uncertainties. The values of the $A_{1/2}$ electrocoupling of the $\Delta(1620)1/2^-$ state from the 
fit of the $W$-interval from 1.51~GeV to 1.61~GeV are different in comparison to the fit results of the two others 
$W$-intervals. We consider the values of the $\Delta(1620)1/2^-$ electrocouplings determined in the $W$-interval 
from 1.56~GeV to 1.66~GeV as the most reliable, since the others $W$-intervals overlap only over part of the 
resonance line width of the $\Delta(1620)1/2^-$. The consistent results on the $\gamma_vpN^*$ electrocouplings 
from the independent analyses of different $W$-intervals strongly support the reliable extraction of these 
fundamental quantities, as well as the capability of the JM model to provide reliable information on the 
$\gamma_vpN^*$ resonance electrocouplings from analysis of the data on exclusive charged double-pion 
electroproduction.

\begin{table}
\begin{center}
\begin{tabular}{|c|c|c|} \hline
Q$^2$,  & $A_{1/2},$         & $S_{1/2}$         \\
GeV$^2$ & GeV$^{-1/2}$*1000  & GeV$^{-1/2}$*1000 \\ \hline
$0.65$  & $21.4 \pm 6.2$     & $25.7 \pm 5.9$ \\
$0.95$  & $29.6 \pm 6.5$     & $25.6 \pm 6.2$ \\
$1.30$  & $42.6 \pm 9.3$     & $29.4 \pm 5.5$ \\ \hline
\end{tabular}
\caption{Electrocouplings of the $N(1440)1/2^+$ resonance determined from this analysis of $\pi^+\pi^-p$ 
electroproduction off the proton~\cite{Ri03} at 1.41~GeV $< W <$ 1.51~GeV within the framework of the 
updated JM model described in Section~\ref{pipipmech}.}
\label{p11el} 
\end{center}
\end{table}

\begin{table}
\begin{center}
\begin{tabular}{|c|c|c|c|} \hline
Q$^2$,  & $A_{1/2},$         & $S_{1/2}$,         & $A_{3/2}$, \\
GeV$^2$ & GeV$^{-1/2}$*1000  & GeV$^{-1/2}$*1000  & GeV$^{-1/2}$*1000 \\ \hline
$0.65$  & $-52.9 \pm 7.5$    & $-29.4 \pm 3.1$    & $42.9 \pm 7.1$ \\
$0.95$  & $-50.8 \pm 7.9$    & $-27.3 \pm 7.1$    & $32.7 \pm 6.2$ \\
$1.30$  & $-39.6 \pm 6.3$    & $-9.8 \pm 2.9$     & $25.2 \pm 3.3$ \\ \hline
\end{tabular}
\caption{Electrocouplings of the $N(1520)3/2^-$ resonance determined from this analysis of $\pi^+\pi^-p$ 
electroproduction off the proton~\cite{Ri03} at 1.46~GeV $< W <$ 1.56~GeV within the framework of the 
updated JM model described in Section~\ref{pipipmech}.}
\label{d13el} 
\end{center}
\end{table}

\begin{table}
\begin{center}
\begin{tabular}{|c|c|c|} \hline
Q$^2$,  & $A_{1/2},$         &  $S_{1/2}$        \\
GeV$^2$ & GeV$^{-1/2}$*1000  &  GeV$^{-1/2}$*1000 \\ \hline
$0.65$  & $15.5 \pm  10.2$   & $-46.3 \pm 3.8$  \\
$0.95$  & $12.5 \pm 5.4$     & $-30.9 \pm 7.0$ \\
$1.30$  & $5.5 \pm 4.4$      & $-17.2 \pm 5.6$ \\ \hline
\end{tabular}
\caption{Electrocouplings of the $\Delta(1620)1/2^-$ resonance determined from this analysis of $\pi^+\pi^-p$ 
electroproduction off the proton~\cite{Ri03} at 1.56~GeV $< W <$ 1.66~GeV within the framework of the 
updated JM model described in Section~\ref{pipipmech}.}
\label{s31el} 
\end{center}
\end{table}

\begin{table*}
\begin{center}
\begin{tabular}{|c|c|c|c|} \hline
Parameter & Current analysis of the CLAS      & Previous analysis \cite{Mo12} of the CLAS & RPP \\
          & $\pi^+\pi^-p$ data \cite{Ri03} at & $\pi^+\pi^-p$ data \cite{Fe09} 0.25 at    &   \\
          & 0.5 GeV$^2$ $< Q^2 < 1.5$~GeV$^2$ & 0.25 ~GeV$^2$ $< Q^2  < 0.6$~GeV$^2$      &   \\ \hline
Breit-Wigner mass, MeV                & 1454 $\pm$ 11  & 1458 $\pm$ 12 & 1420-1470 ($\approx$ 1440) \\
Breit-Wigner width, MeV               &  352 $\pm$ 37  & 363 $\pm$ 39  & 200-450 ($\approx$ 300) \\
$\pi \Delta$ partial decay width, MeV & 120 $\pm$ 41   & 142 $\pm$ 48  & \\
$\pi \Delta$ BF,                      & 20\%-52\%      & 23\%-58\%     & 20\%-30\% \\
$\rho p$ partial decay width, MeV     & 4.9 $\pm$ 2.2  & 6.2 $\pm$ 4.1 & \\
$\rho p$ BF                           & $<$ $~$2.0\%   & $<$ $~$2.0\%  & $<$ $~$8.0\% \\ \hline
\end{tabular}
\caption{Hadronic parameters of the $N(1440)1/2^+$ resonance determined from the CLAS data~\cite{Ri03} on
$\pi^+\pi^-p$ electroproduction off the proton within the framework of the updated JM model described in 
Section~\ref{pipipmech} in comparison with the results of our previous analysis~\cite{Mo12} and RPP~\cite{Rpp12}.}
\label{hpp11} 
\end{center}
\end{table*}

\begin{table*}
\begin{center}
\begin{tabular}{|c|c|c|c|} \hline
Parameter & Current analysis of the CLAS      & Previous analysis \cite{Ri03} of the CLAS & RPP \\
          & $\pi^+\pi^-p$ data \cite{Ri03} at & $\pi^+\pi^-p$ data at                     &     \\
          & 0.5~GeV$^2$ $< Q^2 < 1.5$~GeV$^2$ & 0.25 ~GeV$^2$ $< Q^2 < 0.6$~GeV$^2$       &     \\  \hline
Breit-Wigner mass, MeV                & 1522 $\pm$ 5 & 1521 $\pm$ 4 & 1515-1525 ($\approx$ 1520) \\
Breit-Wigner width, MeV               &  125 $\pm$ 4 & 127 $\pm$ 4  & 100-125 ($\approx$ 115) \\
$\pi \Delta$ partial decay width, MeV & 36 $\pm$ 5   & 35 $\pm$ 4   & \\
$\pi \Delta$ BF                       & 25\%-34\%    & 24\%-32\%    & 15\%-25\% \\
$\rho p$ partial decay width, MeV     & 13 $\pm$ 6   & 16 $\pm$ 5   & \\
$\rho p$ BF                           & 4.8\%-16\%   & 8.4\%-17\%   & 15\%-25\% \\ \hline
\end{tabular}
\caption{Hadronic parameters of the $N(1520)3/2^-$ resonance determined from the CLAS data~\cite{Ri03} on
$\pi^+\pi^-p$ electroproduction off the proton within the framework of the updated JM model described in 
Section~\ref{pipipmech} in comparison with the results of our previous analysis~\cite{Mo12} and RPP~\cite{Rpp12}.}
\label{hpd13} 
\end{center}
\end{table*}

\begin{table*}
\begin{center}
\begin{tabular}{|c|c|c|} \hline
Parameter & Current analysis of the CLAS      & RPP \\
          & $\pi^+\pi^-p$ data~\cite{Ri03} at &     \\ 
          & 0.5~GeV$^2$ $< Q^2 < 1.5$~GeV$^2$ &     \\ \hline
Breit-Wigner mass, MeV                & 1631 $\pm$ 12 & 1600-1660 ($\approx$ 1630) \\
Breit-Wigner width, MeV               &  148 $\pm$ 10 & 130-150 ($\approx$ 140) \\
$\pi \Delta$ partial decay width, MeV & 66 $\pm$ 23   & \\
$\pi \Delta$ BF,                      & 27\%-64\%     & 30\%-60\% \\
$\rho p$ partial decay width, MeV     & 70 $\pm$ 21   & \\
$\rho p$ BF                           & 31\%-63\%     & 7\%-25\% \\ \hline
\end{tabular}
\caption{Hadronic parameters of the $\Delta(1620)1/2^-$ resonance determined from the CLAS data~\cite{Ri03} 
on $\pi^+\pi^-p$ electroproduction off the proton within the framework of the updated JM model described in 
Section~\ref{pipipmech} in comparison with RPP~\cite{Rpp12}.}
\label{hps31} 
\end{center}
\end{table*}

The final results on the $N(1440)1/2^+$, $N(1520)3/2^-$, and $\Delta(1620)1/2^-$ electrocouplings are listed 
in Tables~\ref{p11el},~\ref{d13el}, and \ref{s31el}. They were determined from the fit of the CLAS data~\cite{Ri03} 
in the $W$-intervals given in the captions of Tables~\ref{p11el},~\ref{d13el}, and \ref{s31el} covering the three 
$Q^2$-bins centered at 0.65~GeV$^2$, 0.95~GeV$^2$, and 1.30~GeV$^2$. The intervals over $W$ within which the 
resonance electrocouplings were extracted were determined by the requirement that the selected $W$ intervals 
overlap the area of masses below and above the central resonance mass values. The resonance electrocoupling 
uncertainties reflect both the experimental data uncertainties and the systematic uncertainties imposed by 
the extraction model. 

The $A_{1/2}$ electrocouplings of the $N(1440)1/2^+$ state are positive and increase with $Q^2$, supporting 
the zero crossing observed for this electroexcitation amplitude in our previous analyses of the CLAS $N\pi$ 
and $N\pi\pi$ electroproduction data~\cite{Az09,Mo12}. The $A_{1/2}$ electrocouplings of the $N(1520)3/2^-$ 
state are negative and increase with photon virtualities, confirming the local minimum at $Q^2 \approx 0.45$~GeV$^2$ 
observed in our previous analyses~\cite{Az09,Mo12}. The electroexcitation of the $\Delta(1620)1/2^-$ resonance 
is dominated by longitudinal electrocouplings in the entire area of photon virtualities covered in our analysis, 
0.5~GeV$^2$ $< Q^2 <$ 1.5~GeV$^2$.

In this analysis we also obtained the hadronic decay widths of the $N(1440)1/2^+$, $N(1520)3/2^-$, and 
$\Delta(1620)1/2^-$ resonances to the $\pi \Delta$ and $\rho p$ final states. These parameters were 
determined from the $\pi^+\pi^-p$ electroproduction data~\cite{Ri03} under simultaneous variations of the
resonance masses, $\gamma_vpN^*$ electrocouplings, and hadronic decay widths to the $\pi \Delta$ and $\rho N$ 
final states under the requirement of $Q^2$-independence of the resonance masses and hadronic decay parameters. 

The $N(1440)1/2^+$ and $N(1520)3/2^-$ masses, as well as the branching fractions for the decays to the 
$\pi \Delta$ and $\rho N$ final states extracted in the fit of the data~\cite{Ri03}, are given in 
Tables~\ref{hpp11} and \ref{hpd13} in comparison with the results of our previous analysis~\cite{Mo12} of the 
CLAS $\pi^+\pi^-p$ electroproduction data~\cite{Fe09} carried out at smaller $W$ and $Q^2$. The results of our 
current analysis on the $N(1440)1/2^+$ and $N(1520)3/2^-$ masses, and their total and partial hadronic decay 
widths to the $\pi \Delta$ and $\rho p$ final states are consistent. A successful description of the CLAS 
$\pi^+\pi^-p$ electroproduction data over different and wide ranges of photon virtualities, 0.25~GeV$^2$ 
$< Q^2 <$ 0.6~GeV$^2$ (previous analysis~\cite{Mo12}) 0.5~GeV$^2$ $< Q^2 <$ 1.5~GeV$^2$ (current analysis), 
strongly support the reliable separation of the resonant and non-resonant contributions achieved within the 
framework of the JM model and the credible extraction of the $N(1440)1/2^+$ and $N(1520)3/2^-$ resonance 
parameters. Both the current and previous analyses of the CLAS $\pi^+\pi^-p$ electroproduction data suggest 
that the $\rho p$ hadronic decay widths of the $N(1440)1/2^+$ and $N(1520)3/2^-$ resonances are smaller than
those reported in the RPP, and that the $\pi \Delta$ hadronic decay widths of the $N(1520)3/2^-$ are larger than 
those reported in RPP~\cite{Rpp12}. The successful description of the CLAS $\pi^+\pi^-p$ electroproduction data
\cite{Fe09,Ri03} in a wide area of $Q^2$ from 0.25~GeV$^2$ to 1.5~GeV$^2$ achieved with $Q^2$-independent 
resonance hadronic parameters, makes the results presented in Tables~\ref{hpp11} and \ref{hpd13} reliable. They 
offer new information on the hadronic decays of the $N(1440)1/2^+$ and $N(1520)3/2^-$ resonances to the $\pi \Delta$ 
and $\rho N$ final states that may be considered as input for the upcoming RPP edition. 

For the first time the hadronic decay parameters of the $\Delta(1620)1/2^-$ resonance listed in Table~\ref{hps31}, 
have become available from the analysis of the $\pi^+\pi^-p$ electroproduction data. The mass, total width, and 
the branching fractions for decays of the $\Delta(1620)1/2^-$ to the $\pi \Delta$ final states obtained in our 
analysis are in good agreement with the RPP results~\cite{Rpp12}. The current analysis suggests much 
larger values of the branching fractions for decays of the $\Delta(1620)1/2^-$ to the $\rho p$ final states in 
comparison with those presented in RPP. A successful description of the CLAS $\pi^+\pi^-p$ electroproduction 
data~\cite{Ri03} with $Q^2$ independent values of the $\Delta(1620)1/2^-$ hadronic decay widths strongly supports 
the branching fraction values listed in Table~\ref{hps31}. The large values determined for the branching fraction 
for decays of the $\Delta(1620)1/2^-$ to the $\rho p$ final states represent an interesting and unexpected 
result, since the $\Delta(1620)1/2^-$ state is located in the sub-threshold area for $\rho p$ electroproduction 
off the proton. The absence of $\rho$ peaks in the data on the $\pi^+\pi^-$ invariant mass distributions at 
$W \approx 1.6$~GeV, in combination with large $\rho p$ hadronic decays of the $\Delta(1620)1/2^-$, impose 
restrictions on the upper limits of the $A_{1/2}$ electrocouplings for the $\Delta(1620)1/2^-$ state, making their 
absolute values much smaller than those of the $S_{1/2}$ electrocouplings.

\begin{figure*}[htp]
\begin{center}
\includegraphics[width=14.5cm]{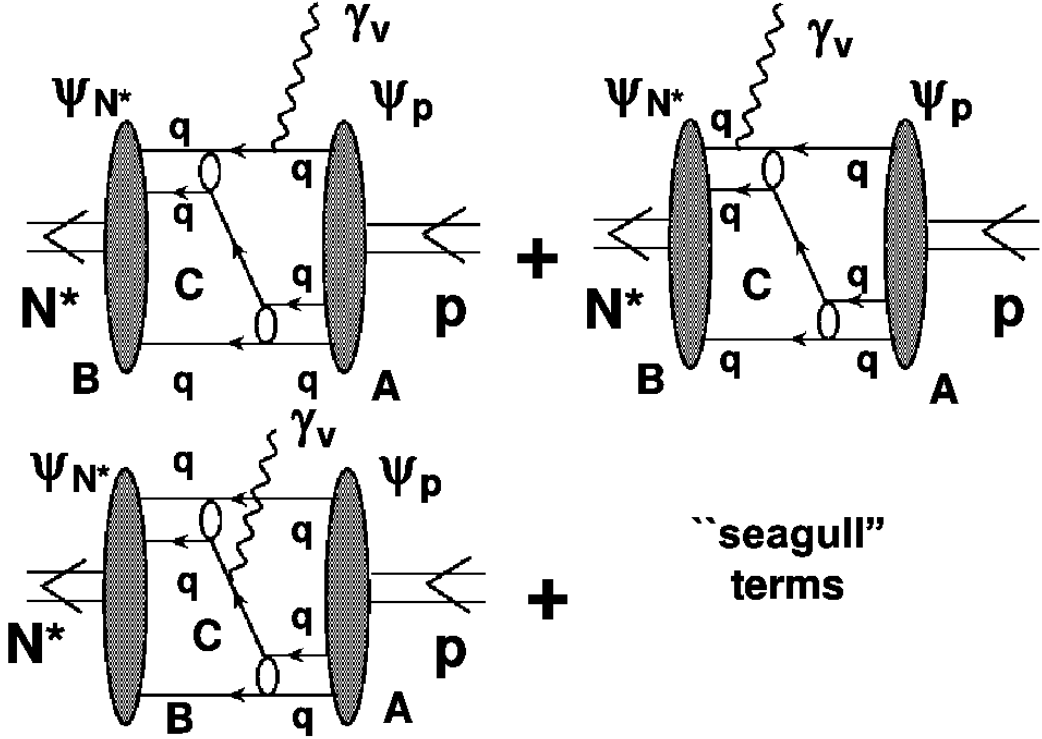}
\vspace{-0.1cm}
\caption{Description of the resonance electroexcitation amplitudes within the framework of DSEQCD
\cite{Cr14,Cr15,Cr15a}: A) the amplitude for the transition $p \to$ three dressed quarks or the ground state 
wave function $\psi_p$, B) the amplitude for the transition three dressed quarks $\to N^*$ or the excited 
nucleon state wave function $\psi_{N^*}$, C) the amplitude that describes the interaction between the virtual 
photon and three dressed quarks bound by the non-perturbative strong interaction between pairs of correlated 
quarks (di-quark) and by the dressed quark exchange between the di-quark pair and third quark. The virtual photon 
interactions with the quark and di-quark currents are shown on the left and top right diagrams, respectively. 
The di-quark currents incorporate the transitions between di-quarks of the same and different quantum numbers. 
The full $N \rightarrow N^*$ transition amplitude can be found in Fig.~C1 of Ref.~\cite{Cr15}.} 
\label{diagdse}
\end{center}
\end{figure*}

\begin{figure*}[htp]
\begin{center}
\includegraphics[width=8.5cm]{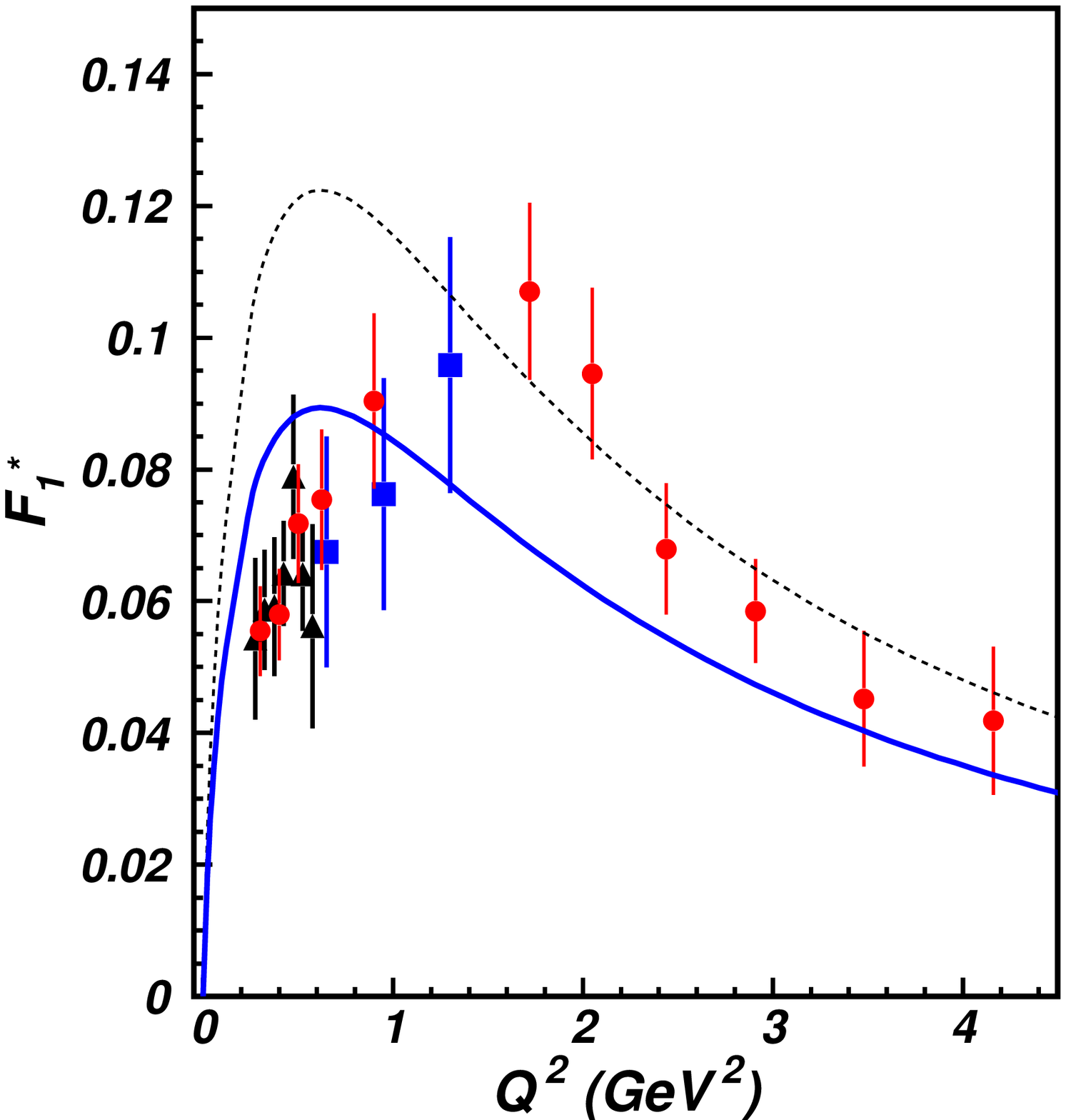}
\includegraphics[width=8.5cm]{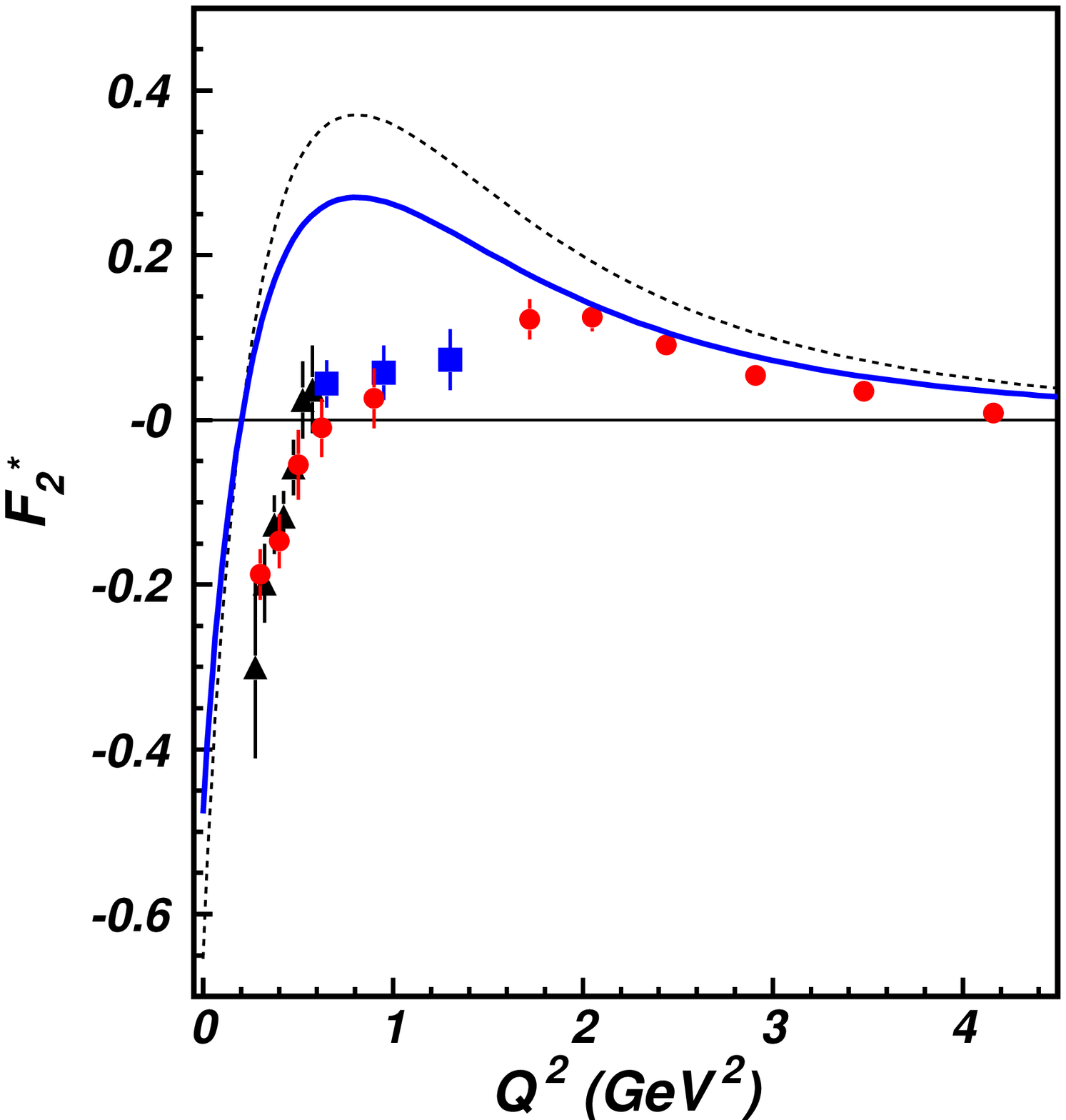}
\vspace{-0.1cm}
\caption{(Color Online) The $F_1^*$ and $F_2^*$ transition $p \to N(1440)1/2^+$ form factors: experimental 
results from analyses of the CLAS data on $N\pi$~\cite{Az09} (red circles) and $\pi^+\pi^-p$~\cite{Mo12} 
(black triangles) electroproduction off the proton and the results of this present work (blue squares). The
data are shown in comparison with DSEQCD evaluations~\cite{Cr15a} start from the QCD Lagrangian (black 
dashed line) and after accounting for the meson-baryon cloud contributions as described in Section~\ref{impp11d13} 
(blue thick solid line).}
\label{p11datdse}
\end{center}
\end{figure*} 

\section{Impact on Studies of the $N^*$ Structure from the New CLAS Results}
\label{impact}

In this section we discuss the impact of the new CLAS results on the $N(1440)1/2^+$, $N(1520)3/2^-$, and
$\Delta(1620)1/2^-$ electrocouplings and their partial hadronic decay widths to the $\pi \Delta$ and 
$\rho N$ final states on the contemporary understanding of the structure of these states. We will also 
outline new possibilities for hadron structure theory to employ these experimental results in order 
to explore how the dynamical properties of three constituent dressed quarks inside the resonance quark core 
emerge from QCD. 

\subsection{$N(1440)1/2^+$ and $N(1520)3/2^-$ Resonances} 
\label{impp11d13}  

Previous studies of the $N(1440)1/2^+$ and $N(1520)3/2^-$ resonances with the CLAS detector~\cite{Az09,Mo12} 
have provided the dominant part of the world-wide information available on their electrocouplings in a wide 
range of photon virtualities 0.25~GeV$^2$ $< Q^2 <$ 5.0~GeV$^2$. This paper extends the CLAS results on the
$N(1440)1/2^+$ and $N(1520)3/2^-$ $\gamma_vpN^*$ electrocouplings in the range of photon virtualities from
0.5~GeV$^2$ to 1.5~GeV$^2$ where there is limited availability of data. Previous studies of $\pi^+\pi^-p$ 
electroproduction~\cite{Mo12} have allowed us to determine the $N(1440)1/2^+$ and $N(1520)3/2^-$ partial 
decay widths to the $\pi \Delta$ and $\rho p$ final states. Our current studies confirmed the previous 
results~\cite{Mo12} on these hadronic decays. Currently the $N(1440)1/2^+$ and $N(1520)3/2^-$ states, 
together with the $\Delta(1232)3/2^+$ and $N(1535)1/2^-$ resonances~\cite{Bu12}, represent the most explored 
excited nucleon states. Detailed information on the electrocouplings of these states that are available for 
the first time from CLAS, have already provided a profound impact on the contemporary understanding of the 
nucleon resonance structure~\cite{Bu12,Az13,Mo12}. 

Recent progress in the studies of resonance structure achieved within the framework of the Dyson-Schwinger 
Equations of QCD (DSEQCD)~\cite{Cr14,Cr15,Cr15a} has allowed us for the first time to interpret the 
experimental results on the nucleon elastic form factors, as well as the magnetic $p \to \Delta$ and 
$p \to N(1440)1/2^+$ Dirac ($F_1^*$) and Pauli ($F_2^*$) transition from factors start from the QCD 
Lagrangian. Currently this approach is capable of evaluating the contributions from the quark core of three 
dressed quarks to the nucleon elastic and $p \to N^*$ transition form factors. DSEQCD approaches describe 
the ground and excited nucleons as bound systems of three dressed quarks that represent the complex objects 
generated non-perturbatively from an infinite number of QCD quarks and gauge gluons. The  dynamical properties 
of dressed quarks, the momentum dependent mass $M(p)$ and form factors, that enter into the quark electromagnetic 
current, are determined start from the QCD Lagrangian employing the towers of gap equations for quarks and 
gluons~\cite{Cr14}. The ground and excited nucleon state masses and the transition amplitudes, $p \to$ three 
dressed quarks (the ground state wave function) and three dressed quarks $\to N^*$ (the excited nucleon state 
wave function), are obtained in a Poincar$\acute{\rm e}$ covariant approach employing Faddeev equations for the 
three dressed quarks. The non-perturbative interactions between the three dressed quarks are reduced to a 
quark-quark interaction that generates pairs of correlated quarks, the so-called dynamical di-quark, and 
dressed quark exchanges between the di-quark pair and third quark shown in the parts labeled ``C'' in 
Fig.~\ref{diagdse}~\cite{Cr14,Cr15c}. The ground and excited nucleon state masses emerge as poles in the 
energy dependence of the amplitude with the respective spin-parity that comes from the Faddeev equation 
solution. The ground/excited nucleon state wave functions represent the residues of the Faddeev equation 
solutions at the respective pole positions. The resonance electroexcitation amplitudes, depicted in 
Fig.~\ref{diagdse}, are evaluated as the product of three amplitudes: A) ground state $p \to$ three dressed 
quarks, B) three dressed quarks $\to$ resonance $N^*$, and C) interaction between real/virtual photons and 
the three dressed quarks. The latter part C is described mostly by real/virtual photon couplings to the dressed quark 
and di-quark pair currents. All details on the evaluations of resonance electroexcitation amplitudes can be 
found in Refs.~\cite{Cr15,Cr15a}.

The resonance electroexcitation amplitudes shown in Fig.~\ref{diagdse} should be sensitive to the momentum 
dependence of the dressed quark mass $M(p)$, since it affects all quark propagators and dressed quark currents. 
Moreover, it was shown in Refs.~\cite{Cr15,Cr15a} that the momentum dependence of the dressed quark mass has a 
pronounced influence on the wave functions of the ground and excited nucleon states. DSEQCD studies of 
experimental results on elastic nucleon form factors~\cite{Cr13} confirmed these expectations and revealed 
considerable sensitivity of the nucleon elastic form factors to the momentum dependence of the dressed quark 
mass function. It was found that the location of the zero crossing for the ratio $\mu_pG_E/G_M(Q^2)$ is 
determined by the derivative of the dressed quark mass function $M(p)$. 

\begin{figure*}[htp]
\begin{center}
\includegraphics[width=8.5cm]{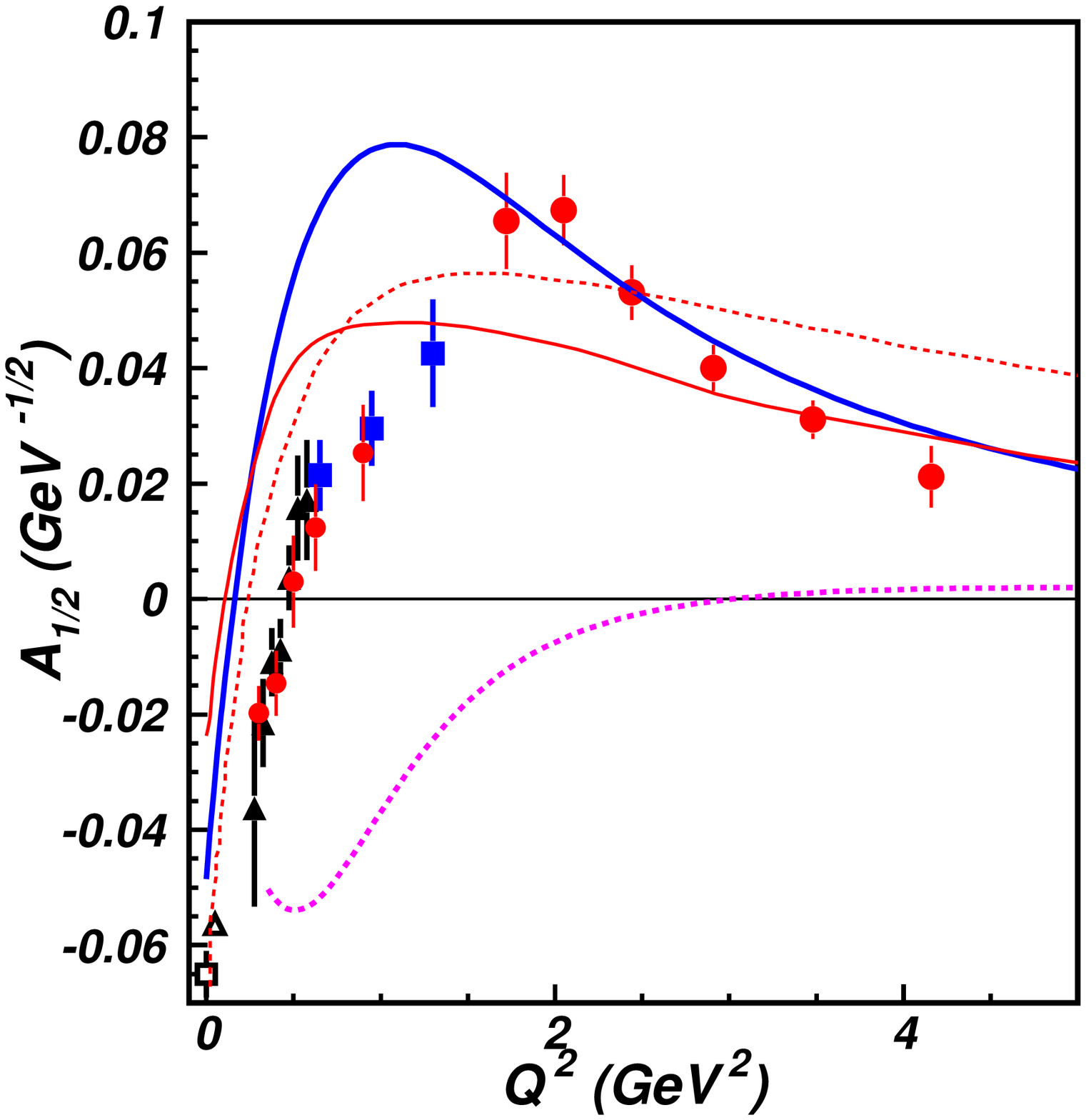}
\includegraphics[width=8.5cm]{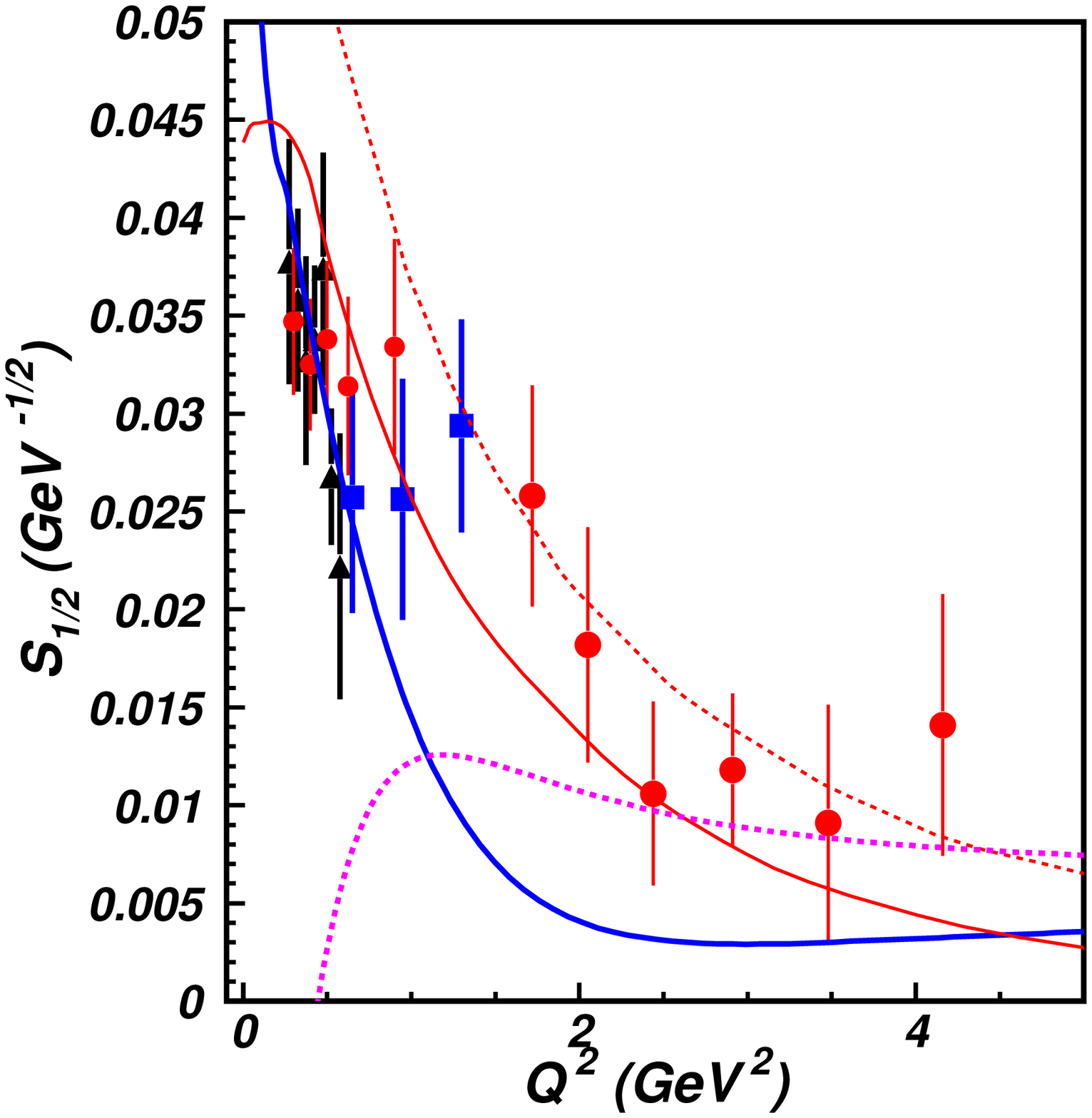}
\vspace{-0.1cm}
\caption{(Color Online) The $A_{1/2}$ and $S_{1/2}$ $\gamma_vpN^*$ electrocouplings of the $N(1440)1/2^+$ 
resonance: experimental results from analyses of the CLAS data on $N\pi$~\cite{Az09} (red circles) and 
$\pi^+\pi^-p$~\cite{Mo12} (black triangles) electroproduction off the proton and the results of this 
work (blue squares). The data are shown in comparison with the DSEQCD evaluations~\cite{Cr15a} (blue thick 
solid) and the results from constituent quark models that account for the contributions from both the quark core 
and the meson-baryon cloud:~\cite{Az12} (thin red solid) and~\cite{Ob14} (thin red dashed). The calculations of 
thin red line includes pion loops and a parametrization of the running quark mass, and the calculations of 
dashed red line contains $N\sigma$ contributions and fixed quark mass. The meson-baryon cloud contributions 
obtained from the experimental data (see Section~\ref{impp11d13}) are shown by the magenta thick dashed lines.}
\label{p11a12s12dseqm}
\end{center}
\end{figure*} 

The need to employ a 
momentum-dependent dressed quark mass function was conclusively demonstrated in the studies of the 
$N \to \Delta$ magnetic transition form factor within the DSEQCD framework~\cite{Cr14}.
Computations employing a dressed quark with a momentum-independent mass generated by simplified contact 
quark-quark interactions were able to describe the experimental results only in a very limited range of 
photon virtualities $Q^2 < 3.0$~GeV$^2$. Instead, the DSEQCD evaluation with running quark mass successfully 
reproduced the experimental data at $Q^2 > 1.0$~GeV$^2$ in the entire range of photon virtualities covered by 
measurements reaching up to 8.0~GeV$^2$. 

The recent DSEQCD studies of the $N(1440)1/2^+$ resonance electroexcitation~\cite{Cr15a} derive from a realistic 
quark-quark interaction that generates a momentum-dependent dressed quark mass function. The evaluated 
contributions from the quark core to the Dirac $F_1^*$ and to the Pauli $F_2^*$ $p \to N(1440)1/2^+$ transition  
form factors are shown in Fig.~\ref{p11datdse} by the dashed lines in comparison with the CLAS experimental 
results published in Refs.~\cite{Az09,Mo12}, as well as with those obtained in this present work. DSEQCD 
reasonably reproduces the experimental results for $Q^2 > 2.5$~GeV$^2$. However, a pronounced disagreement for 
$Q^2 < 1.0$~GeV$^2$, in particular, for the Pauli $F_2^*$ form factor, suggests significant contributions 
from degrees of freedom other than the quark core, presumably the meson-baryon cloud found in the global 
multi-channel analysis of exclusive meson photo-, electro-, and hadroproduction data~\cite{Lee08}. These 
contributions are still beyond the scope of DSEQCD studies~\cite{Cr15a}. However, we have to account for the 
fraction of the meson-baryon degrees of freedom in the wave functions of the ground and excited nucleon states. 
We choose to estimate this contribution by multiplying the $p \to N(1440)1/2^+$ transition form factors computed 
within the DSEQCD approach~\cite{Cr15a} by a common $Q^2$-independent factor fit to the data for $Q^2 > 3.0$~GeV$^2$, 
where the meson-baryon cloud contributions are much smaller than those from the quark core. The fit value of this 
factor of 0.73 is consistent with the results of a recent advanced light front quark model~\cite{Az12}, which 
employs the parameterization of running quark mass function in spirit of DSEQCD~\cite{Cr14}. The $p \to N(1440)1/2^+$ transition form factors obtained in 
this way are shown in Fig.~\ref{p11datdse} by the solid blue lines. A good description of the experimental 
results for $Q^2 > 1.5$~GeV$^2$ is achieved within the entire range of photon virtualities covered by the 
measurements.

\begin{figure*}[htp]
\begin{center}
\includegraphics[width=5.8cm]{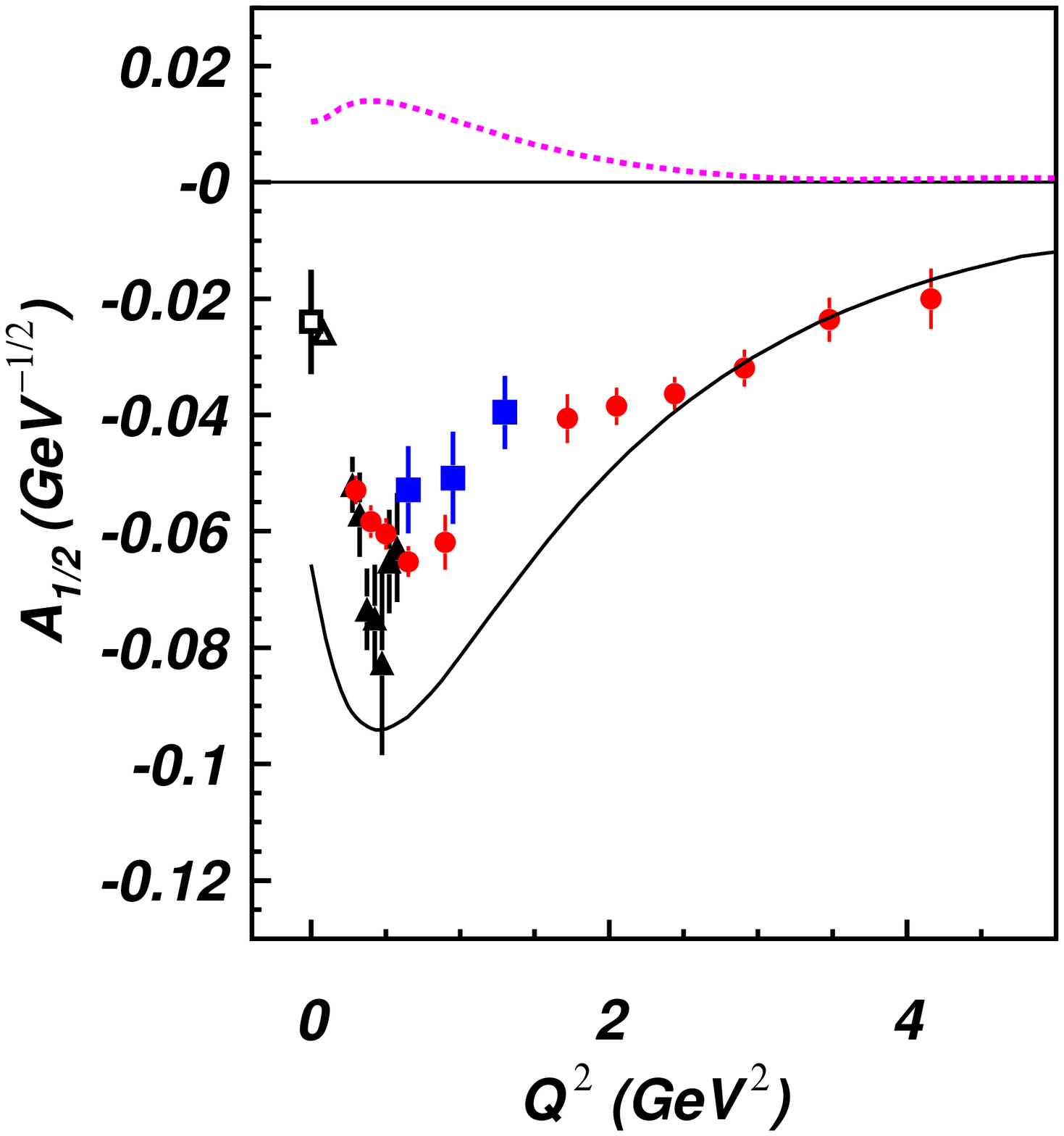}
\includegraphics[width=5.8cm]{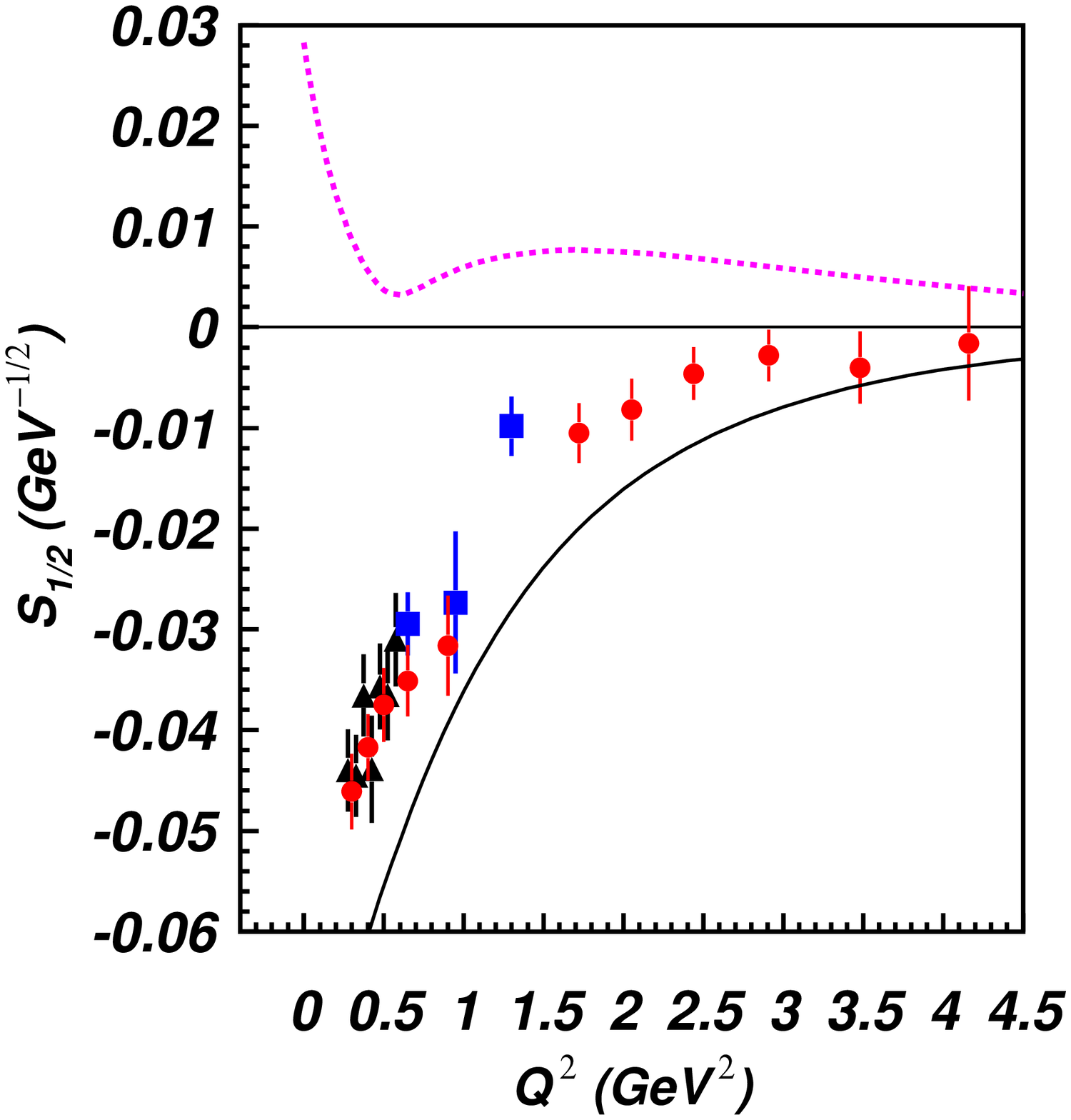}
\includegraphics[width=5.8cm]{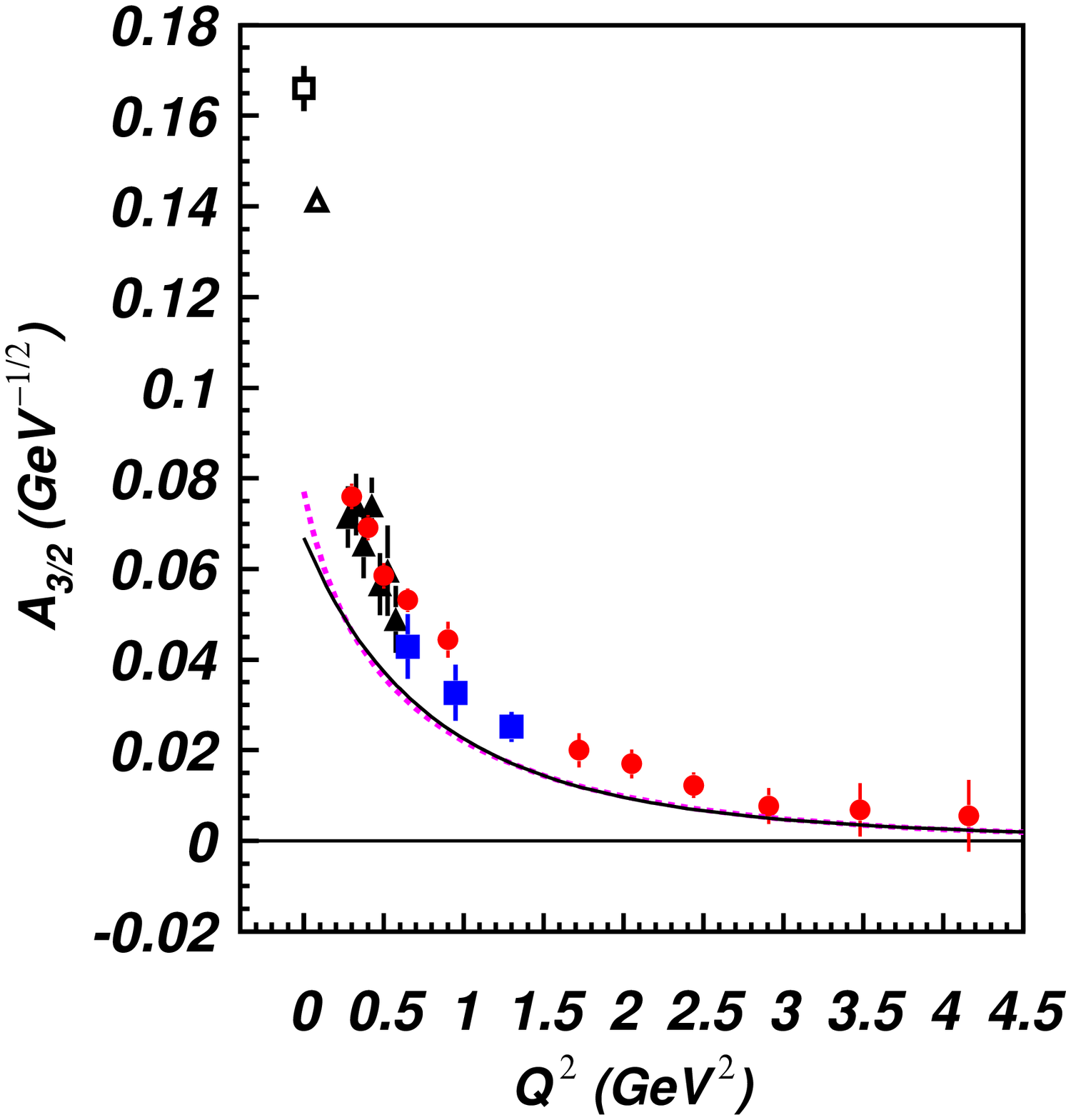}
\vspace{-0.1cm}
\caption{(Color Online) The $A_{1/2}$, $S_{1/2}$, and $A_{3/2}$ $\gamma_vpN^*$ electrocouplings of the
$N(1520)3/2^-$ resonance: experimental results from analyses of the CLAS data on $N\pi$~\cite{Az09} (red 
circles) and $\pi^+\pi^-p$~\cite{Mo12} (black triangles) electroproduction off the proton and the results 
of this present work (blue squares). The data are shown in comparison with the predictions of the 
hypercentral quark model~\cite{Sa12} (thin black solid). The meson-baryon cloud contributions obtained in a
global multi-channel $N\pi$ photo-, electro-, and hadroproduction data analysis~\cite{Lee08} are shown by the
thick dashed magenta lines.}
\label{d13elcoupl}
\end{center}
\end{figure*} 

The dressed quark mass function used in the DSEQCD computations of the $p \to N(1440)1/2^+$ transition form 
factors~\cite{Cr15a} is {\it exactly the same} as that employed in the previous evaluations of the nucleon 
elastic and magnetic $p \to \Delta$ transition form factors~\cite{Cr15,Cr13}. The $\Delta(1232)3/2^+$ and 
$N(1440)1/2^+$ excited nucleon states have a distinctively different structure: spin-flavor flip for the
$\Delta(1232)3/2^+$ and the first radial excitation of three dressed quarks for the $N(1440)1/2^+$. A successful 
description of the elastic and transition form factors to nucleon resonances of distinctively different 
structure achieved with the same dressed quark mass function strongly underlines:
\begin{itemize}
\item the relevance of dynamical dressed quarks with the properties predicted by the DSEQCD approach~\cite{Cr14} 
as constituents of the quark cores for the structure of the ground and excited nucleon states;
\item the capability of the DSEQCD approach~\cite{Cr15,Cr15a} to map out the dressed quark mass function from 
the experimental results on the $Q^2$-evolution of the nucleon elastic and $p \to N^*$ transition form 
factors ($\gamma_vpN^*$ electrocouplings) from a combined analysis.
\end{itemize} 
Consistent results on the momentum dependence of the dressed quark mass function obtained from independent 
analyses of nucleon elastic and $p \to N^*$ form factors, i.e. $\gamma_vpN^*$ electrocouplings for the transition 
to excited nucleons with different quantum numbers, are critical in order to prove the reliable access to this 
fundamental quantity.

DSEQCD analyses~\cite{Cr15,Cr15a} of the CLAS results on the $p \to \Delta$ and $p \to N(1440)1/2^+$ transition 
form factors (the latter shown in Fig.~\ref{p11a12s12dseqm}) have demonstrated the capability of accessing the dressed 
quark mass function from the experimental data for the first time. Studies of the dressed quark mass function 
will address the most challenging and still open problems of the Standard Model on the nature of the dominant 
part of the hadron mass, quark-gluon confinement, its emergence from QCD, and its relation to dynamical chiral 
symmetry breaking, which is expected to be the source of more than 98\% of the hadron mass in universe~\cite{Cr14}.

Recent advances in the development of the constituent quark models make it possible to extend the $Q^2$-range 
for a better description of the $\gamma_vpN^*$ electrocouplings in comparison with DSEQCD approaches, taking 
into account both contributions from the quark core and the meson-baryon cloud. The two models~\cite{Az12,Ob14} 
describe the structure of the $N(1440)1/2^+$ resonance as an interplay between the 
contributions from the inner core of three dressed quarks in the first radial excitation and an external 
meson-baryon cloud. Both approaches treat the quark core contributions within the light front framework. 
The first model~\cite{Az12} employs a phenomenological momentum-dependent dressed quark mass motivated by 
the DSEQCD results~\cite{Cr15,Cr15a}, while the second~\cite{Ob14} employs constituent quarks of 
momentum-independent mass. The meson baryon cloud is modeled by $\pi N$ loops in the first approach~\cite{Az12}, 
while the $\sigma p$ loops are employed in the second approach~\cite{Ob14}. The CLAS experimental results on the
$A_{1/2}$ and $S_{1/2}$ $\gamma_vpN^*$ electrocouplings of the $N(1440)1/2^+$ resonance are shown in 
Fig.~\ref{p11a12s12dseqm} in comparison with the expectations from DSEQCD~\cite{Cr15a} and from the 
aforementioned two advanced constituent quark models~\cite{Az12,Ob14}. Accounting for the meson-baryon cloud 
contributions allowed us to considerably improve the description of the experimental data at $Q^2 < 2.0$~GeV$^2$, 
confirming the relevance of meson-baryon degrees of freedom in the $N(1440)1/2^+$ structure at these distances 
that had previously been established in multi-channel analyses of exclusive meson photo-, electro-, and 
hadroproduction experimental data~\cite{Lee08}.

The CLAS results on the $\gamma_vpN^*$ electrocouplings of the $N(1520)3/2^-$ resonance are shown in 
Fig.~\ref{d13elcoupl}. The currently available models for the description of the structure of this state 
account for quark core contributions only. The quark core contributions to the $\gamma_vpN^*$ 
electrocouplings of most well-established excited nucleon states were explored within the framework of  
two conceptually different approaches: a) hypercentral constituent quark model~\cite{Sa12} and b) 
Bethe-Salpeter approach that employs structureless constituent quarks with momentum-independent mass and an
instanton quark-quark interaction~\cite{Met12}. The hypercentral constituent quark model provides a reasonable 
description of the experimental results at $Q^2 > 1.0$~GeV$^2$ as shown in Fig.~\ref{d13elcoupl}. At smaller 
photon virtualities there are substantial discrepancies between the model~\cite{Sa12} and the CLAS results. 
A similar observation comes from the comparison of the CLAS results with the Bethe-Salpeter approach
\cite{Met12}. Estimates for the contributions from the meson-baryon cloud to the structure of the 
$N(1520)3/2^-$ resonance were obtained in Ref.~\cite{Lee08} from a global multi-channel analysis of the 
experimental data on exclusive pion photo-, electro-, and hadroproduction. The absolute values of the 
meson-baryon cloud shown in Fig.~\ref{d13elcoupl} are maximal at small photon virtualities where discrepancies 
between the quark model expectations and the experimental data are largest. Hence, the meson-baryon cloud 
contributions may explain the difference between the CLAS data and the quark model expectations for the 
$\gamma_vpN^*$ electrocouplings of the $N(1520)3/2^-$ resonance. The aforementioned studies of the CLAS data 
in Fig.~\ref{d13elcoupl} suggest that the structure of the $N(1520)3/2^-$ resonance arises from the contributions 
from the inner core of three dressed quarks in the first orbital excitation with $L=1$ and the external 
meson-baryon cloud. The contributions from the meson-baryon cloud are strongly dependent on the helicity of 
the $N^*$ electroexcitation amplitudes. They decrease with photon virtuality $Q^2$.  

\begin{figure*}[htp]
\begin{center}
\includegraphics[width=8.5cm]{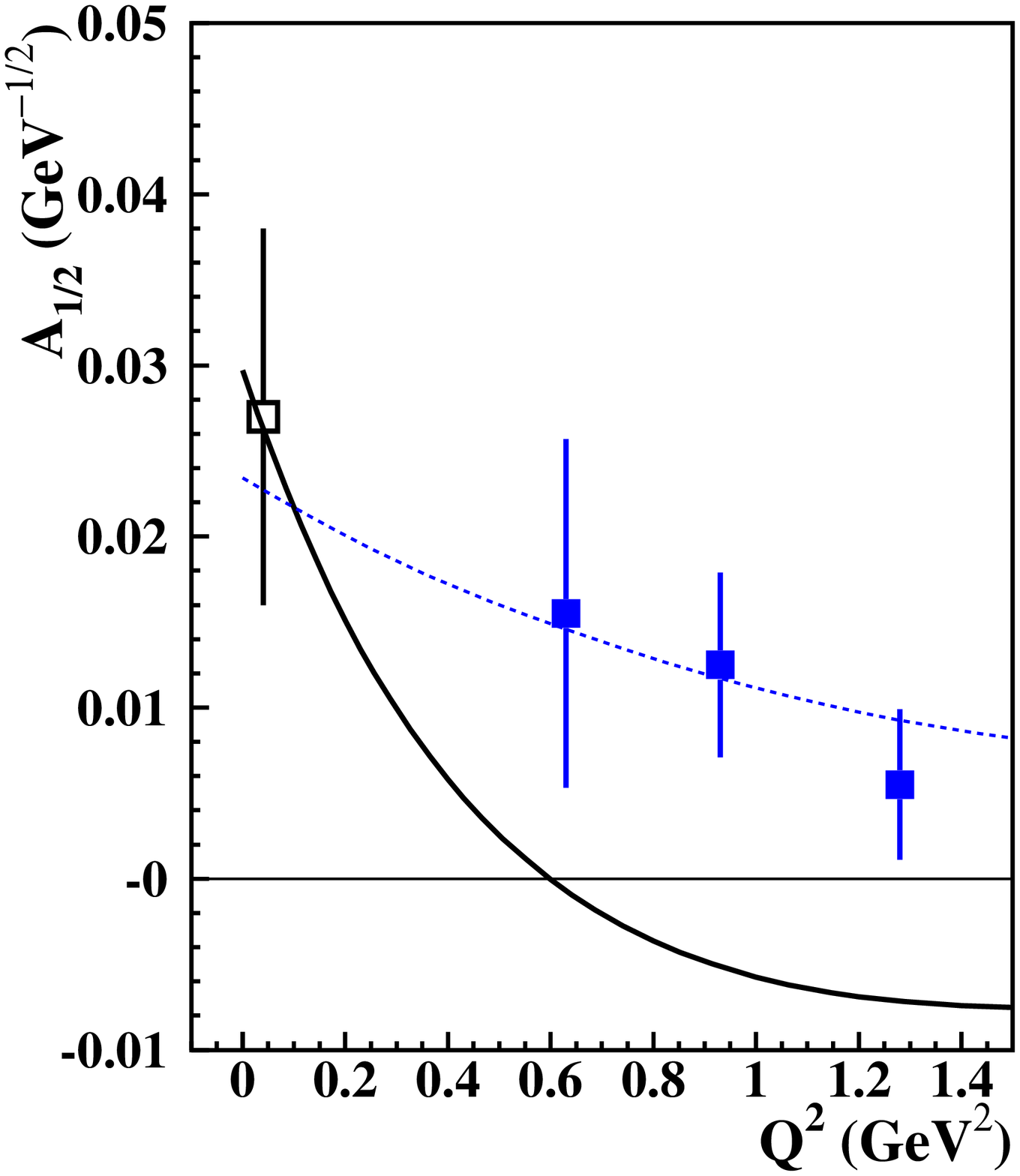}
\includegraphics[width=8.5cm]{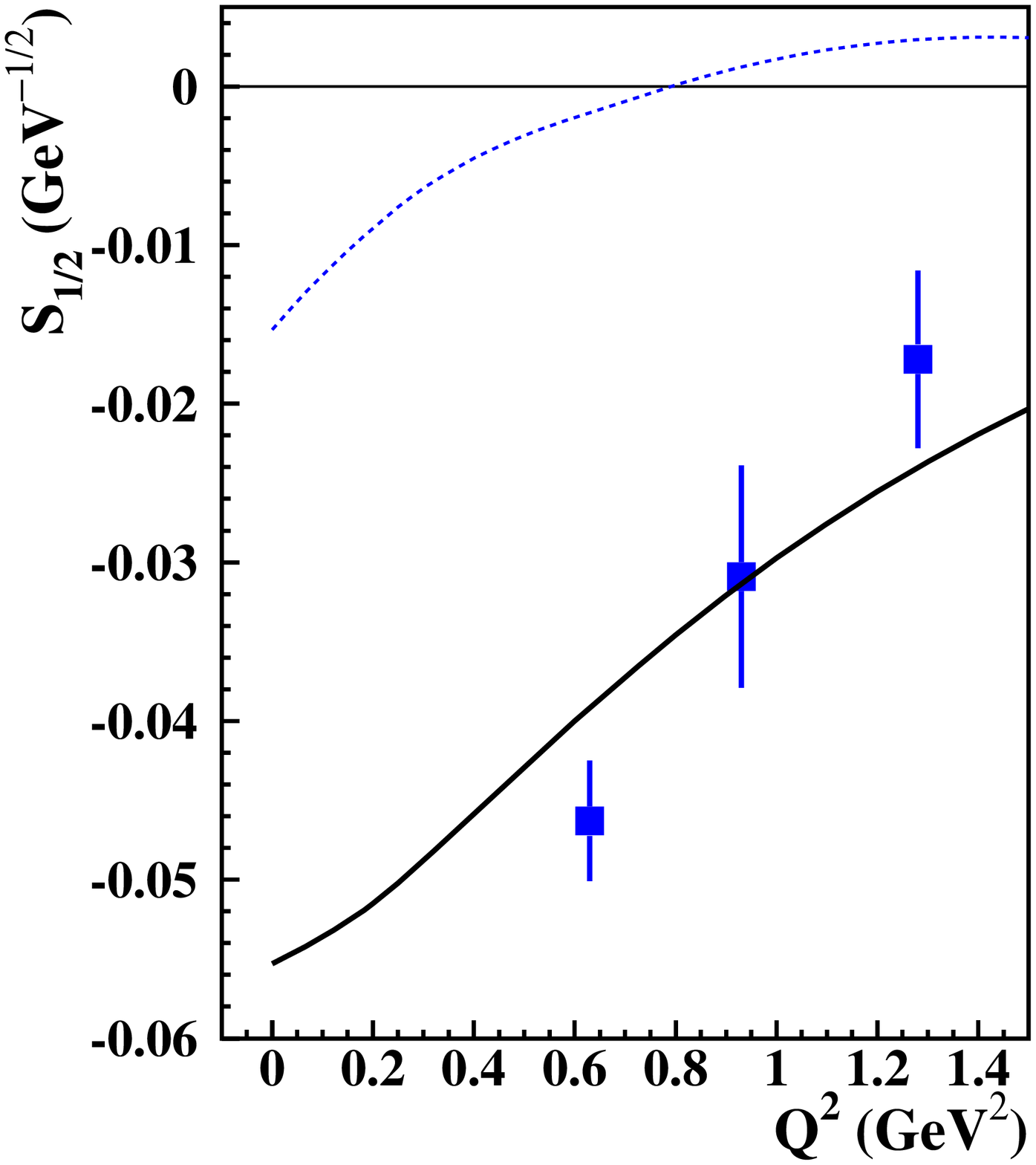}
\vspace{-0.1cm}
\caption{(Color Online) The first results on the $A_{1/2}$ and $S_{1/2}$ $\gamma_vpN^*$ electrocouplings of 
the $\Delta(1620)1/2^-$ resonance from the CLAS data on $\pi^+\pi^-p$ electroproduction off the proton
\cite{Ri03} in comparison with a hypercentral constituent quark model~\cite{Sa12} (thick black solid lines) 
and the Bethe-Salpeter approach~\cite{Met12} (blue dashed lines). The photocoupling value is taken from the
RPP~\cite{Rpp12}.} 
\label{s31qm}
\end{center} 
\end{figure*}

The CLAS data on the $\gamma_vpN^*$ electrocouplings of the $N(1440)1/2^+$ and $N(1520)3/2^-$ resonances, 
together with the results from previous studies~\cite{Az09} and recent analyses of the $N^*$ electroexcitations 
in the third resonance region~\cite{Park15,Az15a}, strongly suggest that the structure of nucleon resonances 
for $Q^2 < 5.0$~GeV$^2$ is determined by a complex interplay between the inner core of three dressed quarks 
bound to the states with the quantum numbers of the nucleon resonance and the external meson-baryon cloud. The 
quark core fully determines the spins and parities of the resonances, while the meson-baryon cloud affects the
resonance masses, electroexcitation amplitudes, and hadronic decay widths.

Access to the different components in the resonance structure represents a challenging objective. The credible 
DSEQCD evaluation of the quark core contributions to the electrocouplings of the $N(1440)1/2^+$ \cite{Cr15a} 
allows us to estimate the meson-baryon cloud contributions as the difference between the fit of the experimental 
results and the quark core electroexcitation amplitudes from DSEQCD~\cite{Cr15a}. The meson-baryon cloud contributions 
to the electrocouplings of the $N(1440)1/2^+$ state obtained in this way are shown in Fig.~\ref{p11a12s12dseqm} 
by the thick dashed magenta lines. The meson-baryon cloud contributions to the $A_{1/2}$ electrocouplings of 
the $N(1440)1/2^+$ are maximal for $Q^2 < 1.0$~GeV$^2$. At $Q^2 > 1.0$~GeV$^2$ they rapidly decrease with photon 
virtualities and become negligible for $Q^2 > 2.0$~GeV$^2$. The meson-baryon cloud contributions to the 
$S_{1/2}$ electrocouplings of the $N(1440)1/2^+$ show a rather slow $Q^2$-evolution for 
2.0~GeV$^2$ $< Q^2 <$ 5.0~GeV$^2$. The $S_{1/2}$ electrocouplings of the $N(1440)1/2^+$ are proportional to 
the difference 
\begin{equation}
\label{ropers12}
S_{1/2} \sim F_1^*-\frac{Q^2}{(M_R-M_N)^2}F_2^*  ,
\end{equation}
where $M_R$ and $M_N$ are the $N(1440)1/2^+$ and nucleon masses, respectively. For $Q^2 > 2.0$~GeV$^2$, 
the contributions from the quark core almost cancel out, making the $S_{1/2}$ electrocouplings of the
$N(1440)1/2^+$ more sensitive to the meson-baryon cloud contributions for $Q^2 > 2.0$~GeV$^2$.

The analysis of the CLAS data has revealed a substantial dependence of the meson-baryon cloud contributions on the 
quantum numbers of the excited nucleon states and the transition helicity amplitudes. The magnitudes of 
the meson-baryon dressing amplitudes for the $A_{1/2}$ electrocouplings of the $N(1520)3/2^-$ are much smaller 
than for either the $S_{1/2}$ or $A_{3/2}$ electrocouplings (see Fig.~\ref{d13elcoupl}), as well as with 
the $A_{1/2}$ electrocoupling for the $N(1440)1/2^+$ (see Fig.~\ref{p11a12s12dseqm}). This makes the $A_{1/2}$ 
electrocoupling of the $N(1520)3/2^-$ attractive for the studies of quark degrees of freedom in the structure 
of the $N(1520)3/2^-$ resonance. 

Studies of the parton content of excited nucleons have been already initiated by the Regensburg University 
theory group~\cite{Br09,Br14}. Recent developments in the Light-Cone-Sum-Rule (LCSR) approach allowed us 
for the first time to determine the partonic structure of the $N(1535)1/2^-$ resonance~\cite{Br15} from the 
CLAS experimental results on the electrocouplings of this state for $Q^2 > 2.0$~GeV$^2$~\cite{Az09}. The analysis 
of the $N(1520)3/2^-$ electrocouplings within the framework of the LCSR approach were carried out in Ref.~\cite{Al14}. 
However, this approach employs quark distribution amplitudes for the nucleon ground states only. Future LCSR 
evaluations of the $p \to N(1520)3/2^-$ electromagnetic transition amplitudes that incorporate the 
$N(1520)3/2^-$ quark distribution amplitudes are needed in order to explore the partonic structure of the
$N(1520)3/2^-$ resonance.

\subsection{$\Delta(1620)1/2^-$ Resonance}
\label{imps31}

The $\gamma_vpN^*$ electrocouplings and the partial $\pi \Delta$ and $\rho p$ hadronic decay widths (see 
Table~\ref{hps31}) of the $\Delta(1620)1/2^-$ resonance obtained for the first time from CLAS data on $\pi^+\pi^-p$ 
electroproduction off the proton~\cite{Ri03} have revealed very unusual properties of this state (Fig.~\ref{s31qm}). 
Currently it is the only well-established $N^*$ state produced via electroexcitation that is dominated by the 
longitudinal $S_{1/2}$ amplitude for 0.5~GeV$^2$ $< Q^2 <$ 1.5~GeV$^2$. The $\rho p$ channel opens for the central 
$\rho$ mass at the threshold of $W$=1.71~GeV. Despite of the much smaller central mass 1.62~GeV, the 
$\Delta(1620)1/2^+$ state has a large branching fraction for decays to the $\rho p$ final state as listed in 
Table~\ref{hps31}.

The attempts to describe the electrocouplings of the $\Delta(1620)1/2^-$ resonance within the framework of 
the constituent quark models, accounting for the contributions from only three dressed quarks in the first 
orbital excitation that belongs to the [70,1$^-$] $SU(6)$ spin-flavor multiplet, were not successful. As  
shown in Fig~\ref{s31qm}, the hypercentral constituent quark model~\cite{Sa12} does allow for a reasonable
description of the longitudinal electrocouplings, but it underestimates the transverse $A_{1/2}$ 
electrocouplings. Instead, the above-mentioned Bethe-Salpeter approach~\cite{Met12} offers a good description 
of the transverse $A_{1/2}$ electrocoupling of the $\Delta(1620)1/2^-$, but underestimates the longitudinal 
$S_{1/2}$ electrocouplings. The unquenched constituent quark models~\cite{Sa15a} currently employed in the 
studies of mesons offer a promising opportunity to explore both the hadron wave functions and the hadronic 
decays. The extension of these approaches into the baryon sector looks promising in order to understand the nature 
of the $\Delta(1620)1/2^-$ resonance from the combined analysis of the electroexcitation amplitudes and the 
hadronic decays of this resonance. 

The large branching fraction for the hadronic decays to the $\rho p$ final state of the deeply sub-threshold 
$\Delta(1620)1/2^-$ state makes it attractive to search for an admixture of exotic configurations such as 
$qqq(q\bar{q})$ that may facilitate the resonance decays to the $\rho p$ final state. 
  
\section{Summary and Outlook}
\label{concl}

Phenomenological analysis of CLAS data~\cite{Ri03} on $\pi^+\pi^-p$ electroproduction off the proton at 
invariant masses of the final hadron system 1.40~GeV $< W <$ 1.82~GeV and photon virtualities $Q^2$ from 
0.5~GeV$^2$ to 1.5~GeV$^2$ was carried out with the primary objective of determining the $\gamma_vpN^*$ 
resonance electrocouplings and their partial hadronic decay widths to the $\pi \Delta$ and $\rho p$ final states 
for all prominent $N^*$ states with masses below 1.64~GeV. The JM reaction model~\cite{Mo09,Mo12} previously 
employed for the extraction of the resonance parameters from $\pi^+\pi^-p$ electroproduction data~\cite{Fe09} was 
further developed for extraction of the resonance parameters in a wider area of $W$ and $Q^2$. In order to 
describe the data~\cite{Ri03} on the final hadron distributions over the $\alpha_i$ angles for $W > 1.5$~GeV, 
the phases of the direct double-pion electroproduction amplitudes were implemented and fit to the measured nine 
one-fold differential cross sections. The updated JM model provides a good description of all available CLAS 
data on $\pi^+\pi^-p$ electroproduction off the proton at 1.40~GeV $< W <$ 1.82~GeV and $Q^2$ from 0.5~GeV$^2$ 
to 1.5~GeV$^2$. The achieved quality of the data fit~\cite{Ri03} is comparable to that obtained in reaction 
models employed previously for extraction of the resonance electrocouplings from CLAS data on $N\pi$
\cite{Az09,Park15} and $\pi^+\pi^-p$~\cite{Mo12} electroproduction off the proton. The contributions to  
charged double-pion electroproduction off the proton from all relevant meson-baryon channels and direct double 
pion production mechanisms determined from CLAS data within the framework of the updated JM model, shown 
in Fig.~\ref{integsec}, are of interest for the future modeling of different exclusive meson electroproduction 
channels that are relevant in the resonance region. These results can also be used in global multi-channel 
analyses aimed at extraction of the resonance parameters from all available data on exclusive meson 
photo-, electro-, and hadroproduction.    

The $\gamma_vpN^*$ electrocouplings of the $N(1440)1/2^+$ and $N(1520)3/2^-$ resonances were determined from the
exclusive charged double-pion electroproduction cross sections measured with CLAS at $Q^2$ from 0.5~GeV$^2$ to 
1.5~GeV$^2$. Consistent values of the $N(1440)1/2^+$ and $N(1520)3/2^-$ electrocouplings obtained in 
independent analyses of three $W$-intervals, where the non-resonant contributions are different, strongly 
support the reliable extraction of these fundamental quantities. Furthermore, the hadronic decay widths of 
these resonances to the $\pi \Delta$ and $\rho p$ final states obtained in our analysis are consistent with 
those previously determined in this exclusive channel at smaller photon virtualities 
$Q^2 < 0.55$~GeV$^2$~\cite{Mo12}. Successful description of the CLAS $\pi^+\pi^- p$ electroproduction data
\cite{Fe09,Ri03} in a wide range of photon virtualities from 0.25~GeV$^2$ to 1.5~GeV$^2$ with $Q^2$-independent 
hadronic decay widths of the contributing resonances, supports a reliable separation between the resonant and 
non-resonant contributions achieved in the updated JM model and confirm reliable extraction of the resonance 
parameters. The $\Delta(1620)1/2^-$ resonance decays preferentially to the $N\pi\pi$ final state. The 
$\pi^+\pi^-p$ exclusive electroproduction off the proton represents the major source of information on the
electrocouplings of this resonance. Our studies provide for the first time information on the $\gamma_vpN^*$ 
electrocouplings and the $\pi \Delta$ and $\rho p$ partial hadronic decay widths of the $\Delta(1620)1/2^-$ 
resonance. 

Due to the recent progress in DSEQCD studies of excited nucleon states~\cite{Cr14,Cr15}, the first evaluations 
of the $p \to N(1440)1/2^+$ Dirac $F_1^*$ and Pauli $F_2^*$ transition form factors starting from the QCD 
Lagrangian have recently become available~\cite{Cr15a}. A good description of the CLAS experimental results 
was obtained at $Q^2 > 2.0$~GeV$^2$ in the DSEQCD approach. In this application the same momentum-dependent 
dressed quark mass function was employed that was also
used in the previous DSEQCD computations of the nucleon elastic~\cite{Cr13} and  
magnetic $p \to \Delta$ transition form factors~\cite{Cr15}. A successful description of 
the nucleon elastic and electromagnetic transition form factors to excited nucleon states of distinctly 
different structure strongly supports a reliable access to the dressed quark mass function achieved in the 
analysis~\cite{Cr15a}. Mapping out the dressed quark mass function from available and future data on 
$p \to N^*$ transition form factors will address the most challenging and still open problems of the Standard 
Model on the nature of the dominant part of the hadron mass, quark-gluon confinement, and their emergence from 
QCD~\cite{Az13,Cr14}. These prospects motivate the future studies of the excited nucleon state structure at 
high photon virtualities from 5~GeV$^2$ to 12~GeV$^2$ with the CLAS12 detector after the completion of the 
Jefferson Lab 12~GeV upgrade~\cite{Az13,Go12,Ca14,Temple15}. 

Analyses of the experimental results on the $\gamma_vpN^*$ electrocouplings of the $N(1440)1/2^+$, 
$N(1520)3/2^-$, and $\Delta(1620)1/2^-$ resonances in the entire range of photon virtualities covered by the 
measurements employing the DSEQCD approach~\cite{Cr15,Cr15a}, advanced quark models~\cite{Az12,Met12,Ob14}, and 
a global multi-channel analysis~\cite{Lee10,Lee091,Lee08}, have convincingly demonstrated that their structure 
at $Q^2 < 5.0$~GeV$^2$ is determined by a complex interplay between the inner core of three dressed quarks and 
the external meson-baryon cloud. A successful description of the quark core contributions to the electrocouplings 
of the $N(1440)1/2^+$ resonance within the framework of DSEQCD~\cite{Cr15a} makes it possible to outline  
meson-baryon cloud contributions for this state at the resonant point ($W=M_{N^*}$) from the experimental results 
on the $\gamma_vpN^*$ electrocouplings. We observed pronounced differences for the meson-baryon cloud contributions 
to different electroexcitation amplitudes and their strong dependence on the quantum numbers of the excited nucleon 
state and photon virtuality. In particular, small contributions from the meson-baryon cloud to the $A_{1/2}$ 
electrocouplings of the $N(1520)3/2^-$ make this resonance attractive for the exploration of its quark components. 
The studies of resonance electrocouplings over the full spectrum of excited nucleon states of different quantum 
numbers are critical in order to explore different components in the $N^*$ structure.
  
Available for the first time, $\Delta(1620)1/2^-$ resonance electrocouplings and hadronic decay widths to the 
$\pi \Delta$ and $\rho p$ final states have demonstrated a rather peculiar behavior. The $\Delta(1620)1/2^-$ 
state is the only known resonance produced via electroexcitation that is dominated by the longitudinal $S_{1/2}$ 
electrocoupling in a wide range of photon virtualities 0.5~GeV$^2$ $< Q^2 <$ 1.5~GeV$^2$. Furthermore, the
$\Delta(1620)1/2^-$ resonance  has a large branching fraction (above 30\%) for the decay into the $\rho N$ final 
states. This is a rather unusual feature for decays of a resonance located in the deeply sub-threshold region for 
the production of the $\rho p$ final state. Failures in describing the $\Delta(1620)1/2^-$ electrocouplings within 
the framework of quark models that account for the contributions of the quark core only~\cite{Sa12,Met12}, indicate 
that the structure of this state can be more complex than that assumed in quark models described by the orbital 
excitation of three quarks with the total orbital momentum $L=1$. Further experimental data are needed in order to 
establish the nature of this state. In the near term future, new CLAS results on the $\Delta(1620)1/2^-$ 
electrocouplings are expected at photon virtualities from 0.3~GeV$^2$ to 1.0~GeV$^2$ with a much finer $Q^2$-binning 
\cite{Fe12}. The results on the $\Delta(1620)1/2^-$ electrocouplings from CLAS data on charged double-pion 
electroproduction off the proton eventually will also be extended in $Q^2$ up to 5.0~GeV$^2$. Further developments in 
hadron structure theory that will allow us to perform a combined analysis of the resonance electrocouplings and 
hadronic decay widths are critical in order to understand the nature of the $\Delta(1620)1/2^-$ state. A search for 
contributions from exotic $qqq(q\bar{q})$ configurations to the structure of this state that may facilitate the decays 
of the $\Delta(1620)1/2^-$ resonance to the $\rho p$ final state are of particular interest. 
  
\section{Acknowledgments}

We would like to acknowledge the outstanding efforts of the staff of the Accelerator and the Physics Divisions 
at Jefferson Lab that made this evaluation of the $N(1440)1/2^+$, $N(1520)3/2^-$, and $\Delta(1620)1/2^-$ 
electrocouplings and hadronic decay parameters possible. We are grateful to I.G. Aznauryan, V.M. Braun, 
I.C. Cl\"{o}et, M.M. Giannini, T-S.~H. Lee, M.R. Pennington, C.D. Roberts, E. Santopinto, J. Segovia, and 
A.P. Szczepaniak for theoretical support and helpful discussions. This work was supported in part by the U.S. 
Department of Energy and the National Science Foundation, the Skobeltsyn Institute of Nuclear Physics and the
Physics Department at Moscow State University, Ohio University, and the University of South Carolina. The 
Southeastern Universities Research Association (SURA) operates the Thomas Jefferson National Accelerator 
Facility for the United States Department of Energy under contract DE-AC05-84ER40150.

\include{appendix1}
\end{document}